\pdfoutput=1
\documentclass[usenatbib]{mn2e}
\usepackage{apjfonts}
\usepackage{amssymb}
\usepackage{amsmath}
\usepackage{ctable}
\usepackage{fixltx2e} 

\newcommand{\be}{\begin{equation}}
\newcommand{\ee}{\end{equation}}

\newcommand{\rhogas}{\rho_{\rm g}}
\newcommand{\grain}{_{\rm p}}
\newcommand{\rhograin}{\rho\grain}

\newcommand{\mach}{\mathcal{M}}
\newcommand{\machcompressive}{\mach_{\rm c}}
\newcommand{\cs}{c_{s}}
\newcommand{\tstop}{t_{\rm s}}
\newcommand{\taustop}{\tau_{\rm s}}
\newcommand{\eddy}{_{e}}

\newcommand{\Veddy}{v\eddy}
\newcommand{\Teddy}{t\eddy}
\newcommand{\vdrift}{v_{\rm drift}}
\newcommand{\tdrift}{t_{\rm cross}}
\newcommand{\Ndim}{N_{\rm d}}

\newcommand{\vk}{V_{K}}
\newcommand{\deltat}{\delta t}
\newcommand{\DN}{\Delta N}
\newcommand{\deltarhonoabs}{\delta \ln{\rho}}
\newcommand{\deltarho}{|\deltarhonoabs|}
\newcommand{\deltarhomean}{\langle\deltarhonoabs\rangle}
\newcommand{\scalevar}{\lambda}
\newcommand{\velslope}{\zeta_{1}}
\newcommand{\deltadim}{C_{\infty}}
\newcommand{\Lcrit}{\scalevar_{\rm crit}}
\newcommand{\Lmax}{\scalevar_{\rm max}}
\newcommand{\Leddy}{\scalevar\eddy}
\newcommand{\rhoratio}{\tilde{\rho}}
\newcommand{\deltazero}{|\delta_{0}|}
\newcommand{\taustopeddy}{\tilde{\tau}_{\rm s}}
\newcommand{\taustopP}{\tau_{\rm s,\,\rho}}
\newcommand{\taustopeddyP}{\tilde{\tau}_{\rm s,\,\rho}}

\newcommand\plotonesize[2]
 {\centering \leavevmode \includegraphics[width={#2\columnwidth}]{#1}}

\newcommand{\acknowledgments}{\begin{small}\section*{Acknowledgments}\end{small}}
\newcommand\altaffilmark[1]{$^{#1}$}
\newcommand\altaffiltext[1]{$^{#1}$}
\title[Turbulent Grain Clustering]{A Simple Phenomenological Model for Grain Clustering in Turbulence\vspace{-0.5cm}}

\author[Hopkins]{
\parbox[t]{\textwidth}{ 
Philip F. Hopkins\altaffilmark{1,2}\thanks{E-mail:phopkins@caltech.edu}} 
\vspace*{6pt} \\
\altaffiltext{1}{TAPIR, Mailcode 350-17, California Institute of Technology, Pasadena, CA 91125, USA} \\
\altaffiltext{2}{Department of Astronomy and Theoretical Astrophysics Center, University of California Berkeley, Berkeley, CA 94720\vspace{-1.1cm}} \\
}

\date{Submitted to MNRAS, August, 2013\vspace{-0.6cm}}
\begin{document}
\maketitle
\label{firstpage}

\begin{abstract}
\vspace{-0.2cm}

We propose a simple model for density fluctuations of aerodynamic grains, embedded in a turbulent, gravitating gas disk. The model combines a calculation for the behavior of a group of grains encountering a single turbulent eddy, with a hierarchical approximation of the eddy statistics. This makes analytic predictions for a range of quantities including: distributions of grain densities, power spectra and correlation functions of fluctuations, and maximum grain densities reached. We predict how these scale as a function of grain drag time $\tstop$, spatial scale, grain-to-gas mass ratio $\rhoratio$, strength of turbulence $\alpha$, and detailed disk properties. We test these against numerical simulations with various turbulence-driving mechanisms. The simulations agree well with the predictions, spanning $\tstop\,\Omega\sim10^{-4}-10$, $\rhoratio\sim0-3$, $\alpha\sim10^{-10}-10^{-2}$. Results from ``turbulent concentration'' simulations and laboratory experiments are also predicted as a special case. Vortices on a wide range of scales disperse and concentrate grains hierarchically. For small grains this is most efficient in eddies with turnover time comparable to the stopping time, but fluctuations are also damped by local gas-grain drift. For large grains, shear and gravity lead to a much broader range of eddy scales driving fluctuations, with most power on the largest scales. The grain density distribution has a log-Poisson shape, with fluctuations for large grains up to factors $\gtrsim1000$. We provide simple analytic expressions for the predictions, and discuss implications for planetesimal formation, grain growth, and the structure of turbulence.

\end{abstract}

\begin{keywords}
planets and satellites: formation --- protoplanetary discs --- accretion, accretion disks --- hydrodynamics --- instabilities --- turbulence
\vspace{-1.0cm}
\end{keywords}

\vspace{-1.1cm}
\section{Introduction}
\label{sec:intro}

Dust grains and aerodynamic particles are fundamental in astrophysics. These determine the attenuation and absorption of light in the interstellar medium (ISM), interaction with radiative forces and regulation of cooling, and form the building blocks of planetesimals. Of particular importance is the question of grain clustering and clumping -- fluctuations in the local volume-average number/mass density of grains $\rhograin$ -- in turbulent gas. 

Much attention has been paid to the specific question of grain density fluctuations and grain concentration in proto-planetary disks. In general, turbulence sets a ``lower limit'' to the degree to which grains can settle into a razor-thin sub-layer; and this has generally been regarded as a barrier to planetesimal formation \citep[though see][and references therein]{goodman.pindor:2000.secular.drag.instabilities.grains,lyra:2009.semianalytic.planet.form.model.grain.settling,lee:2010.grain.settling.vs.grav.instability,chiang:2010.planetesimal.formation.review}. However, it is also well-established that the number density of solid grains can fluctuate by multiple orders of magnitude when ``stirred'' by turbulence, even in media where the turbulence is highly sub-sonic and the gas is nearly incompressible \citep[see e.g.][]{bracco:1999.keplerian.largescale.grain.density.sims,cuzzi:2001.grain.concentration.chondrules,johansen:2007.streaming.instab.sims,carballido:2008.large.grain.clustering.disk.sims,bai:2010.grain.streaming.sims.test,bai:2010.streaming.instability,bai:2010.grain.streaming.vs.diskparams,pan:2011.grain.clustering.midstokes.sims}. This can occur via self-excitation of turbulent motions in the ``streaming'' instability \citep{youdin.goodman:2005.streaming.instability.derivation}, or in externally driven turbulence, such as that excited by the magneto-rotational instability (MRI), global gravitational instabilities, or convection \citep{dittrich:2013.grain.clustering.mri.disk.sims,jalali:2013.streaming.instability.largescales}. Direct numerical experiments have shown that the magnitude of these fluctuations depends on the parameter $\taustop=\tstop\,\Omega$, the ratio of the gas ``stopping'' time (friction/drag timescale) $\tstop$ to the orbital time $\Omega^{-1}$, with the most dramatic fluctuations around $\taustop\sim1$. These experiments have also demonstrated that the magnitude of clustering depends on the volume-averaged ratio of solids-to-gas ($\rhoratio\equiv \rhograin/\rhogas$), and basic properties of the turbulence (such as the Mach number). These have provided key insights and motivated considerable work studying these instabilities; however, the fraction of the relevant parameter space spanned by direct simulations is limited. Moreover, it is impossible to simulate anything close to the full dynamic range of turbulence in these systems: the ``top scales'' of the system are $\Lmax\sim$\,AU, while the viscous/dissipation scales $\scalevar_{\nu}$ of the turbulence are $\scalevar_{\nu}\sim$\,m (Reynolds numbers $Re\sim10^{6}-10^{9}$, under typical circumstances). Reliably modeling $Re\gtrsim10^{4}$ remains challenging in state-of-the-art simulations \citep[see e.g.][]{federrath:2013.intermittency.vs.numerics}. Clearly, some analytic model (even a very approximate one) for these fluctuations would be tremendously helpful.

The question of ``preferential concentration'' of aerodynamic particles is actually much more well-studied in the terrestrial turbulence literature. There both laboratory experiments \citep{squires:1991.grain.concentration.experiments,fessler:1994.grain.concentration.experiments,rouson:2001.grain.concentration.experiment,gualtieri:2009.anisotropic.grain.clustering.experiments,monchaux:2010.grain.concentration.experiments.voronoi} and numerical simulations \citep{cuzzi:2001.grain.concentration.chondrules,yoshimoto:2007.grain.clustering.selfsimilar.inertial.range,hogan:2007.grain.clustering.cascade.model,bec:2009.caustics.intermittency.key.to.largegrain.clustering,pan:2011.grain.clustering.midstokes.sims,monchaux:2012.grain.concentration.experiment.review} have long observed that very small grains, with Stokes numbers $St\equiv \tstop/\Teddy(\scalevar_{\nu})\sim1$ (ratio of stopping time to eddy turnover time at the viscous scale) can experience order-of-magnitude density fluctuations at small scales (at/below the viscous scale). Considerable analytic progress has been made understanding this regime: demonstrating, for example, that even incompressible gas turbulence is unstable to the growth of inhomogeneities in grain density \citep{elperin:1996:grain.clustering.instability,elperin:1998.grain.clustering.instability.rotation}, and predicting the behavior of the small-scale grain-grain correlation function using simple models of gaussian random-field turbulence \citep{sigurgeirsson:2002.grain.markovian.concentration.toymodel,bec:2007.grain.clustering.markovian.flow}. But extrapolation to the astrophysically relevant regime is difficult for several reasons: the Reynolds numbers of interest are much larger, and as a result the Stokes numbers are also generally much larger (in the limit where grains do not cluster below the viscous/dissipation scale because $\tstop\gg \Teddy(\Lmax)$), placing the interesting physics well in the inertial range of turbulence, and rotation/shear, external gravity, and coherent (non-random field) structures appear critical (at least on large scales). This parameter space has not been well-studied, and at least some predictions (e.g.\ those in \citet{sigurgeirsson:2002.grain.markovian.concentration.toymodel,bec:2008.markovian.grain.clustering.model,zaichik:2009.grain.clustering.theory.randomfield.review}) would naively lead one to estimate much smaller fluctuations than are recorded in the experiments above. 

However, these studies still contribute some critical insights. They have repeatedly shown that grain density fluctuations are tightly coupled to the local vorticity field: grains are ``flung out'' of regions of high vorticity by centrifugal forces, and collect in the ``interstices'' (regions of high strain ``between'' vortices). Studies of the correlation functions and scaling behavior of higher Stokes-number particles suggest that, in the inertial range (ignoring gravity and shear), the same dynamics apply, but with the scale-free replacement of a ``local Stokes number'' $\tstop/\Teddy$, i.e.\ what matters for the dynamics on a given scale are the vortices of that scale, and similar concentration effects can occur whenever the eddy turnover time is comparable to the stopping time \citep[e.g.][]{yoshimoto:2007.grain.clustering.selfsimilar.inertial.range,bec:2008.markovian.grain.clustering.model,wilkinson:2010.randomfield.correlation.grains.weak,gustavsson:2012.grain.clustering.randomflow.lowstokes}. Several authors have pointed out that this critically links grain density fluctuations to the phenomenon of intermittency and discrete, time-coherent structures (vortices) on scales larger than the Kolmogorov scale in turbulence \citep[see][and references therein]{bec:2009.caustics.intermittency.key.to.largegrain.clustering,olla:2010.grain.preferential.concentration.randomfield.notes}. In particular, \citet{cuzzi:2001.grain.concentration.chondrules} argue that grain density fluctuations behave in a multi-fractal manner: multi-fractal scaling is a key signature of well-tested, simple geometric models for turbulence \citep[e.g.][]{sheleveque:structure.functions,boldyrev:2002.structfn.model,schmidt:2008.turb.structure.fns}. In these models, the statistics of turbulence are approximated by regarding the turbulent field as a hierarchical collection of ``stretched'' singular, coherent structures (e.g.\ vortices) on different scales \citep{dubrulle:logpoisson,shewaymire:logpoisson,chainais:2006.inf.divisible.cascade.review}. Such statistical models have been well-tested as a description of the {\em gas} turbulence statistics \citep[including gas density fluctuations; see e.g.][]{burlaga:1992.multifractal.solar.wind.density.velocity,sorriso-valvo:1999.solar.wind.intermittency.vs.time,budaev:2008.tokamak.plasma.turb.pdfs.intermittency,shezhang:2009.sheleveque.structfn.review,hopkins:2012.intermittent.turb.density.pdfs}. However, only first steps have been taken to link them to grain density fluctuations: for example, in the phenomenological cascade model fit to simulations in \citet{hogan:2007.grain.clustering.cascade.model}. 

In this paper, we use these theoretical and experimental insights to build a simple, phenomenological model which attempts to ``bridge'' between the well-studied regime of small-scale turbulence and that of large, astrophysical particles in shearing, gravitating disks. The key concepts are based on the work above: we first assume that grain density fluctuations are driven by coherent eddies, for which we can calculate the perturbation owing to a single eddy with a given scale. Building on \citet{cuzzi:2001.grain.concentration.chondrules} and others, we then attach this calculation to some simple fractal-like (self-similar) assumptions for the statistics of eddies. This allows us to make predictions for a wide range of quantities, which we compare to simulations and experiments.

\begin{footnotesize}
\ctable[
  caption={{\normalsize Important Variables \&\ Key Equations Derived in This Paper}\label{tbl:defns}},center,star
  ]{lll}{
}{
\hline\hline
\multicolumn{1}{l}{Variable} &
\multicolumn{1}{l}{Definition} & 
\multicolumn{1}{l}{Eq.} \\ 
\hline
$\rhogas$, $\cs$ & mid-plane gas density and sound speed & -- \\
$R$, $\Omega_{R}$, $\vk$ & distance from center of gravitational potential, Keplerian orbital frequency at $R$, and Keplerian velocity ($\vk\equiv \Omega_{R}\,R$) & -- \\
$\Leddy$, $\Veddy$, $\mach\eddy$, $\Teddy$ & characteristic spatial scale, velocity, Mach number ($\mach\eddy\equiv |\Veddy|/\cs$) and turnover time ($\Teddy\equiv \Leddy/|\Veddy|$) of a turbulent eddy & -- \\
$\Lmax$, $\Veddy(\Lmax)$, $\alpha$ & maximum or ``top''/driving scale of turbulence, with eddy velocity $\Veddy(\Lmax) \equiv \alpha^{1/2}\,\cs$ & --\\
$\scalevar_{\nu}$, $Re$, $St$ & viscous/Kolmogorov or ``bottom'' scale of turbulence; Reynolds number $Re\equiv(\Lmax/\scalevar_{\nu})^{4/3}$; and Stokes $St\equiv \tstop/\Teddy(\scalevar_{\nu})$
& --\\
$\rhoratio$ & mean ratio of the volume-average density of solids to gas, in the midplane ($\rhoratio \equiv \langle \rhograin \rangle / \langle \rhogas \rangle$) & -- \\
$\taustop$ & dimensionless particle stopping time ($\taustop\equiv \tstop\,\Omega_{R}$) & \ref{eqn:eom.peculiar} \\
$\taustopeddy$ & ratio of particle stopping time to eddy turnover time ($\taustopeddy\equiv \tstop/\Teddy = \taustopeddy(\Lmax)\,(\Leddy/\Lmax)^{1-\velslope}$) & -- \\
$\eta,\,\Pi$ & difference between the mean gas circular velocity and Keplerian ($\eta\,\vk \equiv \vk - \langle V_{\rm gas} \rangle$; 
$\Pi \equiv \eta\,\vk/\cs$) & \ref{eqn:eta} \\
$\vdrift$ & mean grain-gas relative drift velocity: 
$\vdrift \equiv {2\,\eta\,\vk\,\taustop\,[{(1+\rhoratio)^{2}+\taustop^{2}/4}}]^{1/2}\,[{\taustop^{2}+(1+\rhoratio)^{2}}]^{-1}$ 
& \ref{eqn:vdrift} \\
$\deltadim$ & filling factor of eddies: $\deltadim\sim1-2$ is plausible 
 & \ref{eqn:deltaN} \\
$\velslope$ & scaling of one-point gas eddy velocity statistics, $\langle|\Veddy|\rangle\propto \Leddy^{\velslope}$ & \ref{eqn:veddy.scaling} \\
 & in the multi-fractal models used: $\displaystyle \velslope \approx \frac{1}{9} + 2\,{\Bigl[}1 - {\Bigl(}\frac{2}{3} {\Bigr)}^{1/3} {\Bigr]}$ & \ref{eqn:velslope} \\
$\Ndim$ & ``wrapping dimension'' of the singular eddy structures driving density fluctuations & \ref{eqn:shrink.eddy} \\
 & ($\Ndim=2$ for simple vortices in the disk plane) & \\
\hline
\\
\hline
\multicolumn{1}{l}{--} &
\multicolumn{1}{l}{Useful variables for Equations below:} & 
\multicolumn{1}{l}{Eq.} \\  
& &  \\
$\beta$ 
& $\displaystyle \beta \equiv  \frac{|\Veddy(\Lmax)|}{|\vdrift|} = 
\frac{|\Veddy(\Lmax)|\,[(1+\rhoratio)^{2}+\taustop^{2}]}{2\,\eta\,\vk\,\taustop\,[{(1+\rhoratio)^{2}+\taustop^{2}/4}]^{1/2}} = 
\frac{(1+\rhoratio)^{2}+\taustop^{2}}{2\,\taustop\,[{(1+\rhoratio)^{2}+\taustop^{2}/4}]^{1/2}}\,{\Bigl(}\frac{\alpha^{1/2}}{\Pi}{\Bigr)}$
& \ref{eqn:v.laminar} \\
& &   \\
$\deltarhomean$ 
& $\displaystyle \deltarhomean \equiv -\frac{\Ndim\,\varpi(\taustop,\,\taustopeddy)}{1+h(\Leddy)^{-1}} $ & \\
& & \\
& $\displaystyle h(\Leddy)\equiv -\taustopeddy\,\ln{{\Bigl[}1 - \frac{(\Leddy/\Lmax)}{\taustopeddy(\Lmax)\,g(\Leddy)^{1/2}} {\Bigr]}}$, \ \ \ \ \ 
$\displaystyle g(\Leddy)\equiv \frac{1}{\beta^{2}} + \taustopeddy(\Lmax)\,\ln{{\Bigl[} \frac{1+\taustopeddy(\Lmax)^{-1}}{1+\taustopeddy^{-1}} {\Bigr]}}$, \ 
& \ref{eqn:g.timescale.function} \\
& & \\
\multicolumn{1}{l}{} &
\multicolumn{1}{l}{(Approximation for $\varpi$: for exact solution see Appendix \ref{sec:appendix:exact}):} & 
\\  
& & \\
& $\displaystyle \varpi = {\rm MAX}{\Bigl[}\varpi_{1},\ \varpi_{0}\equiv 2\,\taustopP\,(1 + \taustopP^{2})^{-1} {\Bigr]}$\ \ \ \ \ \ \ \ \ \ 
[\ $\taustopP\equiv {\taustop}\,({1+\rhoratio})$\ \ \ \ \ $\taustopeddyP\equiv\taustopeddy\,(1+\rhoratio)$\ ]
& \ref{eqn:exact}  \\
& & \\
& $0 = 
16\,\taustopeddyP^{3}\,\varpi_{1}^{4} + 
32\,\taustopeddyP^{2}\,\varpi_{1}^{3} + 
\taustopeddyP\,(20+7\,\taustopP^{2})\,\varpi_{1}^{2}   
+ 4\,(1 + \taustopP^{2} - 3\,\taustopP\,\taustopeddyP)\,\varpi_{1} - 
4\,(\taustopeddyP+2\,\taustopP)$ & \ref{eqn:varpi.full} \\
& & \\
\hline
\multicolumn{1}{l}{$\rho_{\rm p,\,max}$:} &
\multicolumn{1}{l}{Maximum local density of grains $\rhograin$:} & 
\multicolumn{1}{l}{Eq.} \\  
& &  \\
& $\displaystyle \ln{{\Bigl(}\frac{\rho_{\rm p,\,max}}{\langle \rhograin \rangle} {\Bigr)}}  
= \deltadim\,\int_{\scalevar=0}^{\Lmax}\,[1-\exp{(-\deltarho)}]\,{\rm d}\ln{\scalevar}$ 
& \ref{eqn:rhomax} \\
& &  \\
\hline
\multicolumn{1}{l}{$\displaystyle \Delta^{2}(k) = \frac{{\rm d}S_{\ln{\rho}}}{{\rm d}\ln{\scalevar}}$:} &
\multicolumn{1}{l}{(Volume-Weighted) Grain log-density power spectrum (versus scale $\scalevar$):} & 
\multicolumn{1}{l}{Eq.} \\  
& &  \\
& $\displaystyle \Delta_{\ln{\rho}}^{2}{\Big(}k\equiv\frac{1}{\scalevar}{\Bigr)} 
= \deltadim\,\deltarho^{2}$
& \ref{eqn:pwrspec} \\
& &  \\
\hline
\multicolumn{1}{l}{$P_{V}(\ln{\rhograin})$:} &
\multicolumn{1}{l}{(Volume-weighted) Distribution of Grain Densities $\rhograin$:} & 
\multicolumn{1}{l}{Eq.} \\  
& &  \\
& $\displaystyle 
P_{V}(\ln{\rhograin})\,{\rm d}\ln{\rhograin} \approx 
\frac{(S^{-1}\,\mu^{2})^{m^{\prime}}\,\exp{(-S^{-1}\,\mu^{2})}}{\Gamma(m^{\prime}+1)} \, \frac{\mu}{S}\, {\rm d}\ln{\rhograin}$
& \ref{eqn:pdf.S}-\ref{eqn:deltarho.int} \\
& &  \\
& $\displaystyle m^{\prime} \equiv \frac{\mu}{S}\,{\Bigl\{} \frac{\mu^{2}}{S}\,{\Bigl[}1-\exp{{\Bigl(}-\frac{S}{\mu} {\Bigr)}}{\Bigr]}
- \ln{{\Bigl(}\frac{\rhograin}{\langle \rhograin \rangle} {\Bigr)}} {\Bigr\}}$, 
\ \ \ \  $\displaystyle \mu \equiv \deltadim\,\int\deltarho\,{\rm d}\ln{\scalevar}$, 
\ \ \ \ $\displaystyle S \equiv \deltadim\,\int\deltarho^{2}\,{\rm d}\ln{\scalevar}$
& \\
& &  \\
\multicolumn{1}{l}{$P_{M}(\ln{\rhograin})$:} &
\multicolumn{1}{l}{(Mass/particle-weighted) Distribution of Grain Densities $\rhograin$:} & 
\multicolumn{1}{l}{\,} \\  
& &  \\
& $\displaystyle 
P_{M}(\ln{\rhograin})\,{\rm d}\ln{\rhograin} = \rhograin\,P_{V}(\ln{\rhograin})\,{\rm d}\ln{\rhograin}$
& -- \\
& &  \\
\hline\hline\\
& &  \\
& &  \\
& &  \\
& &  \\
& &  \\
& &  \\
& &  \\
& &  \\
}
\end{footnotesize}

\begin{footnotesize}
\ctable[
  caption={{\normalsize Approximations for Large Scales and/or Grains (Appendices~\ref{sec:appendix:exact}-\ref{sec:appendix:largescale})}\label{tbl:largescale}},center,star
  ]{lll}{
}{
\hline\hline
\multicolumn{1}{l}{--} &
\multicolumn{1}{l}{Useful variables:} & 
\multicolumn{1}{l}{} \\  
& &  \\
$\varpi,\,\deltazero,\,\Lcrit$ & $\displaystyle \varpi \sim 2\,\phi\,\frac{\taustopP}{1 + \taustopP^{2}}$\ \ \ ($\phi\sim0.8$), \ \ \ \ \ \ \ \ \ \ \ \ 
$\displaystyle \deltazero \equiv \Ndim\,\varpi \sim 2\,\Ndim\,\phi\,\frac{\taustopP}{1 + \taustopP^{2}}$\ , \ \ \ \ \ \ \ \ \ \ \ \ 
$\displaystyle \Lcrit \equiv \beta^{-1/\velslope}\,\Lmax$\ , \ \ \ \ \ \ \ \ \ \ \ \ 
$\displaystyle \zeta_{1}=0.36$
 &  \\
& &  \\
$\deltarhomean$ & $\displaystyle \langle\deltarho\rangle = \frac{\Ndim\,\varpi}{1+h(\Leddy)^{-1}} 
\sim \frac{2\,\Ndim\,\phi}{\taustopP+\taustopP^{-1}}\,{\Bigl[}1 + \beta^{-1}\,{\Bigl(} \frac{\Leddy}{\Lmax} {\Bigr)}^{-\velslope} {\Bigr]}^{-1}
= \deltazero\,{\Bigl[}1 + (\Leddy/\Lcrit)^{-\velslope}{\Bigr]}^{-1}
$ & \\
& & \\
\hline
\multicolumn{1}{l}{$\rho_{\rm p,\,max}$:} &
\multicolumn{1}{l}{Maximum local density of grains $\rhograin$:} & 
\multicolumn{1}{l}{} \\  
& &  \\
& $\displaystyle \ln{{\Bigl(}\frac{\rho_{\rm p,\,max}(\scalevar\rightarrow0)}{\langle \rhograin \rangle} {\Bigr)}}  
\sim \frac{\deltadim}{\velslope}\,\frac{\deltazero}{1+\deltazero}\,
\ln{{\Bigl[}1 + \beta\,(1+\deltazero) + \beta^{3/2}\{(1 + \deltazero^{2})^{1/2}-1\}
{\Bigr]}}$ 
&   \\
& &  \\
& $\displaystyle \rho_{\rm p,\,max}(\scalevar) \propto \scalevar^{-\gamma}$, with $\displaystyle \gamma \sim 
\begin{cases}
{\displaystyle \deltadim\,[1 - \exp{(-\deltazero)]}}\ \ \ \ \hfill {\tiny (\scalevar \gg \Lcrit)} \\ 
\\
{\displaystyle \deltadim\,\deltazero\,(\scalevar/\Lcrit)^{\velslope}}\ \ \ \ \hfill {\tiny (\scalevar \ll \Lcrit)} \\ 
\end{cases}
$ &  \\
& &  \\
\hline
\multicolumn{1}{l}{$\displaystyle \Delta^{2}(k)$:} &
\multicolumn{1}{l}{(Volume-Weighted) Grain linear-density and log-density power spectrum (versus scale $\scalevar$):} & 
\multicolumn{1}{l}{} \\  
& &  \\
& $\displaystyle \Delta_{\ln{\rho}}^{2}{\Big(}k\equiv\frac{1}{\scalevar}{\Bigr)} 
\sim {\deltadim\,\deltazero^{2}}{\Bigl[}{1 + (\scalevar/\Lcrit)^{-\velslope}} {\Bigr]}^{-2} $
&   \\
& &  \\
& $\displaystyle \Delta_{{\rho}}^{2} \sim
\begin{cases}
{\displaystyle \deltadim\,\deltazero^{2}}\ \  \hfill {\tiny (\scalevar \gg \Lcrit,\ \Delta_{\ln{\rho}}^{2}\ll 1)} \\ 
\\
{\displaystyle \deltadim\,\deltazero^{2}\,(\scalevar/\Lcrit)^{2\,\velslope}}\ \ \hfill {\tiny (\scalevar \ll \Lcrit,\ \Delta_{\ln{\rho}}^{2}\ll 1)} \\ 
\end{cases}
$
\ \ \
$\displaystyle\sim
\begin{cases}
{\displaystyle \deltadim\,(\scalevar/\Lmax)^{-\deltadim}}\ \ \hfill {\tiny (\scalevar \gg \Lcrit,\ \deltazero\gg1)} \\ 
\\
{\displaystyle 2\,\deltadim\,e^{\Delta N_{\rm int}}\,\frac{\deltazero}{\deltarho_{\rm int}}\,(\scalevar/\Lcrit)^{\velslope}}\ \ \hfill {\tiny (\scalevar \ll \Lcrit,\ \deltazero\gg1)} \\ 
\end{cases}
$
&  \\
& &  \\
\hline
\multicolumn{1}{l}{$P_{V}(\ln{\rhograin})$:} &
\multicolumn{1}{l}{(Volume-weighted) Distribution of Grain Densities $\rhograin$:} & 
\multicolumn{1}{l}{} \\  
& &  \\
& $\displaystyle 
P_{V}(\ln{\rhograin})\,{\rm d}\ln{\rhograin} \approx 
\frac{(\Delta N_{\rm int})^{m^{\prime}}\,\exp{(-\Delta N_{\rm int})}}{\Gamma(m^{\prime}+1)} \, \frac{{\rm d}\ln{\rhograin}}{\deltarho_{\rm int}}$\ , 
\ \ \ \ \ \ \ \ \ \ 
$\displaystyle m^{\prime} = \deltarho_{\rm int}^{-1}\,{\Bigl\{}\Delta N_{\rm int}\,{\Bigl[}1 - \exp{(-\deltarho_{\rm int})} {\Bigr]} - \ln{{\Bigl(} \frac{\rhograin}{\langle \rhograin \rangle}{\Bigr)}} {\Bigr\}}$
& \\
& &  \\
& $\displaystyle \Delta N_{\rm int} = \frac{\mu^{2}}{S}$, \ \ \ \ $\displaystyle \deltarho_{\rm int} = \frac{S}{\mu}$, 
\ \ \ \ $\displaystyle \mu \sim \frac{\deltadim}{\velslope}\,\deltazero\,\ln{(1+\beta)}$, 
\ \ \ \ $\displaystyle S \sim \frac{\deltadim}{\velslope}\,\deltazero^{2}\,{\Bigl(}\ln{(1+\beta)} - \frac{\beta}{1+\beta}{\Bigr)}$
& \\
& &  \\
\hline\hline\\
}
\end{footnotesize}

\vspace{-0.5cm}
\section{Arbitrarily Small Grains: Pure Gas Density Fluctuations}
\label{sec:smallgrains}

First consider the case where the grains are perfectly coupled to the gas ($\tstop\rightarrow0$), and their volume-average mass density ($\rhograin$, as distinct from the {\em internal} physical density of a single, typical grain) is small compared to the gas density $\rhogas$, so grain density fluctuations simply trace gas density fluctuations.

In both sub-sonic and super-sonic turbulence, the gas experiences density fluctuations directly driven by compressive (longitudinal) velocity fluctuations. 
This leads to the well-known result, in {both} sub-sonic and super-sonic turbulence, that the density PDF becomes approximately log-normal, with a variance that scales as 
$S_{\ln{\rhogas}} = \ln[1+\machcompressive^{2}]$
where $\machcompressive$ is the rms compressive (longitudinal) component of the turbulent Mach number $\mathcal{M}$ (component projected along $\nabla\cdot {\bf v}$; see \citealt{federrath:2008.density.pdf.vs.forcingtype,price:2011.density.mach.vs.forcing,konstantin:mach.compressive.relation,molina:2012.mhd.mach.dispersion.relation,federrath.2015:density.pdf.and.sfr.in.polytropic.turbulence})

However, in sub-sonic turbulence, the gas density fluctuations quickly become small. Simulations of the (very thin) mid-plane dead-zone dust layers typically record $\mach\lesssim0.1$; they confirm that the scaling above holds, but this produces correspondingly small fluctuations in $\rhogas$ \citep[see e.g.][]{johansen:2007.streaming.instab.sims}.\footnote{Note that this does not necessarily mean that Mach numbers in the much larger-scale height gas disk are small, nor that they are unimportant.} Yet these same simulations record orders-of-magnitude fluctuations in $\rhograin$.

\begin{figure}
    \centering
    \plotonesize{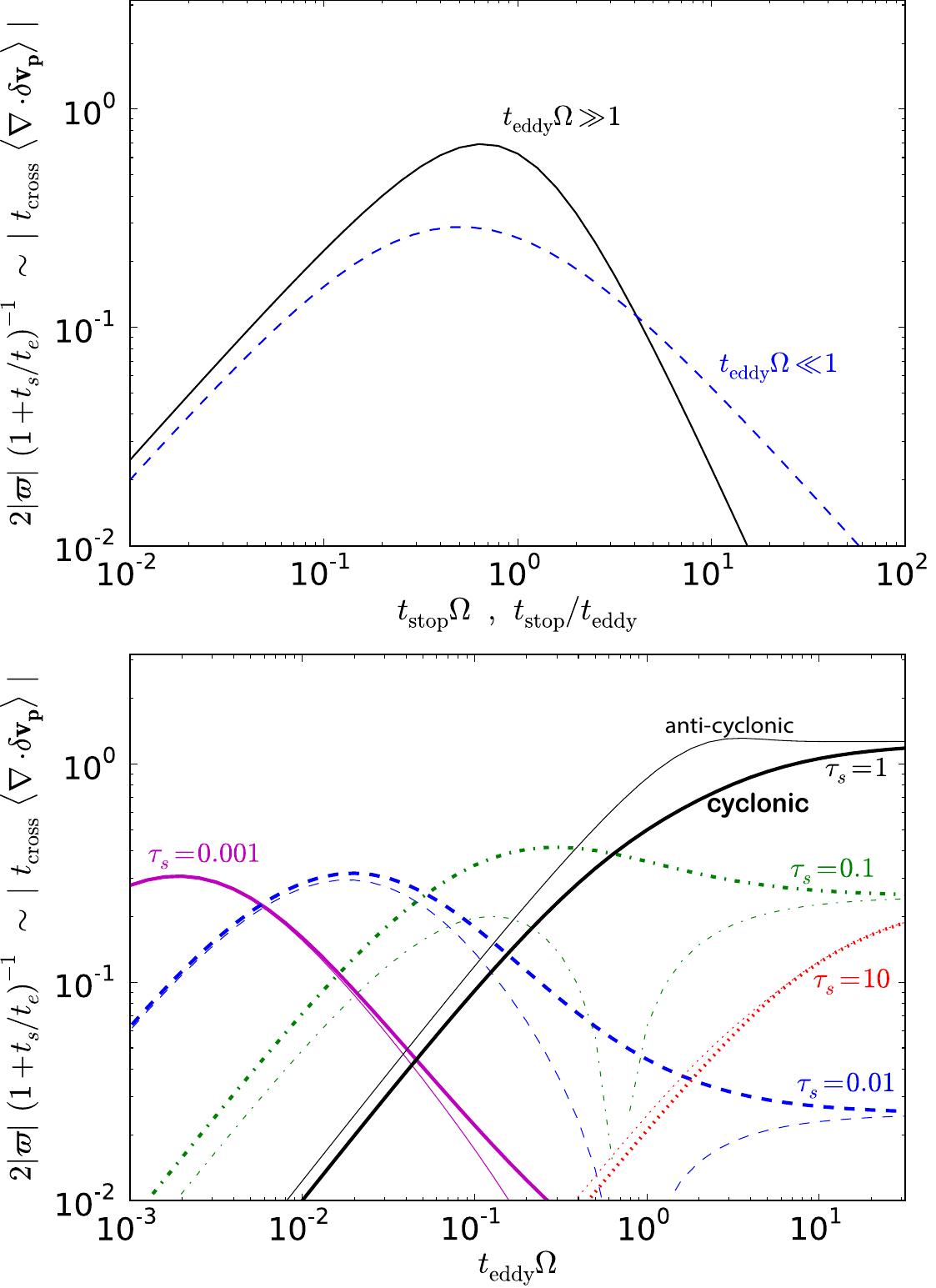}{1.01}
    \vspace{-0.5cm}
    \caption{
    ``Response function'' $\varpi$ defined in \S~\ref{sec:model.encounters}: this is the mean divergence produced in the peculiar grain velocity distribution by a simple vortex with a turnover time $\Teddy$. {\em Top:} Limiting cases. First, small eddies ($\Teddy\ll\Omega^{-1}$), where $\varpi$ is a function of $\taustopeddy\equiv \tstop/\Teddy$ alone ($\propto \taustopeddy$ for $\taustopeddy\ll1$, $\propto (2\taustopeddy)^{-1/2}$ for $\taustopeddy\gg1$). Second, large eddies ($\Teddy\gg\Omega^{-1}$), where $\varpi$ is a function of $\taustop\equiv\tstop\Omega$ alone ($\propto 2\,\taustop$ for $\taustop\ll1$, $\propto 2\,\taustop^{-1}$ for $\taustop\gg1$).     {\em Bottom:} Full solution (Appendix~\ref{sec:appendix:exact}), for different $\taustop$ and $\Teddy$. For small grains $\taustop\ll0.1$, there is a broad ``resonant'' peak around $\tstop\sim\Teddy$ (spanning $0.05\,\tstop\lesssim\Teddy\lesssim10\,\tstop$). On the largest scales ($\Teddy$), the value saturates -- this produces a broader and higher-amplitude ``plateau'' for large grains ($\taustop\gtrsim0.1$). Both cyclonic (thick) and anti-cyclonic (thin) eddies are shown: for the anti-cyclonic cases (eddy angular momentum anti-aligned with $\Omega$), the ``dip'' near $\Teddy\sim\Omega$ comes from a sign change in $\varpi$. For $\Teddy\ll \Omega$ the two cases are identical. For $\Teddy \gg \Omega$ they have opposite signs.
    \label{fig:response}}
\end{figure}

\vspace{-0.5cm}
\section{Partially-Coupled Grains: The Model}
\label{sec:model}

\subsection{The Equations of Motion and Background Flow}
\label{sec:model.background}

Now consider grains with non-zero $\tstop$, in a gaseous medium and some potential field (for now we take this to be a Keplerian disk, the case of greatest interest, but generalize below). Absent grains and turbulence, the gas equilibrium is in circular orbits, at a cyclindrical radius $R$ from the potential center, with orbital frequency $\Omega(R)$. Because of pressure support, the gas does not orbit at exactly the circular velocity $\vk$, but at the reduced speed $V_{\rm gas}$, where 
\be
\label{eqn:eta}
\eta\,\vk \equiv \vk-\langle V_{\rm gas}(R,\,\rhograin=0) \rangle \approx \frac{1}{2\,\rhogas\,\vk}\,\frac{\partial P}{\partial \ln{R}} 
\ee
Define a rotating frame with origin in the disk midplane at $R$, with the $\hat{x}$ axis along the radial direction and $\hat{y}$ axis in the azimuthal (orbital $\phi$) direction; the frame rotates at the circular velocity $\vk(R)$, with the angular momentum vector $\boldsymbol{\Omega}$ oriented along the $\hat{z}$ axis and $\Omega_{R}\equiv \Omega(R)$. 
The local equation of motion for a grain $i$ with stopping time $\tstop$ becomes
\be
\label{eqn:eom.1}
\frac{{\rm d}{\bf v}^{\prime}_{i}}{{\rm d}t} =  2\,{\bf v}^{\prime}_{i}\times{\boldsymbol{\Omega}}_{R} + 3\,\Omega_{R}^{2}\,x_{i}\hat{x} - \Omega_{R}^{2}\,z_{i}\,\hat{z} - \frac{{\bf v}^{\prime}_{i}-{\bf u}^{\prime}}{\tstop}
\ee
where ${\bf u}^{\prime}$ is the gas velocity in the rotating frame. Note that this is a Lagrangian derivative (Eq.~\ref{eqn:eom.1} follows the grain path). With no loss of generality, we can conveniently define velocities relative 
to the linearized Keplerian velocities, ${\bf v} \equiv {\bf v}^{\prime}_{i} + (3/2)\,\Omega_{R}\,x\,\hat{y}$ and ${\bf u} \equiv {\bf u}^{\prime} + (3/2)\,\Omega_{R}\,x\,\hat{y}$. 

\citet{nakagawa:1986.grain.drift.solution} show that for the coupled gas-grain system with dimensionless stopping time $\taustop \equiv \tstop\,\Omega_{R}$ and mid-plane volume-average grain-to-gas mass ratio $\rhoratio \equiv \rhograin/\rhogas$, this leads to a quasi-steady-state equilibrium drift solution for the grains and gas, with grain velocity (in the local rotating frame) $\langle {\bf v} \rangle = {\bf v}^{d} =  v_{x}^{d}\,\hat{x} + v_{y}^{d}\,\hat{y}$ and gas velocity $\langle {\bf u} \rangle = {\bf u}^{d} =  u_{x}^{d}\,\hat{x} + u_{y}^{d}\,\hat{y}$:
\begin{align}
\label{eqn:vdrift}
v_{x}^{d} & = -\frac{2\,\taustop}{\taustop^{2}+(1+\rhoratio)^{2}}\,\eta\,\vk \\ 
v_{y}^{d} &= -\frac{1+\rhoratio}{\taustop^{2}+(1+\rhoratio)^{2}}\,\eta\,\vk \\ 
u_{x}^{d} &= +\frac{2\,\taustop\,\rhoratio}{\taustop^{2}+(1+\rhoratio)^{2}}\,\eta\,\vk \\ 
\label{eqn:vdrift.last}
u_{y}^{d} &= -\frac{\taustop^{2}+(1+\rhoratio)}{\taustop^{2}+(1+\rhoratio)^{2}}\,\eta\,\vk \\
|\vdrift| &= |{\bf v}^{d}-{\bf u}^{d}| = \frac{2\,\taustop\,\sqrt{(1+\rhoratio)^{2}+\taustop^{2}/4}}{\taustop^{2}+(1+\rhoratio)^{2}}\,\eta\,\vk
\end{align}
So now define the ``peculiar'' grain/gas velocity relative to the steady-state solution, ${\bf v}\equiv {\bf v}^{d} + \delta {\bf v}$ and ${\bf u} = {\bf u}^{d} + \delta {\bf u}$. Insert these definitions into Eq.~\ref{eqn:eom.1}, and -- since the turbulent velocities are much smaller than Keplerian\footnote{We show below that this is internally consistent, but this amounts to the assumption that the individual eddy sizes {\em within the dust layer} are small compared to the (full) gas disk gradient scale length, which is easily satisfied in realistic systems.} -- expand $\eta=\eta(R)$ and $\vk(R)$ to leading order in $x/R$. We then obtain 
\begin{align}
\label{eqn:eom.peculiar}
\delta\dot{v}_{x} &\approx 2\,\Omega_{R}\,\delta v_{y} - \frac{\delta v_{x} - \delta u_{x}}{\tstop} \\ 
\delta\dot{v}_{y} &\approx -\frac{1}{2}\,\Omega_{R}\,\delta v_{x} - \frac{\delta v_{y} - \delta u_{y}}{\tstop} 
\end{align}


The $\hat{z}$ component of Eq.~\ref{eqn:eom.1} forms a completely separable equation which is simply that of a damped harmonic oscillator. Thus retaining it has no effect on our derivation below.

\begin{figure}
    \centering
    \plotonesize{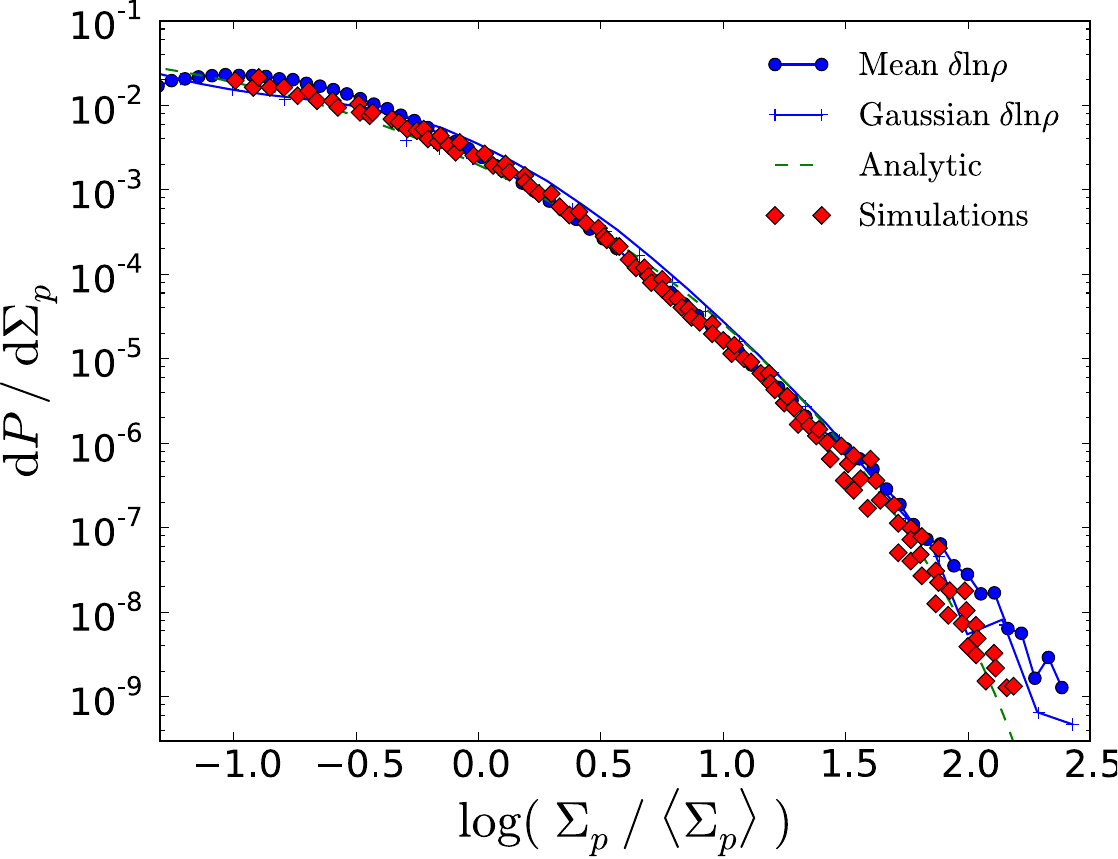}{1.01}
    \vspace{-0.6cm}
    \caption{Predicted grain density distribution in numerical simulations of MRI-driven turbulence with $\taustop=1$ and $\rhoratio=0$ (no grain-gas back-reaction). The exact prediction from our Monte-Carlo method, given the simulation parameters, is shown either assuming vortices with fixed $\Teddy$ each produce the same, mean multiplicative effect (``mean $\deltarhonoabs$'') or draw from a Gaussian distribution (``Gaussian $\deltarhonoabs$''). We also show the simple closed-form fitting function (``Analytic'') derived for fluctuations on large scales (Table~\ref{tbl:largescale}). This all assumes our ``default'' model ($N_{d}=2$ dimensional vortices, $C_\infty=2$, and a random cyclonic/anticylonic distribution). We compare the simulation results from \citet{dittrich:2013.grain.clustering.mri.disk.sims}. The agreement is very good; the simulations are not able to distinguish the (very similar) ``mean $\deltarhonoabs$'' and ``Gaussian $\deltarhonoabs$'' models.
    \label{fig:grain.rho.mri}}
\end{figure}

\begin{figure}
    \centering
    \plotonesize{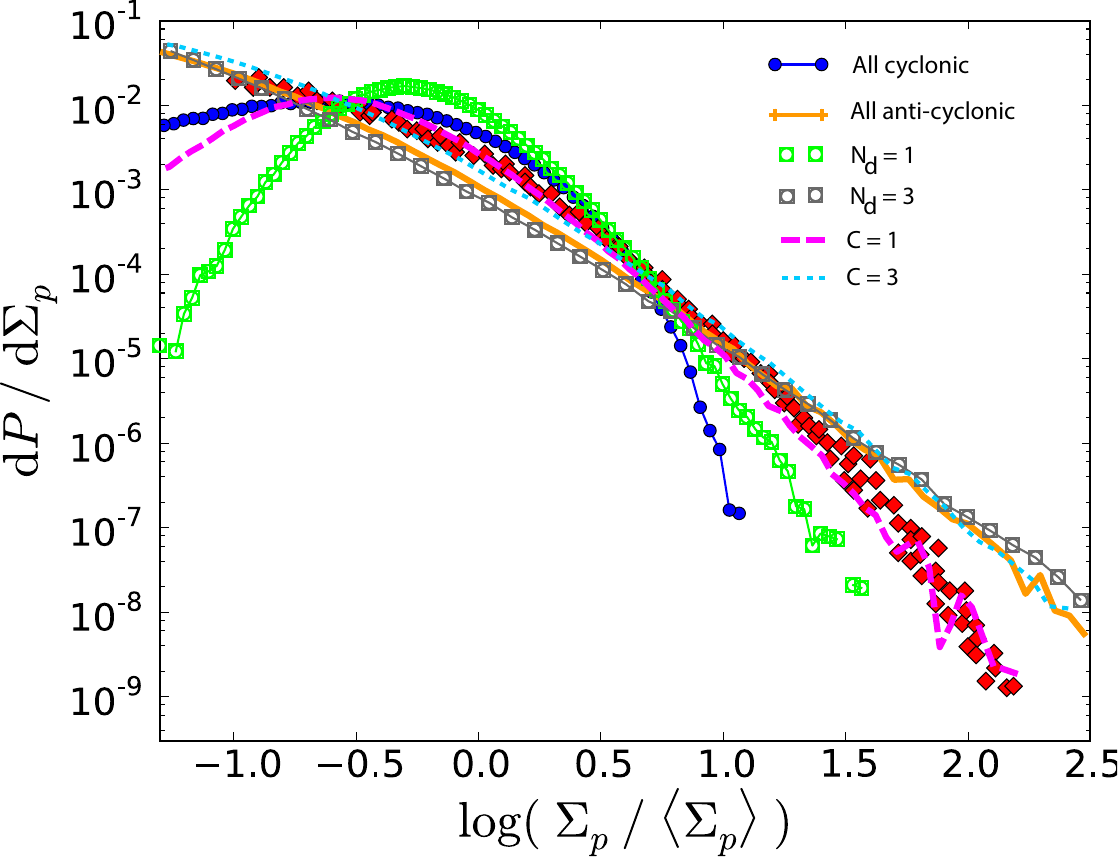}{1.01}
    \vspace{-0.6cm}
    \caption{Effect of model choices on the predicted distribution in Fig.~\ref{fig:grain.rho.mri}. If we assume all vortices are cyclonic, the distribution is too broad at low densities and cuts off too sharply at high densities. But assuming all vortices are anti-cyclonic over-predicts the high-density tail (these would also predict net angular momentum in the vortices, which is not allowed in our assumptions). Decreasing/increasing the assumed wrapping dimension of eddies decreases/increases the predicted scatter correspondingly. Interestingly, the effective ``filling factor'' $C_\infty$ of eddies is only weakly constrained: a range $C_\infty\sim0.5-2$ is broadly consistent with the distribution, though much larger values $C_\infty\gtrsim 3$  are ruled out.
    \label{fig:grain.rho.mri.examples}}
\end{figure}

\vspace{-0.5cm}
\subsection{Encounters Between Grains and Individual Turbulent Structures}
\label{sec:model.encounters}


\subsubsection{A Toy Model}
\label{sec:model.encounters:toy}

Now consider an idealized encounter between a grain group with $\taustop$ and single, coherent turbulent eddy. We'll first illustrate the key dynamics with a purely heuristic model, then follow with a rigorous derivation (for which the key equations are given in Table~\ref{tbl:defns}).

Define the eddy coherence length $\Leddy$, and some characteristic peculiar velocity difference across $\Leddy$ of $\delta u = \Veddy = \mach\eddy\,\cs$, so the eddy turnover time can be defined as $\Teddy = \Leddy/|\Veddy|$. 
In inertial-range turbulence, we expect these to scale as power laws, so define 
\begin{align}
\label{eqn:veddy.scaling}
\langle |\Veddy| \rangle &=\langle  \mach\eddy \rangle\,\cs = {|} \Veddy(\Lmax) {|}\,{\Bigl(}\frac{\Leddy}{\Lmax} {\Bigr)}^{\velslope} \propto \Leddy^{\velslope} \\ 
\langle \Teddy \rangle &\equiv \frac{\Leddy}{|\Veddy|} = \Teddy(\Lmax)\,{\Bigl(} \frac{\Leddy}{\Lmax} {\Bigr)}^{1-\velslope}
\end{align}
It is also convenient to define the dimensionless stopping time relative to either the orbital frequency or eddy turnover time: 
\begin{align}
\taustop\equiv\tstop\,\Omega  \ \ \ \ \ \ \ \ , \ \ \ \ \ \ \ \ \taustopeddy\equiv \tstop/\Teddy
\end{align}
For now, we will assume $\rhograin\ll\rhogas$, so that the back-reaction of the grains on gas can be neglected.

Consider a grain with $\tstop\ll \Teddy$, in a sufficiently small eddy that we can ignore the shear/gravity terms across it. Typical eddies are two-dimensional vortices, so the grain is quickly accelerated to the eddy velocity $\Veddy$ in an approximately circular orbit. This produces a centrifugal acceleration $a_{\rm cen}=\delta v_{\theta}^{2}/r \sim \Veddy^{2}/\Leddy = |\Veddy|/\Teddy$, which is balanced by pressure forces for the gas but causes the grain to drift radially out from the eddy center, at the approximate ``terminal velocity'' where this is balanced by the drag acceleration $\sim \delta v_{r}/\tstop$, so $\delta v_{r} \sim \tstop\,\Veddy^{2}/\Leddy = (\tstop/\Teddy)\,|\Veddy|$. If, instead, the eddy is sufficiently large, expansion of the centrifugal force gives $a_{\rm cen}\sim 2\,\Omega\,\Veddy$ (the $2\,{\bf v}\times{\boldsymbol{\Omega}}$ term in Eq.~\ref{eqn:eom.1}) -- i.e.\ the global centrifugal force sets a ``floor'' here, so the terminal velocity is $\delta v_{r} \sim 2\,(\tstop/\Omega^{-1})\,|\Veddy|$. 

A real eddy has a velocity gradient across itself. Assume for simplicity that the gradient is strongest across a single dimension, so to first order the local velocity perturbation scales as $\delta u_{y} \approx (\Veddy/\Leddy)\,x$ (where $x=0$ is the center of the eddy), while $\delta u_{x}\sim$\,constant. If grains have time to come to their terminal velocity while still inside the eddy ($\tstop\ll \Teddy$), then by the arguments above the relative perturbed velocity of two grains on opposite ``sides'' ($\pm x$) of the eddy will be $\delta v_{x} \sim 2\,x\,(\Veddy/\Leddy)\,(\tstop/\Teddy)$ or $\sim 2\,x\,(\Veddy/\Leddy)\,\taustop$ (for small $\Teddy\ll \Omega^{-1}$ and large $\Teddy\gg \Omega^{-1}$, respectively). But if the grains do not have time to reach terminal velocity ($\tstop\gg \Teddy$), we can consider them to be sitting ``in place'' experiencing an approximately constant drag acceleration $\approx \Veddy/\tstop$ for a time $\sim \Teddy$, so the velocity difference across the eddy at time $\sim \Teddy$ is just $\sim 2\,x\,(\Veddy/\Leddy)\,(\tstop/\Teddy)^{-1}$.


The grain density is determined by the continuity equation $\partial\rhograin/\partial t + \nabla\cdot(\rhograin\,\delta {\bf v})=0$ 
which we can write as 
\begin{align}
\frac{{\rm D}\,\ln{\rhograin}}{{\rm d}t} = {\Bigl(} \frac{\partial}{\partial t} + {\delta {\bf v}\cdot \nabla} {\Bigr)}\,\ln{\rhograin} = 
-\nabla\cdot \delta {\bf v}
\end{align}
where ${\rm D}/{\rm d}t$ is the Lagrangian derivative for a ``grain population.''
So the $\delta v_{x} \propto x$ term means that a population of grains, on encountering this eddy, will expand (be pushed away from the origin of the rotating frame) if $\Veddy>0$. This is just the well-known result that anti-cyclonic vortices ($\Veddy<0$) on the largest scales tend to collect grains, while cyclonic vortices ($\Veddy > 0$) disperse them; note that for the small-scale eddies, the sense is {\em always} dispersal in eddies.\footnote{This description of anti-cyclonic eddies, while common, is actually somewhat misleading. Grains always preferentially avoid regions with high absolute value of vorticity $|\boldsymbol{\omega}|\sim |{\bf v}_{e}\,\Leddy^{-1} + {\boldsymbol{\Omega}}|$. It is simply that very large eddies ($\Teddy \gtrsim\Omega^{-1}$, with $\Teddy=\Leddy/|{\bf v}_{e}|$) which are locally anti-cyclonic and in-plane ($\hat{\bf v}_{e} = -\hat{\boldsymbol{\Omega}}$) have lower $|\boldsymbol{\omega}|$ than the mean (${\bf v}_{e}=0$) Keplerian flow; so grains concentrate there by being dispersed out of higher-$|\boldsymbol{\omega}|$ regions.}

Note that above, if the eddy velocity gradient is just one-dimensional, the grain population is preferentially dispersed in one dimension; however when the eddies are vortices in two-dimensions, the flow is radial (along each dimension). In general, non-zero $\nabla\cdot\delta{\bf v}$ will occur along $\Ndim$ dimensions, where $\Ndim$ is the number of dimensions along which the eddy flow is locally ``wrapped.'' For the expected case of simple vortices this is an integer $\Ndim=2$, but for eddies with complicated structure, or a population of eddies, this can take any non-integer value between zero and the total spatial dimension.

So, on encountering an eddy of scale $\Leddy$, the Lagrangian population of grains with initial extent $\scalevar = \Leddy$ will shrink or grow in scale according to 
\begin{align}
\label{eqn:shrink.eddy}
\frac{{\rm D}\ln{\rhograin}}{{\rm d}t} &\sim 
\begin{cases}
{\displaystyle \frac{\Ndim\,\taustopeddy}{\Teddy\,(1+\mathcal{O}(\taustopeddy^{2}))]} + \mathcal{O}(\taustopeddy\,\eta^{2}) + \mathcal{O}(\Teddy\,\Omega)}\ \ \ \   \hfill {\tiny ({\rm small}\ \ \Teddy)}\\ 
\\
{\displaystyle \frac{2\,\Ndim\,\taustop}{\Teddy\,(1+\taustop^{2})} + \mathcal{O}(\taustop\,\eta^{2}) + \mathcal{O}(\Teddy\,\Omega)^{-1}} \hfill {\tiny ({\rm large}\ \ \Teddy)}\\ 
\end{cases}
\end{align}
We will derive the exact scalings more precisely below, but this simple approximation actually does correctly capture the asymptotic behavior for small and large eddies.

\begin{figure}
    \centering
    \hspace{-0.4cm}
    \plotonesize{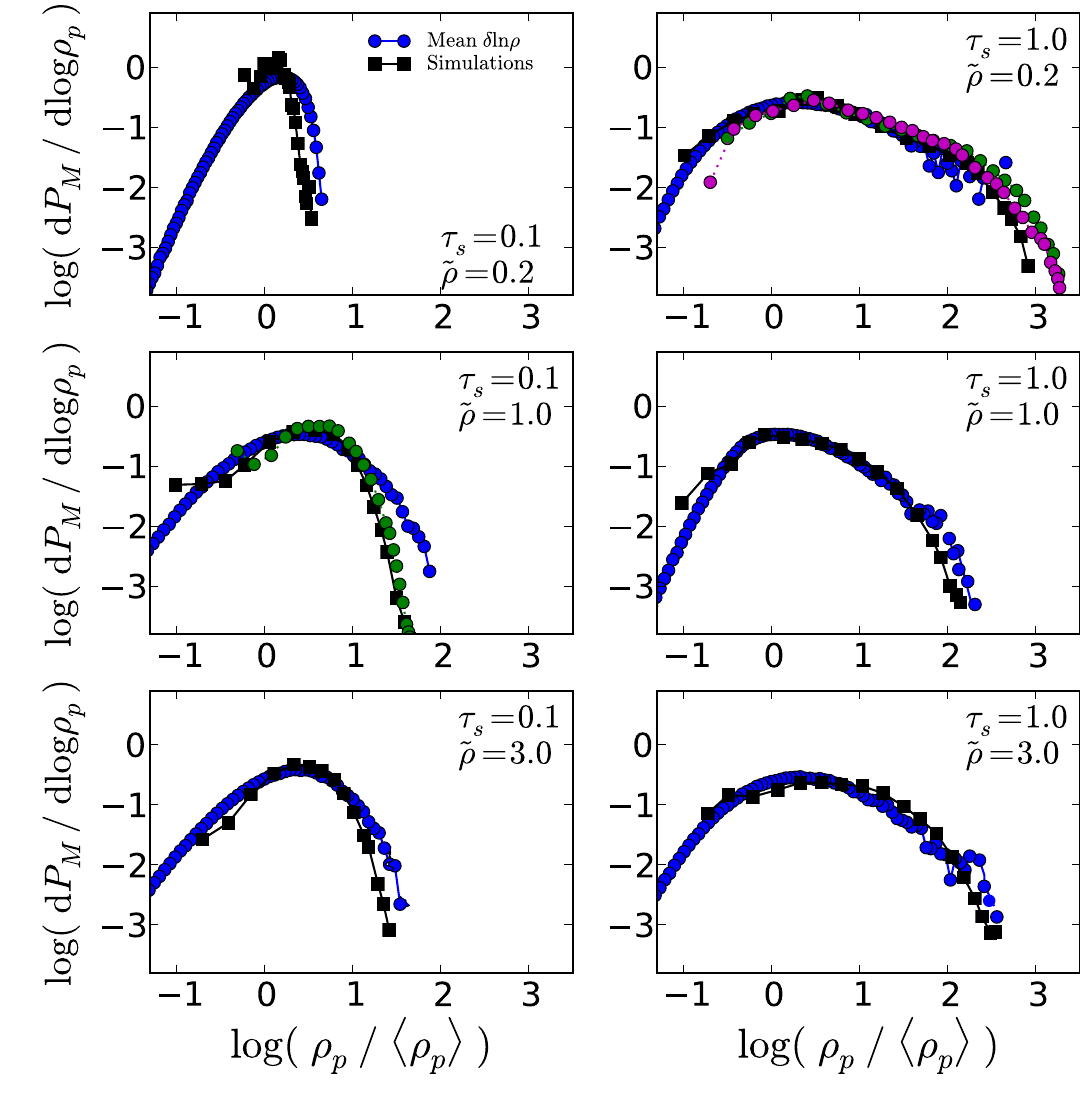}{1.01}
    \vspace{-0.3cm}
    \caption{Grain density distribution, as Fig.~\ref{fig:grain.rho.mri}, for simulations of streaming-instability turbulence with $\taustop=0.1-1$ and $\rhoratio=0.2-3$ (labeled). The simulations are from \citet{johansen:2007.streaming.instab.sims} and \citet{bai:2010.grain.streaming.sims.test}; where multiple simulations with different numerical methods are available we plot them, to represent differences owing purely to numerics. As expected from the stronger response functions on large scales in Fig.~\ref{fig:response}, the PDF width is larger for $\taustop=1$. Increasing $\rhoratio$ broadens the PDF for $\taustop\ll1$, but narrows it when $\gg1$, consistent with our lowest-order estimate that it changes the ``effective stopping time'' as $\taustop\rightarrow\taustop\,(1+\rhoratio)$. All the predictions have some growing discrepancies at the highest densities, probably because our assumption that the grains represent a perturbation on the gas turbulence structure is no longer valid. 
    \label{fig:grain.rho.jy}}
\end{figure}

We now wish to know how long (in an average sense) the perturbation affecting the local grain density distribution in Eq.~\ref{eqn:shrink.eddy} is able to act, which we define as the timescale $\deltat$. This obviously cannot be longer than the eddy coherence time, which is about an eddy turnover time $\Teddy$. Since the eddy will not, in equilibrium, accelerate grains to relative speeds {\em greater} than the flow velocity, it follows that $t_{\rm sink} = |\Leddy/\langle \delta v_{\rm induced} \rangle|$ (the timescale for grains to be fully expelled from the eddy region) is always $>\Teddy$, so this does not limit $\deltat$. However, if the grains have some non-zero initial relative velocity $v_{0}$ with respect to the eddy, and the stopping time is sufficiently long that they are not rapidly decelerated, they can drift through or cross the eddy in finite time $\tdrift \sim \Leddy/|v_{0}|$. If $\tdrift < \tstop$ and $\tdrift < \Teddy$, then $\deltat = \tdrift$ becomes the limitation. If $v_{0}$ is approximately constant (from e.g.\ global drift or much larger eddies), we have $\tdrift \sim \Leddy/|v_{0}| \sim \Teddy\,|\Veddy|/|v_{0}|$ decreasing on small scales. So we expect $\deltat/\Teddy$ is a constant $\approx1$ for large-scale eddies, then turns over as an approximate power law $\propto \Veddy \propto \Leddy^{\velslope}$, for $\Leddy$ below some $\Lcrit$ where $\tdrift<{\rm MIN}(\Teddy,\,\tstop)$.\footnote{It is straightforward to show that this timescale restriction guarantees our earlier assumption (where we dropped higher-order terms in the gradient of $\eta$) is valid (for timescales $<\deltat$ and sub-sonic peculiar velocities).} Again, we'll consider this in detail below.

Together, this defines what we call the ``response function'': the typical density change induced by encounter with an eddy
\begin{align}
\deltarhomean = {\Bigl\langle}\int \frac{{\rm D}\ln{\rhograin}}{{\rm d}t}\,{\rm d}t\,{\Bigr\rangle} \sim {\Bigl\langle}\frac{{\rm D}\ln{\rhograin}}{{\rm d}t} {\Bigr\rangle}\,\deltat
\end{align}

\vspace{-0.2cm}
\subsubsection{Exact Solutions in Turbulence without External Gravity}
\label{sec:encounters:pure.turb}

Now we will derive the previous scalings rigorously.

Consider the behavior of grain density fluctuations in inertial-range turbulence.\footnote{We specifically will assume high Reynolds number $Re\gg1$ and Stokes number $St=\tstop/\Teddy(\scalevar_{\nu})\gg1$, where $\scalevar_{\nu}$ is the viscous scale, so will neglect molecular viscosity/diffusion throughout.} First we derive the ``response function'' above, i.e.\ the effect of an eddy on a grain distribution. Subtracting the bulk background flow, the grain equations of motion (Eq.~\ref{eqn:eom.peculiar}) become the Stokes equations 
\begin{align}
\label{eqn:eom.noshear}
\delta\dot{{\bf v}} &= -\frac{\delta {\bf v} - \delta {\bf u}}{\tstop} 
\end{align}
with the continuity equation $\partial\rhograin/\partial t + \nabla\cdot(\rhograin\,\delta {\bf v})=0$ which we can write as 
\begin{align}
\frac{{\rm D}\,\ln{\rhograin}}{{\rm d}t} = {\Bigl(} \frac{\partial}{\partial t} + {\delta {\bf v}\cdot \nabla} {\Bigr)}\,\ln{\rhograin} = 
-\nabla\cdot \delta {\bf v}
\end{align}
where ${\rm D}/{\rm d}t$ is the Lagrangian derivative for a ``grain population.''

Many theoretical and experimental studies have suggested that the dynamics of incompressible gas turbulence on various scales can be understood by regarding it as a collection of Burgers vortices \citep[see][and references therein]{marcu:1995.grain.burgers.vortex}. The Burgers vortex is an exact solution of the Navier-Stokes equations, and provides a model for vortices on all scales (which can be regarded as ``stretched'' Burgers vortices). In a cylindrical coordinate system centered on the vortex tube, the fluid flow components can be written as $\delta u_{z}=2\,A\,z$, $\delta u_{r}=-A\,r$, $\delta u_{\theta} = (B/2\pi\,r)\,(1-\exp{(-r^{2}/2\,r_{0}^{2}))}$, where $r_{0}$ is the vortex size, $B$ the circulation parameter, and $A = \nu/B$ is the inverse of the ``vortex Reynolds number.'' Since we consider large $Re$, $A\rightarrow0$ for the large-scale vortices, so $\delta u_{z}=\delta u_{r}=0$, and we can specify $\delta u =\delta  u_{\theta} \equiv u_{0}\,(r_{0}/r)\,(1-\exp{[-r^{2}/2\,r_{0}^{2}]})$. 

On scales $\lesssim 1.58\,r_{0}$, $\delta u_{\theta}\propto r + \mathcal{O}(r^{2}/2.5\,r_{0}^{2})$ increases linearly with $r$, before turning over beyond the characteristic scale and decaying to zero. So, since we specifically consider the effects of an eddy on scales {\em within} the eddy size ($\lesssim r_{0}$), we can take $\delta u_{\theta} \propto r$, in which case the eddy is entirely described by the (approximately constant) turnover time $\Teddy$ such that $\delta u_{\theta} \equiv r/\Teddy$. Note that this is now the general form for {\em any} eddy with pure circulation and constant turnover time, so while motivated by the Burgers vortex should represent real eddies on a wide range of scales.

We derive an exact, general solution of the problem here in Appendix~\ref{sec:appendix:exact}. The qualitative behaviors of the solution derived there is not obvious however, so we illustrate it with a slightly simplified derivation (which captures the correct behavior in various limits) here. 

In the vortex plane, the equations of motion (Eq.~\ref{eqn:eom.noshear}) become
\begin{align}
\nonumber \delta \dot{v}_{x} &= \delta\dot{v}_{r^{\prime}}\,\cos{\theta} - \delta\dot{v}_{\theta}\,\sin{\theta} - \dot{\theta}\,(\delta v_{r^{\prime}}\,\sin{\theta} + \delta v_{\theta}\,\cos{\theta}) \\ 
\label{eqn:eom.ideal.1}
&= \tstop^{-1}\,(-\delta v_{r^{\prime}}\,\cos{\theta} + \delta v_{\theta}\,\sin{\theta} - \delta u_{\theta}\,\sin{\theta}  ) \\ 
\nonumber \delta \dot{v}_{y} &= \delta \dot{v}_{r^{\prime}}\,\sin{\theta} + \delta \dot{v}_{\theta}\,\cos{\theta} + \dot{\theta}\,(\delta v_{r^{\prime}}\,\cos{\theta} - \delta v_{\theta}\,\sin{\theta}) \\ 
&= \tstop^{-1}\,(-\delta v_{r^{\prime}}\,\sin{\theta} - \delta v_{\theta}\,\cos{\theta} + \delta u_{\theta}\,\cos{\theta}  ) 
\label{eqn:eom.ideal.2}
\end{align}
with $\dot{\theta}\equiv \delta v_{\theta}/r^{\prime}$. It is straightforward to verify that the peculiar solution is given by $\delta v_{r} = \varpi\,r/\Teddy$ ($\delta v_{r}\propto \delta v_{\theta} \propto u_{\theta} \propto r \propto \exp{(\varpi\,t/\Teddy)}$) with $\varpi$ being a root of 
$\varpi\,(1+\varpi\,\taustopeddy)\,(1+2\,\varpi\,\taustopeddy)^{2}-\taustopeddy=0$, all of which are decaying solutions except the positive real root: 
\begin{align}
\label{eqn:varpi.pureturb}
\varpi &= \frac{-2 + \sqrt{2\,{\Bigl(}1 + \sqrt{1+16\,\taustopeddy^{2}} {\Bigr)}}}{4\,\taustopeddy} \\ 
\varpi &\rightarrow
\begin{cases}
      {\displaystyle \taustopeddy}\ \ \ \ \   \hfill {\tiny (\taustopeddy\ll1)} \\ 
      {\displaystyle (2\,\taustopeddy)^{-1/2}}\ \ \ \ \   \hfill {\tiny (\taustopeddy\gg1)} \
\end{cases}
\end{align}
Because $\delta v_{r}\propto r$ and $\delta v_{\theta}$ is independent of $\theta$, it follows that along this solution
\be
{(}\nabla\cdot \delta {\bf v}{)}_{\rm pec} = \frac{1}{r}\frac{\partial (r\,v_{r})}{\partial r} + \frac{1}{r}\frac{\partial v_{\theta}}{\partial \theta} = 
\Ndim\,\frac{|v_{r}|}{r} = \frac{2\,\varpi}{\Teddy}
\ee

To determine the general solution, we must consider how the eddy evolves in time, since it is able to act on the grains for only finite $\deltat$. To approximate this, consider the simplest top-hat model, $\delta u = \delta u_{\theta}\Theta(0<t<\deltat)$, where $\delta {\bf u}\propto \Theta=0$ at $t<0$ and $t>\delta t$ and $\Theta=1$ ($\delta {\bf u} = \delta u_{\theta}(r)\,\hat{\theta}$) for $0<t<\delta t$. We require the net effect of the eddy on the density field (i.e.\ the late-time result of the perturbation), so we integrate ${\rm d}\ln{\rhograin}/{\rm d}t = -\nabla\cdot{\bf \delta v}$, from the boundary condition $\delta {\bf v}=\delta {\bf v}_{0}$ at $t<0$ until some time much longer than the eddy lifetime $t\rightarrow \infty$. For the simple top-hat form of the eddy lifetime this gives\footnote{Here we note that there are two decaying oscillatory solutions to Eqs.~\ref{eqn:eom.ideal.1}-\ref{eqn:eom.ideal.2} with decay rate $\omega^{\prime} = -1/2\,\taustopeddy$, which correspond to the usual damped modes for grains with drag in a uniform flow.
The general solution is derived by matching the modes to the initial velocities $\delta {\bf v}_{0}$; then the solution at $t=\delta t$ is matched to the solution for the post-eddy field. The exact result is in Appendix~\ref{sec:appendix:exact}; but if we linearize in $\Teddy$, for example, it is straightforward to show that the integral over time ($t\rightarrow\infty$) of $\nabla\cdot \delta {\bf v}$ is exactly the integral of the positive real peculiar solution $|\nabla\cdot{\bf v}|_{\rm pec}$ from $t=0$ to $t=\delta t$.}
\begin{align}
\nonumber
\langle 
\deltarhonoabs \rangle &={\Bigl\langle}  - \int_{t} (\nabla\cdot \delta {\bf v})\,{\rm d}\,t  {\Bigr\rangle}^{\delta u = \delta u_{\theta}\Theta(0<t<\deltat)} \\
&=- | \nabla\cdot \delta {\bf v} |_{\rm pec}\,\deltat = -{2\,\varpi}\,\frac{\deltat}{{\Teddy}}
\end{align}
where the $\langle...\rangle$ denotes an average over an assumed homogeneous, isotropic initial ensemble in position and velocity space.

\begin{figure}
    \centering
    \hspace{-0.2cm}
    \plotonesize{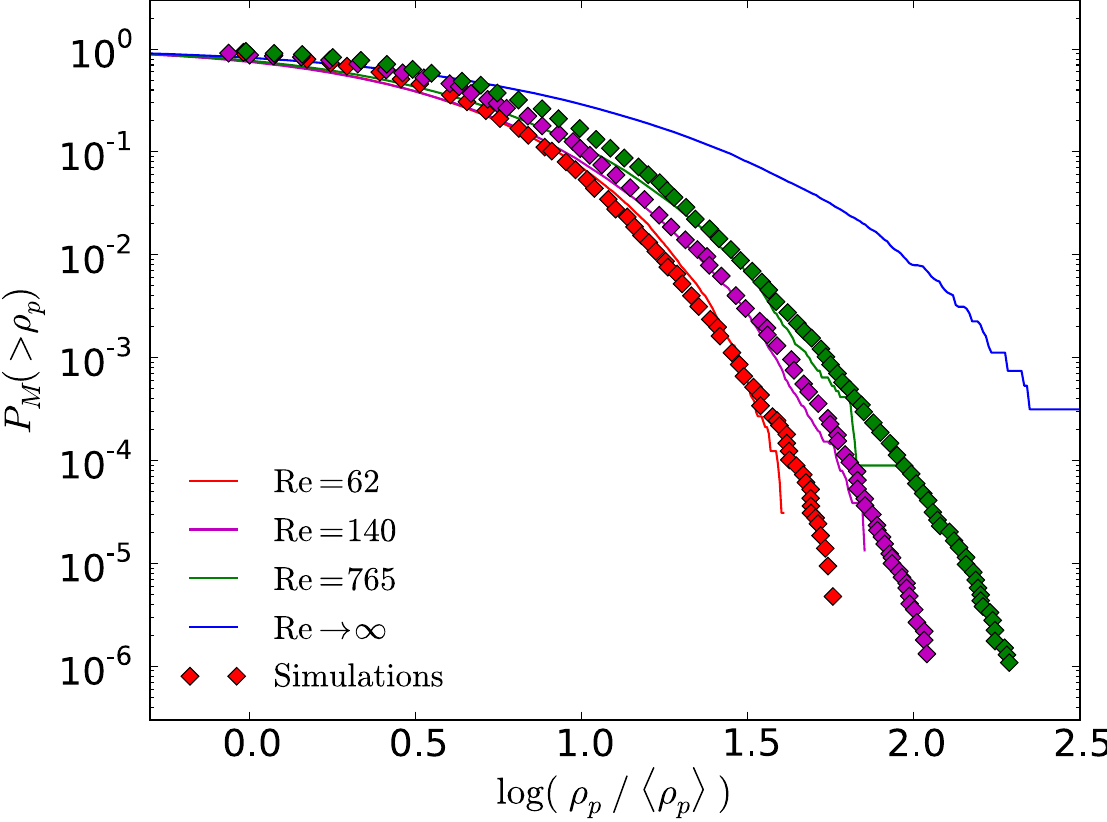}{1.01}
    \vspace{-0.6cm}
    \caption{Grain density distribution, as Fig.~\ref{fig:grain.rho.mri}, for simulations of ``turbulent concentration'' from \citet{hogan:1999.turb.concentration.sims}. Here there is no shear/gravity (see \S~\ref{sec:encounters:pure.turb}), and the flow is simulated from a fixed viscous scale to various Reynolds numbers ($Re=62,\,140,\,765$). In each case the Stokes number is unity ($\tstop\approx \Teddy(\scalevar_{\nu})$, where $\scalevar_{\nu}$ is the viscous scale); this gives $\taustopeddy(\scalevar=\Lmax)$ at the top of the cascade of $\taustopeddy(\Lmax)\approx0.13,\,0.03,\,0$ ($Re=62,\,765,\,\infty$). The PDF width grows logarithmically with $Re$ as we integrate over more of the broad response function in Fig.~\ref{fig:response}. However, the decay in this function at $\taustopeddy\ll1$, and the increase in rms turbulent velocities of grains lowering their eddy crossing times, means that the PDF does not grow indefinitely as $Re\rightarrow\infty$. For comparison, pure uncorrelated (Markovian) fluctuations predict a PDF with dispersion in $\rhograin$ of $\approx1.7$, giving a PDF that falls below the minimum plotted range here at $\log{(\rhograin/\langle\rhograin\rangle)}\approx0.8$ \citep{zaichik:2009.grain.clustering.theory.randomfield.review}.
    \label{fig:grain.rho.tc}}
\end{figure}

Now we need to determine $\delta t$. If the grains are ``trapped'' well within the eddy, this is simply the eddy lifetime $\Teddy$. However, we have not yet accounted for the finite spatial coherence of the eddy. If the grains are moving sufficiently fast and/or if the stopping time is large, they can cross or move ``through'' the eddy (to $r\gg r_{0}\equiv \Leddy = |\Veddy|\,\Teddy$, where the eddy circulation is super-exponentially suppressed so becomes negligible) in a timescale $\tdrift\lesssim \Teddy$. Since we are integrating a rate equation, the full $\deltat$ is simply given by the harmonic mean 
\be
\deltat^{-1} = \Teddy^{-1} + \tdrift^{-1}
\ee 
(capturing both limits above; see \citealt{voelk:1980.grain.relative.velocity.calc,markiewicz:1991.grain.relative.velocity.calc}). The timescale for a grain to cross a distance $\Leddy=|\Veddy|\,\Teddy$ in a smooth flow (constant $\delta{\bf u}$), with initial (relative) grain-gas velocity $|v_{0}|$, is just
\begin{align}
\tdrift &=
\begin{cases}
      {\displaystyle -\tstop\,\ln{{\Bigl[}1 - \frac{\Leddy}{\tstop\,|v_{0}|} {\Bigr]}}}\ \ \ \ \   \hfill {\tiny (|v_{0}|>\Leddy/\tstop)} \\ 
      {\displaystyle \infty}\ \ \ \ \   \hfill {\tiny (|v_{0}|\le\Leddy/\tstop)} \
\end{cases}
\end{align}
Note that for large $\tstop$, this is just the ballistic crossing time $\rightarrow \Leddy/|v_{0}|$, but for small $|v_{0}|<\Leddy/\tstop$ this diverges because the grain is fully stopped and trapped without reaching $\scalevar_{0}$. So now we need to determine $|v_{0}|$, but this is considered in \citet{voelk:1980.grain.relative.velocity.calc} and many subsequent calculations \citep[e.g.][]{markiewicz:1991.grain.relative.velocity.calc,pan:2010.grain.velocity.sims,pan:2013.grain.relative.velocity.calc}. Assuming the turbulence is isotropic and (on long timescales) velocity ``kicks'' from independent eddies are uncorrelated, then 
\be
\langle |{\bf v}_{0}|^{2} \rangle = |{\bf V}_{L}|^{2} + \langle |{\bf V}_{\rm rel}(\Leddy)^{2}| \rangle
\ee
where $V_{L} = |{\bf V}_{L}|$ is the difference in the laminar bulk flow velocity of grains and gas (due to e.g.\ settling or gravity) and $\langle |{\bf V}_{\rm rel}(\Leddy)^{2} |\rangle$ represents the rms grain-eddy velocities (averaged on the eddy scale) due to the turbulence itself. 

For the ``pure turbulence'' case here $V_{L}=0$, and $\langle | {\bf V}_{\rm rel}(\Leddy)^{2}| \rangle$ is derived in \citet{voelk:1980.grain.relative.velocity.calc} as 
\begin{align}
\nonumber
\langle |{\bf V}_{\rm rel}(\Leddy)^{2}| \rangle &= \int_{k(\Lmax)}^{k(\Leddy)} {\rm d}k\,P(k)\,\frac{\tstop}{\tstop + \Teddy(k)} \\ 
&= |\Veddy(\Lmax)|^{2}\,\taustopeddy(\Lmax)\,\ln{{\Bigl[} \frac{1+\taustopeddy(\Lmax)^{-1}}{1 + \taustopeddy(\Leddy)^{-1}} {\Bigr]}}  
\end{align}
where $k$ is the wavenumber and $P(k)=(p-1)\,k^{-p}$ is the velocity power spectrum ($\int {\rm d}k\,P(k)=\langle \delta {\bf u}^{2} \rangle = \Veddy(\Lmax)^{2}$; the closed-form expression here follows for any power-law $P(k)$; see \citealt{ormel:2007.closed.form.grain.rel.velocities}).\footnote{Due to intermittency, the eddy intensity will vary at a given scale, which will in turn lead to non-linear variations in the timescale for 
particle crossing. We consider a heuristic model for this below, but it is worth further investigation.}

After some simple substitution, we now have
\begin{align}
\frac{\deltat}{\Teddy} &= {\Bigl[}1 + {\Bigl(} \frac{\tdrift}{\Teddy} {\Bigr)}^{-1} {\Bigr]}^{-1} \\ 
\frac{\tdrift}{\Teddy} &= -\taustopeddy\,\ln{{\Bigl[}\,1 - \frac{(\Leddy/\Lmax)}{\taustopeddy(\Lmax)\,g_{0}(\Leddy)^{1/2}}\, {\Bigr]}} \\ 
g_{0}(\Leddy) &\equiv \taustopeddy(\Lmax)\,\ln{{\Bigl[}\, \frac{1+\taustopeddy(\Lmax)^{-1}}{1+\taustopeddy(\Leddy)^{-1}}\, {\Bigr]}}
\end{align}
giving a complete description of $\deltarho = 2\,\varpi\,(\deltat/\Teddy)$.

Note that we have implicitly assumed a simple and time-constant structure for the eddies in deriving this; another important effect, discussed in 
\citet{falkovich:2004.intermittent.distrib.heavy.particles}, is that the eddy can stretch and deform the ``parcel'' of particles such that the crossing time varies across the eddy, or even to deform the parcel until its largest dimension is longer than that of the eddy itself. This could revise the timescales above significantly. Our simplified model essentially folds this into the ``effective'' eddy lifetime. If the effect were systematic, it would manifest as a systematic change in the eddy lifetime relative to $\Teddy$, which is ultimately degenerate with the amplitude of the effects we predict (in e.g.\ the parameter $\Ndim$). More likely, it is another cause which in more detail would contribute to a distribution of eddy-crossing times.

\begin{figure}
    \hspace{-0.4cm}
    \plotonesize{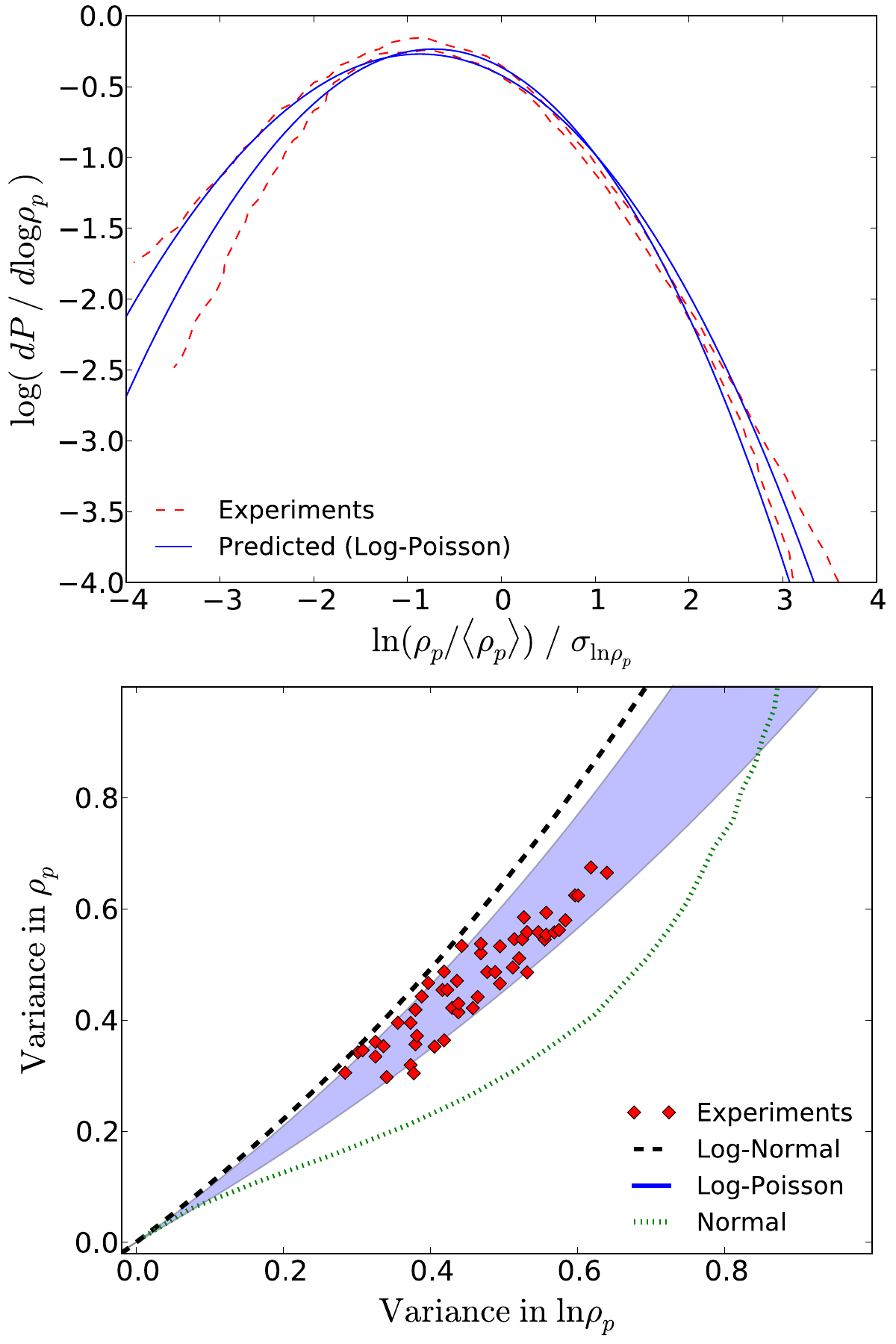}{1.03}
    \vspace{-0.2cm}
    \caption{Density PDF in laboratory experiments of water droplets in wind tunnel turbulence \citep{monchaux:2010.grain.concentration.experiments.voronoi}. {\em Top:} Range of particle density PDFs obtained (dashed), normalized by their variance. We compare the predicted log-Poisson distribution, with the same variance and a range of $\deltarho_{\rm int}\sim0.01-0.2$ corresponding to model predictions. {\em Bottom:} Test of log-normality: We compare the variance in $\rhograin$ to that in $\ln{\rhograin}$ from the same experiments, to the range predicted for log-Poisson PDFs with the predicted range in $\deltarho_{\rm int}$, the prediction from a lognormal distribution, and from a normal (Gaussian) distribution. The experiments favor a log-Poisson distribution as opposed to a linear Gaussian distribution or log-normal.
    \label{fig:nongaussian}}
\end{figure}

\vspace{-0.2cm}
\subsubsection{Solution With Shear}
\label{sec:encounters:shear.solution}

Here we discuss the full solution (including shear) for the density fluctuations for $\Ndim=2$ vortices above. Again, the exact result is derived in Appendix~\ref{sec:appendix:exact}. Consider the same eddies as in \S~\ref{sec:encounters:pure.turb}, but retain the shear terms from Eq.~\ref{eqn:eom.1}. Eq.~\ref{eqn:eom.ideal.1}-\ref{eqn:eom.ideal.2} become
\begin{align}
\nonumber \delta \dot{v}_{x} &= \delta\dot{v}_{r^{\prime}}\,\cos{\theta} - \delta\dot{v}_{\theta}\,\sin{\theta} - \dot{\theta}\,(\delta v_{r^{\prime}}\,\sin{\theta} + \delta v_{\theta}\,\cos{\theta}) \\ 
\nonumber 
&= \tstop^{-1}\,(-\delta v_{r^{\prime}}\,\cos{\theta} + \delta v_{\theta}\,\sin{\theta} - \delta u_{\theta}\,\sin{\theta}  ) \\ 
\label{eqn:eom.shear.1}
&+ 2\,\Omega_{R}\,(\delta v_{r^{\prime}}\,\sin{\theta} + \delta v_{\theta}\,\cos{\theta}) \\ 
\nonumber \delta \dot{v}_{y} &= \delta \dot{v}_{r^{\prime}}\,\sin{\theta} + \delta \dot{v}_{\theta}\,\cos{\theta} + \dot{\theta}\,(\delta v_{r^{\prime}}\,\cos{\theta} - \delta v_{\theta}\,\sin{\theta}) \\ 
\nonumber
&= \tstop^{-1}\,(-\delta v_{r^{\prime}}\,\sin{\theta} - \delta v_{\theta}\,\cos{\theta} + \delta u_{\theta}\,\cos{\theta}  ) \\
&- \frac{1}{2}\,\Omega_{R}\,(\delta v_{r^{\prime}}\,\cos{\theta} - \delta v_{\theta}\,\sin{\theta}) 
\label{eqn:eom.shear.2}
\end{align}
First note that, even when $\Teddy\gg\tstop$, there is no $\theta$-independent solution if we retain all terms (unlike the case in \S~\ref{sec:encounters:pure.turb}, valid at all $\theta$). Because the shear terms break the symmetry of the problem, in the equilibrium solution a grain drifts on an approximately elliptical orbit, with an epicyclic correction to the circular solution which extends the orbit along the shear direction. The exact result in Appendix~\ref{sec:appendix:exact} accounts for this by computing the Jacobian for the distorted ellipse; however we can gain considerable intuition by considering the simplest ($\theta\approx 0$) case. 

With that caveat, we can follow the identical procedure as in \S~\ref{sec:encounters:pure.turb}. In this regime there are two solution branches: the first is again an exponentially growing solution with frequency $=\varpi/\Teddy$ and divergence 
\be
\label{eqn:rhomean.for.fullderiv}
\langle \deltarhonoabs\rangle_{\Leddy} = {\Bigl\langle} -\int_{0}^{\infty} (\nabla\cdot \delta {\bf v})\,{\rm d}t {\Bigr\rangle} \approx -2\,\varpi(\Leddy)\,\frac{\deltat}{\Teddy}
\ee
but now with $\varpi=\varpi_{1}$ given by the positive, real root of 
\begin{align}
\label{eqn:varpi.full}
0 &= \nonumber
16\,\taustopeddy^{3}\,\varpi_{1}^{4} + 
32\,\taustopeddy^{2}\,\varpi_{1}^{3} + 
\taustopeddy\,(20+7\,\taustop^{2})\,\varpi_{1}^{2}  \\ 
&
+ 4\,(1 + \taustop^{2} - 3\,\taustop\,\taustopeddy)\,\varpi_{1} - 
4\,(\taustopeddy+2\,\taustop)
\end{align}
\begin{align}
\varpi_{1} &\rightarrow
\begin{cases}
      {\displaystyle \taustopeddy}\ \ \ \ \ \ \ \ \ \ \   \hfill {\tiny (\taustopeddy\ll1)} \\ 
      {\displaystyle 2\,(\taustop+\taustop^{-1})^{-1}}\ \ \ \ \ \ \ \ \ \ \ \ \ \ \ \ \   \hfill {\tiny (\taustop\gg\taustopeddy)} \
\end{cases}
\end{align}

As expected, on small scales where $\Teddy\ll \Omega^{-1}$ ($\taustopeddy\gg\taustop$), this reduces to the solution for turbulence without shear (Eq.~\ref{eqn:varpi.pureturb}): we can write $\varpi_{1}$ in this limit just in terms of $\taustopeddy$, and it scales as $\taustopeddy$ for $\taustopeddy\ll1$. On sufficiently large scales, $\taustop\gg\taustopeddy$, we recover the solution we estimated in \S~\ref{sec:model.encounters:toy}, the approximately constant $\varpi_{1}\rightarrow 2/(\taustop+\taustop^{-1})$. 

Nominally for $\taustopeddy\gg1$ this gives $\varpi_{1}\rightarrow (2\,\taustopeddy)^{-1/2}$; however, in the full solution there is an additional ``early time'' solution branch we have dropped. Upon first encountering the eddy, the grains have zero mean (peculiar) vorticity, so the coherent $\delta v$ grows with time. At sufficiently early times, $\delta v$ is small and the solution to Eqs.~\ref{eqn:eom.shear.1}-\ref{eqn:eom.shear.2} is obtained by expanding to leading order in $\delta v$. After substitution to eliminate $\delta v_{\theta}$, the system simplifies to $\taustop^{2}\,\delta \ddot{v}_{r} + 2\,\taustop\,\delta \dot{v}_{r} + (1+\taustop^{2})\,\delta v_{r} = 2\,\taustop\,\delta u_{\theta}$. Just as the solution in \S~\ref{sec:encounters:pure.turb} above, this has two decaying oscillatory solutions which do not contribute to the integrated $t\rightarrow\infty$ divergence (since the equations are linearized), and peculiar solution $\delta v_{r} = 2\,\delta u_{\theta}/(\taustop+\taustop^{-1})$, i.e.\ just the ``large scale'' solution from before. This leads to an identical expression (Eq.~\ref{eqn:rhomean.for.fullderiv}) for $\langle \deltarhonoabs \rangle$, but with $\varpi = \varpi_{0} = 2/(\taustop+\taustop^{-1})$. This solution track dominates when $\varpi_{1}<\varpi_{0}$; when $\varpi_{1}>\varpi_{0}$, the $\varpi_{0}$ solution track dominates only for an initial time $t\ll \Teddy$, until (as $\delta v_{r}$ grows) the second-order terms in $\delta v$ become important and $\varpi\rightarrow\varpi_{1}$. Comparing to the exact numerical integration, it is straightforward to verify that general solution for both regimes is qualitatively represented by
\begin{align}
\varpi(\Leddy) = {\rm MAX}{\Bigl[} \varpi_{1},\ \varpi_{0}=2\,(\taustop+\taustop^{-1})^{-1} {\Bigr]}
\end{align}
This is directly analogous to our heuristic estimate in \S~\ref{sec:model.encounters:toy}; global angular momentum sets a ``floor'' in the (second-order) centrifugal force, here represented by $\varpi_{0}$.

Likewise, $\deltat$ obeys the same scalings as in \S~\ref{sec:encounters:pure.turb}, with ${\bf V}_{\rm rel}$ contributed by the turbulence, but now there is a non-zero laminar relative gas-grain flow, given by the equilibrium drift solution ${\bf V}_{L} = {\bf v}^{d} - {\bf u}^{d}$ in Eqs.~\ref{eqn:vdrift}-\ref{eqn:vdrift.last}:
\begin{align}
\label{eqn:v.laminar}
|{\bf V}_{L}|^{2} = \frac{4\,(1+\rhoratio)^{2}\,\taustop^{2} + \taustop^{4}}{[ (1+\rhoratio)^{2} + \taustop^{2} ]^{2}}\,(\eta\,\vk)^{2} = \frac{1}{\beta^{2}}\,|\Veddy(\Lmax)|^{2}
\end{align}
Together with Eqs.~\ref{eqn:rhomean.for.fullderiv}-\ref{eqn:varpi.full}, we can now write 
\begin{align}
\frac{\deltat}{\Teddy} &= {\Bigl[}1 +  {\Bigl(}\frac{\tdrift}{\Teddy} {\Bigl)}^{-1} {\Bigr]}^{-1} \\ 
\frac{\tdrift}{\Teddy} &= -{\taustopeddy}\,\ln{{\Bigl[}1 - \frac{(\Leddy/\Lmax)}{\taustopeddy(\Lmax)\,g(\Leddy)^{1/2}}  {\Bigr]}} 
\end{align}
where 
\begin{align}
\label{eqn:g.timescale.function}
g(\Leddy) &\equiv \frac{\langle |{\bf v}_{0} |^{2} \rangle}{|\Veddy(\Lmax)|^{2}} = g_{0}(\Leddy) + \frac{1}{\beta^{2}} \\
&= \frac{1}{\beta^{2}} + \taustopeddy(\Lmax)\,\ln{{\Bigl[} \frac{1+\taustopeddy(\Lmax)^{-1}}{1 + \taustopeddy(\Leddy)^{-1}} {\Bigr]}} \nonumber
\end{align}

\vspace{-0.5cm}
\subsection{Hierarchical Encounters with Many Structures}
\label{sec:model.hierarchy}

Now, if we assume the {\em gas} turbulence follows some simplified scalings, we can embed our estimates for the behavior in individual eddy encounters into the statistics of the eddies themselves.

Consider the following. We consider a random point ${\bf x}$ in space. Assume that it begins at the mean density. But if there are ``eddies'' present, which intersect the point, then the density will be modified according to our derivation above. So we need to determine ``how many'' eddies of different sizes are present. 

First, we will make the ad-hoc and heuristic assumption that the entire velocity field can be decomposed into a superposition of ``eddies'' of various ``sizes'' $\Leddy$, which are described by the toy model in the previous section. This is a tremendous simplification, but it allows us to phenomenologically model how a complicated velocity field might non-linearly affect grain clustering. Now, consider the scaling of the gas velocity statistics. Given the assumption of an eddy model, then from standard models of turbulent structure we expect $|\Veddy| \propto \Leddy^{\velslope}$ with $\velslope\approx1/3$. In more detail, the cascade models of \citet{shewaymire:logpoisson,dubrulle:logpoisson} predict that the structure functions $\sigma_{p}(\scalevar)=\langle |\Delta {\bf u}(\scalevar)|^{p} \rangle \equiv \langle |\delta{\bf u}({\bf x}) - \delta{\bf u}({\bf x}+{\bf \scalevar})|^{p} \rangle$ scale as power laws $\sigma_{p}(\scalevar)\propto \scalevar^{\zeta_{p}}$ with 
\be
\label{eqn:structfn}
\zeta_{p} = (1-\gamma)\,\frac{p}{3} + c_{g}\,{\Bigl[}1-{\Bigl(}1 - \frac{\gamma}{c_{g}}{\Bigr)}^{p/3}{\Bigr]}
\ee
where $\gamma=2/3$ follows generically from the Navier-Stokes equations (from the Kolmogorov $4/5$ths law), and $c_{g}=2$ follows from geometric arguments and fitting to experimental data.\footnote{In the original \citet{kolmogorov:turbulence} model, $\zeta_{p}=p/3$, giving the familiar $|\Veddy|\propto\Leddy^{1/3}$. The low-order differences between this and the more detailed multi-fractal models are small. So for our purposes, forcing $\velslope=1/3$ instead of Eq.~\ref{eqn:structfn} gives very similar predictions. However, a wide range of experiments favor the scaling in Eq.~\ref{eqn:structfn}.} What we refer to as $\Veddy$ is the one-point function (peculiar eddy velocity differences across the eddy) $p=1$, so we have 
\be
\label{eqn:velslope}
\zeta_{1} = \frac{1}{9} + c_{g}\,{\Bigl[}1 - {\Bigl(}1 - \frac{2}{3\,c_{g}} {\Bigr)}^{1/3} {\Bigr]} \approx 0.36
\ee

If all eddies/structures had a single ``size'' $\Leddy$, then the probability that our point ${\bf x}$ lies within the domain of an eddy is simply given by the volume filling factor of such structures. Since we have already assumed that eddies are statistically independent and discrete in \S~\ref{sec:model.encounters}, the number $m$ of such eddies (within the context of our very simplified assumptions) intercepting a random point must be Poisson-distributed (because this is just a counting exercise): 
\be
\label{eqn:poisson}
P(m) = P_{\DN}(m) = \frac{\DN^{m}}{m!}\,\exp{(-\DN)}
\ee
where the mean $\langle m \rangle \equiv \DN$ is simply related to the eddy filling-factor. 
Note that if the flow is ergodic, we can equivalently consider this the number of structures (e.g.\ vortices) encountered by a Lagrangian parcel over a coherence timescale \citep{cuzzi:2001.grain.concentration.chondrules,hopkins:frag.theory}. As argued in \citet{hopkins:2012.intermittent.turb.density.pdfs} and confirmed by \citet{federrath:2013.intermittency.vs.numerics,federrath.2015:density.pdf.and.sfr.in.polytropic.turbulence}, that assumption leads to a remarkably accurate description of the first-order intermittency corrections to the {\em gas} density PDFs in super-sonic isothermal turbulence.

Of course, in our eddy decomposition, within the inertial range, there is no single ``eddy size.'' Rather there is a hierarchy of all sizes from the Kolmogorov scale (vanishingly small, in the limit we consider) to the driving scale. Consider ``searching'' around our point ${\bf x}$ for eddies which intercept ${\bf x}$: we begin with a ``counting sphere'' of radius $\scalevar_{1}$ centered at ${\bf x}$ and then increase the size of the sphere by a differential interval $\Delta \ln{\scalevar} = \ln{\scalevar_{2}} - \ln{\scalevar_{1}}$. Obviously, we do not care if ``new'' eddies appear within the sphere with sizes $\Leddy \ll \scalevar_{1}$, since they cannot interact with the point ${\bf x}$. Nor do we care about eddies with sizes $\Leddy \gg \scalevar_{1}$, since we only move a tiny fraction across a single such eddy with this extension. What does matter is whether we find eddies with $\Leddy \sim \scalevar_{1}$. In a differential volume element, we expect to find some differential number $\DN_{1}(\Leddy\sim\scalevar_{1})$ of eddies which intercept the point ${\bf x}$.

Now, if the turbulent structure is self-similar (or truly fractal), then the only allowed scaling is $\DN \propto |\Delta \ln{\scalevar}|$ on all scales, i.e.
\be
\label{eqn:deltaN}
\DN = \deltadim\,|\Delta \ln{\scalevar}|
\ee
This is equivalent to the statement that the volume-filling factor of eddies with different sizes $\Leddy$ is constant -- the only possibility if the turbulence is truly self-similar.\footnote{Note, this is only a statement about the filling factor. The eddies can have any dimensionality, in principle. If eddies are, for example, thin filaments (co-dimension $C_{d}=3-1=2$), then if the filling factor is constant, we expect the {\em total} number of eddies we encounter in the differential volume element to increase as $\DN_{\rm tot}\propto \scalevar^{C_{d}}\,,|\Delta \ln{\scalevar}|$. This is the standard expectation in most geometric models of turbulence \citep[see e.g.][]{sheleveque:structure.functions,shezhang:2009.sheleveque.structfn.review}. However, {\em each} of those eddies only has a probability $\propto \scalevar^{-C_{d}}$ of intercepting the specific point ${\bf x}$. Thus, what we care about -- the number of eddies we ``find'' that interact with the point ${\bf x}$ -- scales $\propto |\Delta \ln{\scalevar}|$ as Eq.~\ref{eqn:deltaN}.}

Thus far, this is a purely geometric argument, which follows from the idea that the flow structure is self-similar over the inertial range. For the specific geometric assumptions we have made already about the size and characteristic velocity structure of eddies, then if we do assume the eddies of interest are two-dimensional circular vortices as described above (and that such eddies contain {\em all} the turbulent power) we can infer $\deltadim=2$ by simply normalizing the integrated power. But one could imagine only part of the power is in such structures, or trade off between the assumptions above about the relation between eddy size, turnover time, and $\deltadim$, and hence we will consider variations in this parameter below.

It is now straightforward to combine this with the arguments in \S~\ref{sec:model.encounters} to obtain the predicted grain density statistics. Following \S~\ref{sec:model.encounters}, assume that each encounter in $\Delta\ln{\scalevar}$ produces a multiplicative effect on the density statistics with the mean expected magnitude $\langle \deltarhonoabs\,[\Leddy = \scalevar] \rangle$ on that scale.\footnote{This simplification, that structures produce fluctuations of mean magnitude (given their scale), is substantial; yet for the gas velocity statistics it appears sufficient to capture the power spectrum and PDF shape, and higher-order structure function/correlation statistics to $\gtrsim10$th order in experiments \citep[see][]{shezhang:2009.sheleveque.structfn.review}. However, we will consider below what happens if there is a variation in the fluctuations produced by structures of the same scale.} In a probabilistic sense, as we sample the density statistics about some random point in space -- integrating the effects of all eddies, beginning by counting those with sizes comparable to the largest scales and successively counting smaller and smaller structures -- the statistics on successive scales $\scalevar_{1}$ and $\scalevar_{2}$ are given by 
\begin{align}
\nonumber
\ln{[\rhograin(\scalevar_{2})]} &= \ln{[\rhograin(\scalevar_{1})]} + m\,\langle \deltarhonoabs\,[\Leddy=\scalevar_{1}]\rangle + \epsilon_{0} \\ 
\label{eqn:logpoisson.1}
&= \ln{[\rhograin(\scalevar_{1})]} - m\,\deltarho + \epsilon_{0} 
\end{align}
where $m$ is Poisson-distributed as Eq.~\ref{eqn:poisson}-\ref{eqn:deltaN}, i.e.\ 
\begin{align}
P{\Bigl(}\ln{{\Bigl[}\frac{\rhograin(\scalevar_{2})}{\rhograin(\scalevar_{1})} {\Bigr]}}{\Bigr)}\,{\rm d}\ln{{\Bigl[}\frac{\rhograin(\scalevar_{2})}{\rhograin(\scalevar_{1})} {\Bigr]}} = P(m)\,{\rm d}m 
\end{align}
Mass conservation trivially determines the integration constant 
\be
\epsilon_{0} = \DN\,[1-\exp{(-\deltarho_{\Leddy=\scalevar_{1}})}]
\ee

Physically, this should be interpreted as follows. Beginning at the ``top'' scale $\Lmax$ (where $\rhograin=\langle \rhograin \rangle$ by definition), we can recursively divide the volume about a random point into smaller sub-volumes of size $\scalevar$, each containing a (discrete) number of structures (vortices) with characteristic scale $\sim\scalevar$ (which overlap the ``point'' we are zooming in on). Each such vortex produces another multiplicative effect on the local density field $\deltarhonoabs$ (additive in log-space); we simplify the statistics by assigning each its mean expected effect $\langle \deltarhonoabs(\scalevar,\,\tstop,..) \rangle$. Per \S~\ref{sec:model.encounters}, this effect applies {\em within} the eddy region (dispersing grains from of the eddy center). But ``dispersed'' grains must go somewhere; the $\epsilon_{0}$ term simply represents the mean effect on the density in the interstices {\em between} vortices, created by their expulsion of grains.\footnote{More rigorously, we could calculate the effects of the strain part of the turbulent field on actively ``trapping'' grains, rather than assuming purely ``passive'' trapping via this mass conservation argument. However, calculating the strain field in detail requires a more specific model for the structure of eddies: specifically, for their velocity decay at large radii, {\em and} the characteristic separations between eddies. If, however, we assume a configuration of Burgers vortices separated by lengths comparable to their core radii, then our simple model produces a PDF quite similar to what one would obtain by direct numerical integration, because the characteristic decay length for the eddy vorticity is itself similar to the eddy size and coherence size of the strain regions, $\sim\Leddy$. If we accounted for the exact spatial distribution of eddies, it would also not much change our result so long as the rms separation in space between eddies was $\le\,\Leddy$, which is our ``overlapping'' limit. That said, it is a particularly interesting subject for future work to consider the role of special strain regions which, although they may be rare, can act as ``caustics'' (for example, a ridgeline between two overlapping and counter-rotating eddies) and may be important local sites of extreme concentration. We will discuss the accuracy of this approximation further below.} These qualitative effects are  well-known from simulations and experiments (see references in \S~\ref{sec:intro}); this is their quantitative representation.

\vspace{-0.5cm}
\subsection{Behavior at High Grain Densities}
\label{sec:model.highrho}

Thus far, we have neglected the back-reaction of grains on the gas (our predictions are appropriate when $\rhograin < \rhogas$). To extrapolate to $\rhograin \gtrsim \rhogas$, we require additional assumptions. 

Recall, the background drift solution in Eqs.~\ref{eqn:vdrift}-\ref{eqn:vdrift.last} already accounts for $\rhoratio$. So, after subtracting this flow, the Eqs.~\ref{eqn:eom.peculiar} for peculiar grain motion remain identical; but the gas equation of motion is (dropping the shear terms for simplicity)
\begin{align}
\delta \dot{{\bf u}} &=  - \rhoratio\,{\Bigl(}\frac{\delta {\bf u}-\delta {\bf v}}{\tstop} {\Bigr)} - \frac{1}{\rhogas}\,\nabla \delta P_{\rm g} 
\end{align}
where $\nabla \delta P_{\rm g}$ represents the peculiar hydrodynamic forces. In sub-sonic turbulence, it seems reasonable to make the {\em ansatz} that the back-reaction, while it may distort the flow $\delta {\bf u}$, does not alter the {driving} force ($\nabla \delta P_{\rm g}$) that forms the eddy.\footnote{Although we caution that this cannot be strictly true in the regime of high $\rhoratio$ where the streaming instability operates \citep{goodman.pindor:2000.secular.drag.instabilities.grains,youdin.goodman:2005.streaming.instability.derivation}.} But we know that the ``zero back-reaction'' eddy structure $\delta {\bf  u}_{0}\equiv \delta {\bf u}(\rhoratio=0)$ is, by definition, a solution to the equation 
$\delta \dot{{\bf u}}_{0} = - \rhogas^{-1}\,\nabla \delta P_{\rm g}$. 
So decompose ${\bf u}$ into the sum $\delta{\bf u}\equiv \delta {\bf u}_{0} + \delta {\bf u}^{\prime}$, substitute, and obtain
\begin{align}
\label{eqn:duprime}
\delta \dot{{\bf u}}^{\prime} &=  - \rhoratio\,{\Bigl(}\frac{\delta {\bf u}_{0}+\delta {\bf u}^{\prime}-\delta {\bf v}}{\tstop} {\Bigr)} = 
\rhoratio\,{\Bigl(}\frac{\delta{\bf v}-\delta{\bf u}_{0}}{\tstop} + \frac{\delta {\bf u}^{\prime}}{\tstop}
{\Bigr)}
\end{align}

In the limit $\rhoratio\rightarrow0$, we know $\delta {\bf u}^{\prime}\rightarrow 0$. For $0<\rhoratio\ll1$, we expect $|\delta {\bf u}^{\prime}|\ll \delta {\bf u}$, so we can linearize the equations of motion and obtain $\delta\dot{{\bf u}}^{\prime} \approx (\rhoratio/\tstop)\,(\delta {\bf v} - \delta {\bf u}_{0}) \sim \rhoratio\,{\rm d}(\delta {\bf v} - \delta {\bf u}_{0})/{\rm d}t + \mathcal{O}(\rhoratio^{2})$ (where the latter follows from the grain and gas momentum equations assuming $\langle |\delta {\bf u}_{0}| \rangle > \langle |\delta {\bf v} | \rangle$), so $\delta {\bf u} \approx \rhoratio\,(\delta {\bf v} - \delta {\bf u}_{0} )$. In the limit $\rhoratio\rightarrow\infty$ ($\rhogas/\rhograin = \rhoratio^{-1} \rightarrow 0$), the gas is perfectly dragged by the grains, so $\delta {\bf u}\rightarrow \delta {\bf v}$ and $\delta {\bf u}^{\prime} \rightarrow \delta {\bf v} - \delta {\bf u}_{0}$. Linearizing in this limit in $\rhogas/\rhograin=\rhoratio^{-1}$ similarly gives $\delta {\bf u}^{\prime} \sim (1-\rhoratio^{-1})\,(\delta {\bf v} - \delta {\bf u}_{0}) + \mathcal{O}(\rhoratio^{-2})$. We can simply interpolate between these limits by assuming that, in equilibrium
\begin{align}
\delta {\bf u}^{\prime} \sim \frac{\rhoratio}{1+\rhoratio}\,{\Bigl(} \delta{\bf v} - \delta{\bf u}_{0} {\Bigr)}
\end{align}
We stress that this is {\em not} exact, but it at least gives the correct asymptotic behavior. Inserting this into the equation for $\delta {\bf v}$, we have
\begin{align}
\delta\dot{\bf v} = -\frac{(\delta {\bf v} - [\delta{\bf u}_{0} + \delta {\bf u}^{\prime}])}{\tstop} \rightarrow -\frac{(\delta {\bf v} - \delta{\bf u}_{0})}{\tstop\,(1+\rhoratio)}
\end{align}
But this is our original equation for $\delta {\bf v}$, modulo the substitution $\tstop\rightarrow\tstop(1+\rhoratio)$. So -- given the extremely simple ansatz here -- our derivation of $\varpi$ and previous quantities is identical, but we should replace $\tstop$ with an ``effective'' $t_{\rm s,\,\rho}\equiv\tstop\,(1+\rhoratio)$. In this lowest-order approximation, back-reaction lessens the relative velocities (hence friction strength) by dragging gas with grains, and thus lengthens the ``effective'' stopping time.

The timescale $\delta t$ is of course still limited by the eddy lifetime $\Teddy$, and the crossing time solution we previously derived already accounted for $\rhoratio>0$ (in the drift time), so we do not need to re-derive it. Finally, we will further assume that the back-reaction, while it may distort individual eddies, does not alter their fractal dimensions (hence other statistics like the gas power spectrum). Of course, this cannot be true in detail, and all of these assumptions are questionable in the limit of $\rhoratio\sim1$. Nonetheless, it provides us with a plausible ``guess,'' and allows us to phenomenologically extend our model to simulations with large $\rhoratio$. Below, we discuss the accuracy of these assumptions, and how well this simplistic extension actually performs.

\vspace{-0.5cm}
\section{Predictions}

\begin{figure}
    \centering
    \plotonesize{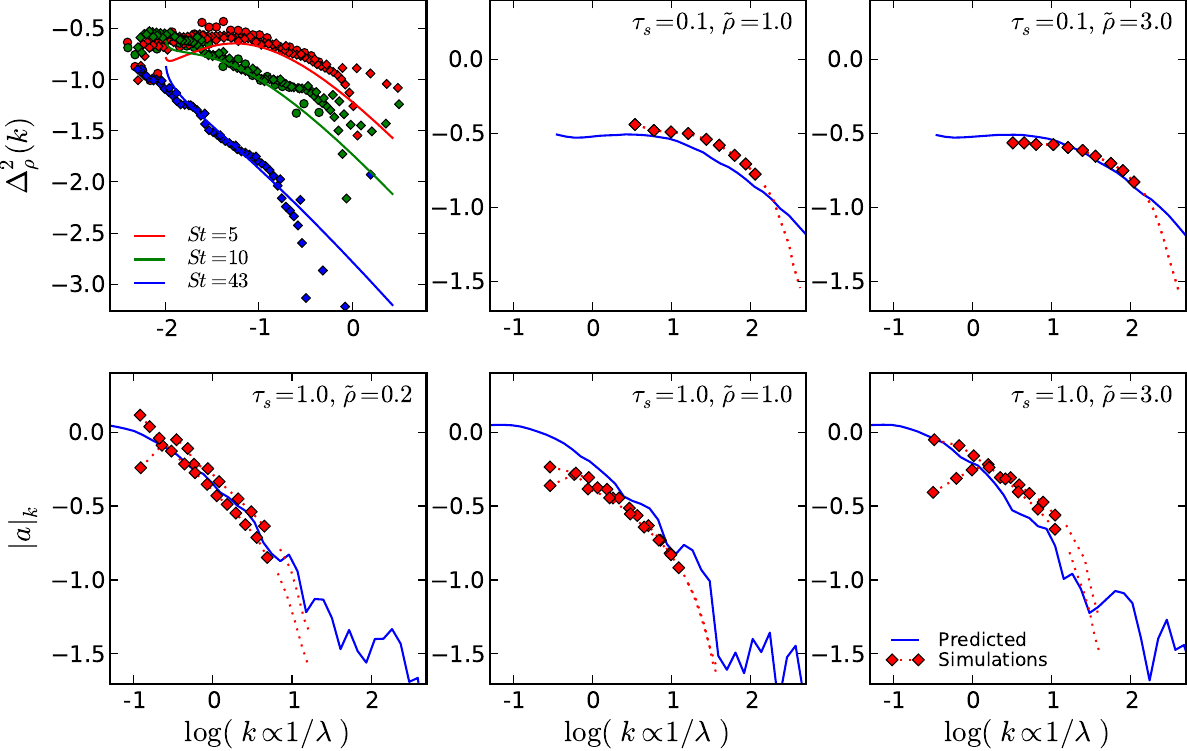}{1.01}
    \vspace{-0.4cm}
    \caption{Power spectra of linear grain density fluctuations. We compare the streaming-instability simulations in Fig.~\ref{fig:grain.rho.jy} for $\taustop=0.1-1$, and the turbulent concentration simulations ({\em top left}) in \citet[][squares]{yoshimoto:2007.grain.clustering.selfsimilar.inertial.range} and \citet[][diamonds]{pan:2011.grain.clustering.midstokes.sims} for Stokes numbers $\sim5,\,10,\,50$. Simulation results within five cells of the resolution limit are shown as dashed lines; the power suppression here is artificial. Agreement is good. For large particles, the power is preferentially concentrated on large scales. For small particles, the power spectrum is quite flat, until a turnover at scales below which $\Teddy\ll \tstop$.
    \label{fig:pwrspec}}
\end{figure}

\subsection{The Shape of the Grain Density Distribution}
\label{sec:pred.rhodist}

The full grain density PDF, averaged on any spatial scale, can now be calculated.

To do, we start on the initial scale $\scalevar=\Lmax$. By definition, since this is the top scale of the turbulence and/or box, the density distribution is a delta function with $\rhograin(\Lmax) = \langle \rhograin \rangle$. Now take a differential step in scale $\ln{\scalevar} \rightarrow \ln{\Lmax} - {\rm d}\ln{\scalevar}$ and convolve this density with the PDF of density changes $P(\ln[\rhograin(\scalevar_{2})/\rhograin(\scalevar_{1})])$ from \S~\ref{sec:model.hierarchy}. Repeat until the desired scale is reached; the one-point density PDF is just this iterated to $\scalevar\rightarrow0$.\footnote{Really we should truncate (or modify) this at the viscous scale $\scalevar_{\nu}$; here we assume large Reynolds number $Re\rightarrow\infty$. However the distinction is important for modest Stokes numbers $St=\tstop/\Teddy(\scalevar_{\nu})$ or simulations with limited resolution (small ``effective'' $St$).}

It is easiest to do this with a Monte-Carlo procedure; each point in a large ensemble represents a random point in space (thus they equally sample volume) with its own independent $\ln{\rho}_{i}$. For each step in scale $\Delta\ln{\scalevar}$, draw $m=m_{i}$ for each point from the appropriate Poisson distribution for the step (Eq.~\ref{eqn:poisson}), and use $|\deltarhonoabs(\scalevar)|$ to calculate the change in $\ln{\rho}_{i}$ (Eq.~\ref{eqn:logpoisson.1}), and repeat until the desired scale is reached. 

Recall that for the largest eddies $\deltarho\approx 2\,\Ndim/(\taustop+\taustop^{-1})$ is approximately constant. In that case, the log-Poisson distribution in \S~\ref{sec:model.hierarchy} is scale-invariant and infinitely divisible, meaning that the integrated PDF of $\ln{\rho}$ is {\em also} exactly a log-Poisson distribution on all scales, with the same $\deltarho\sim$\,constant, and $m$ drawn from a Poisson distribution with the integrated $\Delta N = \deltadim\,\ln{(\Lmax/\scalevar)}$. 

However, if $\deltarho$ depends on scale (as it does on small scales), then the convolved distribution is not exactly log-Poisson. But it is quite accurately approximated by a log-Poisson with the same mean and variance as the exact convolved distribution \citep[see][]{stewart:2006.gamma.function.convolution.approximation}. These quantities add linearly with scale. Over the differential interval in scale ${\rm d}\ln{\scalevar}$ ($\DN=\deltadim\,{\rm d}\ln{\scalevar}$), the added variance ($\Delta S$) in ${\Delta}\ln{\rho} = \ln{(\rhograin[\scalevar_{2}]/\rhograin[\scalevar_{1}])}$ is $\Delta S = \DN\,\deltarho^{2}$. So the exact integrated variance in the final volume-weighted $\ln{\rho}$ distribution is 
\begin{align}
\label{eqn:pdf.S}
S_{\ln{\rho},\,V}(\scalevar) &= \int \frac{{\rm d}S_{\ln{\rho},\,V}}{{\rm d}\ln{\scalevar}}\,{\rm d}\ln{\scalevar} = \int_{\scalevar}^{\Lmax} \DN\,|\deltarhonoabs|^{2} \\
&= \int_{\scalevar}^{\Lmax} \deltadim\,|\deltarhonoabs(\scalevar)|^{2}\,{\rm d}\ln{\scalevar} \nonumber  
\end{align}
And the integrated first moment (subtracting the $\epsilon_{0}$ term) is 
\begin{align}
\label{eqn:pdf.mu}
\mu &= \int \DN\,\deltarho = \int_{\scalevar}^{\Lmax}\deltadim\,|\deltarhonoabs(\scalevar)|\,{\rm d}\ln{\scalevar}
\end{align}

The approximate integrated PDF on a scale $\scalevar$ is then given by 
\begin{align}
P_{V}&(\ln{\rhograin})\,{\rm d}\ln{\rhograin} \approx \frac{\Delta N_{\rm int}^{m}\,\exp{(-\Delta N_{\rm int})}}{\Gamma(m+1)}\,\frac{{\rm d}\ln{\rhograin}}{\deltarho_{\rm int}} \\ 
\nonumber
m &= \deltarho_{\rm int}^{-1}\,{\Bigl\{}\Delta N_{\rm int}\,{\Bigl[}1 - \exp{(-\deltarho_{\rm int})} {\Bigr]} - \ln{{\Bigl(} \frac{\rhograin}{\langle \rhograin \rangle}{\Bigr)}} {\Bigr\}}
\end{align}
which is just the log-Poisson distribution (Eq.~\ref{eqn:logpoisson.1}) with 
\begin{align}
\Delta N \rightarrow \Delta N_{\rm int} &\equiv \frac{\mu^{2}}{S_{\ln{\rho},\,V}} \\ 
\label{eqn:deltarho.int}
\deltarho \rightarrow |\deltarhonoabs|_{\rm int} &\equiv \frac{S_{\ln{\rho},\,V}}{\mu}
\end{align}
Note $\Delta N_{\rm int}$ is now $\sim \deltadim\,\langle \ln{(\Lmax/\scalevar)} \rangle$, where $\langle...\rangle$ denotes an average over integration weighted by $\deltarho$ (i.e.\ the ``effective'' dynamic range of the cascade which contributes to fluctuations). And $\deltarho_{\rm int}$ similarly reflects a variance-weighted mean. 


This determines the volumetric grain density distribution, i.e.\ the probability, per unit volume, of a given mean grain density $\rhograin=M_{\rm p}(V)/V$ within that volume $V$
\be
P_{V}(\ln{\rhograin}) = \frac{{\rm d}P_{\rm Vol}(\ln{\rhograin})}{{\rm d}\ln{\rhograin}}
\ee
This is trivially related to the mass-weighted grain density distribution $P_{M}$, or equivalently the Lagrangian grain density distribution (distribution of grain densities at the location of each grain, rather than at random locations in the volume): 
\be
P_{M}(\ln{\rhograin}) = \frac{{\rm d}P_{\rm Mass}(\ln{\rhograin})}{{\rm d}\ln{\rhograin}} = \rhograin\,P_{V}(\ln{\rhograin})
\ee

Note that as $\Delta N_{\rm int}\rightarrow\infty$, this distribution becomes log-normal. This is generically a consequence of the central limit theorem, for sufficiently large number of independent multiplicative events in the density field.

\vspace{-0.5cm}
\subsection{The Grain Density Power Spectrum}
\label{sec:pred.pwrspec}

The power spectrum of a given quantity is closely related to the real-space variance as a function of scale. Specifically, the variance in some field smoothed with an isotropic real-space window function of size $\scalevar$ is related to the power spectrum by
\be
S(\scalevar) = \int {\rm d}^{3}{\bf k}\,P({\bf k})\,|W({\bf k},\,\scalevar)|^{2}
\ee
where $W$ is the window function. If we isotropically average, and adopt for convenience a window function which is a Fourier-space top-hat\footnote{We treat the isotropically-averaged, Fourier-space top-hat case purely for convenience, because it is usually measured and more relevant on small scales. This is not the same as assuming the power spectrum is intrinsically isotropic or that Fourier modes are uncoupled.} we obtain
\be
S(\scalevar) = \int_{k=1/\scalevar}^{\infty}\,\Delta^{2}(k)\,{\rm d}\ln{k}
\ee
where $\Delta^{2}(k)$ is now defined as the isotropic, dimensionless power spectrum, and is related to $S(\scalevar)$ by 
\be
\frac{{\rm d}S}{{\rm d}\ln{\scalevar}} = \Delta^{2}(k[\scalevar])
\ee

But we know how the variance ``runs'' as a function of scale, for the logarithmic density distribution. Specifically, for the distribution in Eq.~\ref{eqn:logpoisson.1}, over some differential interval in scale ${\rm d}\ln{\scalevar}$ corresponding to $\DN=\deltadim\,{\rm d}\ln{\scalevar}$, the variance in $\ln{\rho}$ is just $\Delta S = \DN\,\deltarho^{2}$ (and this adds linearly in scale). So 
\begin{align}
\label{eqn:pwrspec}
\Delta^{2}_{\ln{\rho}}(k) &= \frac{ {\rm d}\,S_{\ln{\rho}} }{ {\rm d}\ln{\scalevar} } = 
\deltadim\,\deltarho^{2} = \deltadim\,{\Bigl[} \Ndim\,\varpi(\scalevar)\,\frac{\deltat}{\Teddy} {\Bigr]}^{2}
\end{align}
Recall the turnover in $(\deltat/\Teddy)$ below $\Lcrit$ (\S~\ref{sec:model.encounters}), which leads to a two-power law behavior in $\Delta^{2}$: 
\begin{align}
\Delta_{\ln{\rho}}^{2}(k) \propto 
\begin{cases}
      {\displaystyle {\rm constant}}\ \ \ \ \   \hfill {\tiny (\scalevar \gg \Lcrit)} \\ 
      \\
      {\displaystyle \Veddy^{2} \propto k^{-2\velslope}}\ \ \ \ \  \hfill {\tiny (\scalevar \ll \Lcrit)} \
\end{cases}
\end{align}
i.e.\ we predict a turnover/break in the power spectrum at a characteristic scale $\Lcrit$ (defined in \S~\ref{sec:model.encounters} as the scale where the timescale for grains to cross an eddy is shorter than the stopping time). On large scales where eddy turnover times are long, the logarithmic statistics are nearly scale-free, but on small scales, where eddy turnover times are short compared to the stopping and drift times, the variance is suppressed.

The power spectrum for the {\em linear} density field $\rho$ is similarly trivially determined as: 
\be
\Delta_{\rho}^{2} = \frac{{\rm d}S_{\rho}}{{\rm d}\ln{\scalevar}}
\ee
However $S_{\rho}$ is not so trivially analytically tractable, since the total variance does not sum simply in {\em linear}\footnote{This is a general point discussed at length in \citet{hopkins:frag.theory}, Appendices~F-G; it is not, in general, possible to construct a non-trivial field distribution that is simultaneously scale-invariant under linear-space and logarithmic-space convolutions. However, as shown therein, the compound log-Poisson cascade is {\em approximately} so, to leading order in the expansion of the logarithm.} $\rho$. But it is straightforward to construct $S_{\rho}$, by simply using Eq.~\ref{eqn:logpoisson.1} to build the density PDF at each scale, directly calculating the variance in the linear $\rho$, and then differentiating. If the density PDF is approximately log-Poisson, then we can have 
\begin{align}
\label{eqn:Srho.approx}
S_{\rho} \approx \exp{{\Bigl\{} \Delta N_{\rm int}\,{\Bigl(}1 - e^{-\deltarho_{\rm int}}{\Bigr)}^{2} {\Bigr\}}} - 1
\end{align}
This leads to the somewhat cumbersome expression for $\Delta^{2}_{\rho}$: 
\begin{align}
\nonumber
\Delta^{2}_{\rho} \approx&\, \deltadim\,S_{\rho}\,\frac{\deltarho_{\rm int}}{|\deltarhonoabs(\scalevar)|^{2}}\,
e^{-2\,\deltarho_{\rm int}}\,{\Bigl(}e^{\deltarho_{\rm int}} -1 {\Bigr)}\, \\
\nonumber
&
\times{\Bigl [} 2\,\deltarho_{\rm int}\,{\Bigl(}e^{\deltarho_{\rm int}} -1 + |\deltarhonoabs(\scalevar)| {\Bigr)} \\ 
& - 2\,\deltarho_{\rm int}^{2} 
 - |\deltarhonoabs(\scalevar)|\,{\Bigl(}e^{\deltarho_{\rm int}} -1 {\Bigr)}
{\Bigr]}
\end{align}
But the limits are easily understood: if $\deltarho_{\rm int}\approx|\deltarhonoabs(\scalevar)|$, $\Delta^{2}_{\rho}\rightarrow \deltadim\,[1 - \exp{(-\deltarho)}]^{2}\,S_{\rho}$: this just reflects the scaling from the ``number of structures.'' If $\deltarho\ll1$, this further becomes $\Delta^{2}_{\rho}\sim\deltadim\,\deltarho^{2} = \Delta^{2}_{\ln{\rho}}$, since for small fluctuations the linear and logarithmic descriptions are identical. If $\deltarho\gtrsim1$ is large, $\Delta^{2}_{\rho}\sim \deltadim\,\exp{(\Delta N_{\rm int})} \sim \deltadim\,(\scalevar/\Lmax)^{-\deltadim}$ is a power-law, whose scaling (slope $\deltadim\sim2$) only depends on geometric scaling of the ``number of structures'' (fractal dimension occupied by vortices).

\vspace{-0.5cm}
\subsection{Correlation Functions}

The (isotropically averaged) autocorrelation function $\xi(r)$ is 
\begin{align}
\xi(r) &\equiv \frac{1}{\langle\rhograin\rangle^{2}}\,\langle(\rhograin({\bf x})-\langle\rhograin\rangle)\,(\rhograin({\bf x})-\langle\rhograin\rangle)\rangle
\end{align}
equivalently, this is the excess probability of finding a number of grains in a volume element ${\rm d}V$ at a distance $r$ from a given particle (not a random point in space)\footnote{Since we assume a uniform grain population, we treat grain mass and number densities as equivalent.}
\begin{align}
\langle {\rm d}N_{\rm p}(r,\,r+{\rm d}r) \rangle &= \langle n_{\rm p} \rangle\,{\rm d}V\,[1 + \xi(r)]
\end{align}

$\xi(r)$ is directly related to the variance $\langle (\rhograin[r]-\langle \rhograin\rangle)^{2}\rangle$ of the linear density field $\rhograin[r]$  averaged on the scale $r$, by
\begin{align}
\label{eqn:corrfn.r}
\frac{1}{V(r)}\int_{V(r)}\,\xi({\bf r}^{\prime})\,{\rm d}^{3}\,{\bf r^{\prime}} &= \frac{\langle (\rhograin(r)-\langle\rhograin\rangle)^{2} \rangle}{\langle \rhograin \rangle^{2}} \equiv S_{\rho}  
\end{align}
\citep{peebles:1993.cosmology.textbook}.\footnote{We assume the absolute number of grains is large so we can neglect Poisson fluctuations.} So the correlation function contains the same statistical information as the density power spectrum; and if we calculate $S_{\rho}$ above it is straightforward to determine $\xi(r)$ by Eq.~\ref{eqn:corrfn.r}. 

Note that if $\xi(r)$ is a power-law, $S_{\rho}(r)\sim \xi(r)$. And if $\deltarho_{\rm int}\ll1$ ($\taustop\ll1$), Eq.~\ref{eqn:Srho.approx} simply becomes $S_{\rho}\sim \Delta N_{\rm int}\,\deltarho_{\rm int}^{2}$. On the largest scales $\Teddy\gtrsim\Omega^{-1}$, $\deltarho\sim$\,constant so $\xi(r)$ rises weakly (with $\Delta N_{\rm int}$) with decreasing $r$ as $1+\xi(r) \propto \ln{(1/r)}$; approaching scales $\Teddy\sim\tstop$, $\deltarho \propto \taustopeddy$ rises so $\xi(r)\propto \taustopeddy^{2} \propto \scalevar^{-2\,(1-\velslope)}$ rises as a power law with a slope near unity; finally on small scales $\Teddy\lesssim \tstop$, $\deltarho$ falls rapidly, so $\deltarho_{\rm int}$ and $\Delta N_{\rm int}$ converge and $\xi(r)\rightarrow$\,constant.

\begin{figure}
    \centering
    \hspace{-0.2cm}
    \plotonesize{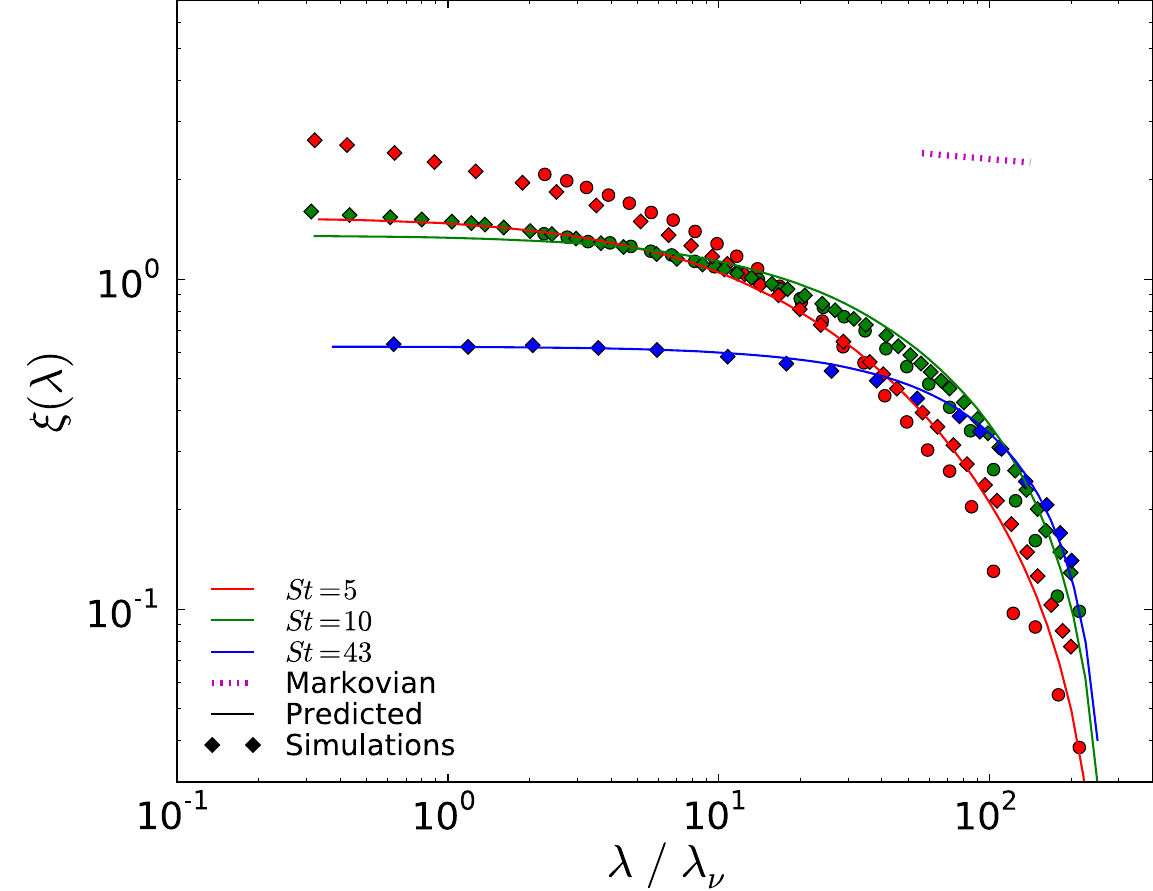}{1.01}
    \vspace{-0.5cm}
    \caption{Radial grain correlation functions, for the same simulations in Fig.~\ref{fig:pwrspec}. The predictions agree well at $St\gg1$ and/or scales $\scalevar\gg\scalevar_{\nu}$, but the clustering is under-predicted at $\scalevar\lesssim\scalevar_{\nu}$ for $St\sim1$, owing to non-inertial range effects we do not include. The shallow dotted line shows the slope predicted for Markovian (pure random-field) fluctuations (the amplitude is below the range plotted) in the inertial range ($\scalevar\gg\scalevar_{\nu}$), following \citet{bec:2007.grain.clustering.markovian.flow}: uncorrelated/incoherent fluctuations lead to negligible clustering when $St\gg1$.
    \label{fig:correlation.functions}}
\end{figure}

\vspace{-0.5cm}
\subsection{Maximum Grain Densities}
\label{sec:pred.rhomax}

Using the predicted grain density PDFs, we can predict the maximum grain densities that will arise under various conditions.

In Eq.~\ref{eqn:logpoisson.1}, note that there is, in fact, a maximum density, given by $m=0$ (and $\epsilon_{0}=0$) on all scales. This is approximately the density where the distributions ``cut off'' in Figs.~\ref{fig:grain.rho.mri}-\ref{fig:grain.rho.jy}, steeper than a Gaussian. It is straightforward to estimate this using $\rhograin(\Lmax)=\langle \rhograin \rangle$, and taking $m=0$ on all scales:
\begin{align}
\label{eqn:rhomax}
\ln{{\Bigl(} \frac{\rho_{\rm p,\,max}[\scalevar]}{\langle\rhograin\rangle}{\Bigr)}} &= \int_{\scalevar}^{\Lmax} \epsilon_{0} \\
&=
\nonumber
\deltadim\,\int_{\scalevar}^{\Lmax}
{\Bigl[}1 - \exp{{(}-{\deltarho}{)}} {\Bigr]}
\,{\rm d}\ln{\scalevar}
\end{align}

Trivially, we see
\begin{align}
\frac{{\rm d}\ln{\rho_{\rm p,\,max}}}{{\rm d}\ln{\scalevar}} = -\deltadim\,{\Bigl[}1 - \exp{{(}-{|\deltarhonoabs(\scalevar)|}{)}} {\Bigr]}
\end{align}
i.e.\ $\rho_{\rm p,\,max}$ behaves {\em locally} over some scale range in $\scalevar$ as a power-law $\rho_{\rm p,\,max} \propto \scalevar^{-\gamma}$ with slope $\gamma \equiv \deltadim\,[1 - \exp{(-\deltarho)}]$. When $\deltarho\ll1$ is small, $\gamma\sim \deltadim\,\deltarho$ is also small, so $\rho_{\rm p,\, max}$ grows slowly. For sufficiently large $\deltarho\gtrsim1$, $\gamma\sim\deltadim\sim2$ saturates at a value determined by the fractal filling factor of vortices -- $\rho_{\rm p,\,max}$ grows rapidly with scale, in a power-law fashion with slope $\sim2$ determined by the density of structures in turbulence.

\begin{figure}
    \centering
    \plotonesize{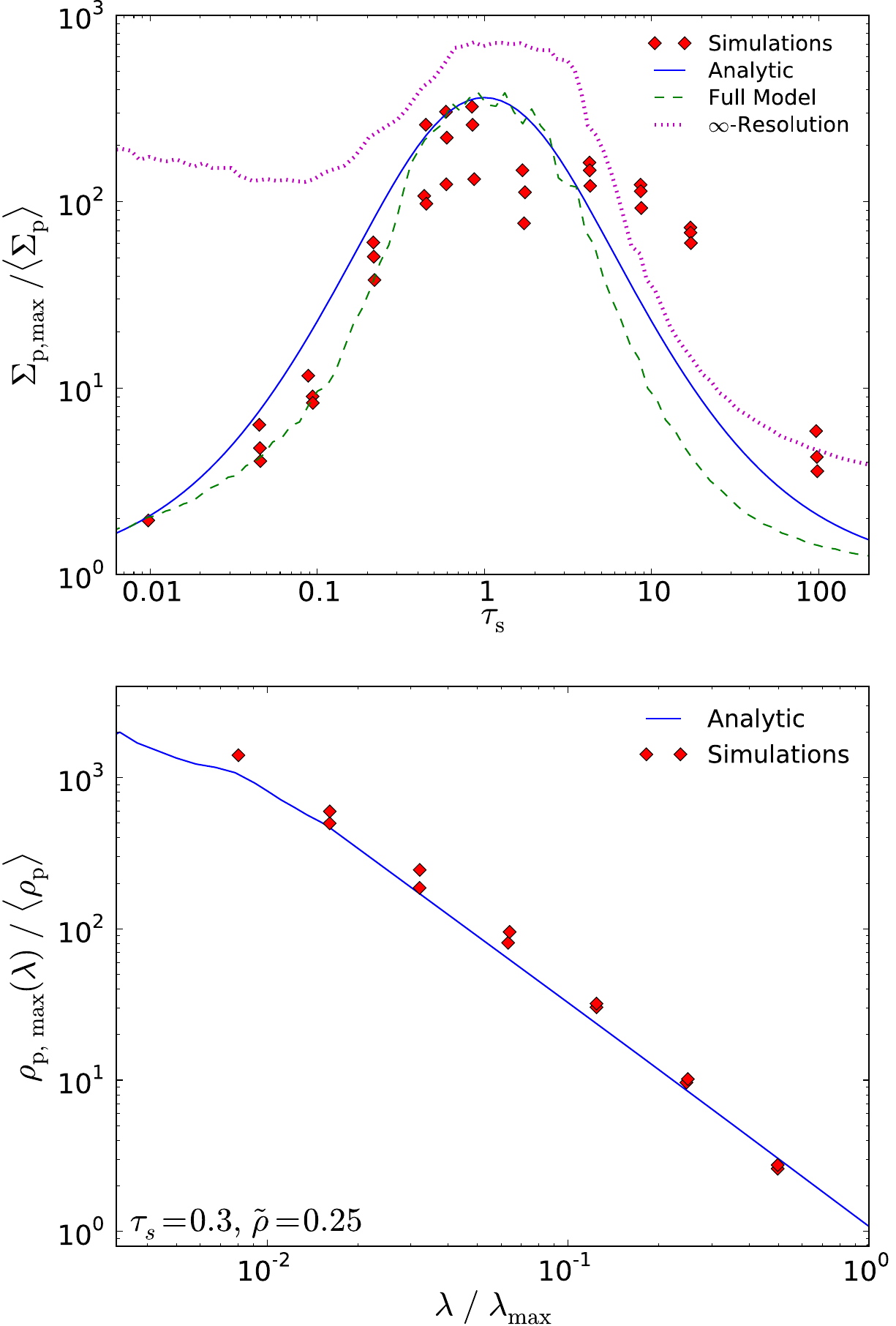}{0.98}
    \vspace{-0.15cm}
    \caption{{\em Top:}
    Maximum grain density measured in the MRI simulations from Fig.~\ref{fig:grain.rho.mri}, as a function of population stopping time $\taustop$. Different simulations with various box sizes and resolution are plotted, the predictions should form an ``upper envelope.'' We compare our simple approximation for large eddies and full prediction, given the (finite) simulation resolution and turbulence properties. These agree very well up to $\taustop\sim$ a few, though they under-predict fluctuations when $\taustop\gg1$. On large scales, maximum densities increase rapidly with $\taustop$ up to $\taustop\sim1$.
    We also show the prediction if the simulation were infinitely high-resolution (densities measured on arbitrarily small scales). In this regime, $\taustop\ll1$ grains also show large fluctuations; however, these high densities are manifest on very small scales. Roughly, convergence to this solution requires resolving eddies with $\Teddy\gtrsim0.05\,\tstop$; for the smallest $\tstop$ and given simulation turbulence properties and box sizes here, this would require a minimum $\sim(10^{6})^{3}$-cell simulation. 
    {\em Bottom:} Maximum grain density in simulations from \citet{johansen:2012.grain.clustering.with.particle.collisions}, averaged on different smoothing scales $\scalevar$. For large grains, the predicted maximum increases on smaller smoothing scales $\propto \scalevar^{-(1-2)}$.
    \label{fig:rho.max}}
\end{figure}

\vspace{-0.5cm}
\section{Comparison with Simulations and Experiments}

Going forward, unless otherwise specified we will assume the ``default'' $\deltadim=2$ and $\Ndim=2$. The values $\rhoratio$ and $\taustop$ are necessarily specified for each experiment. With these values, we only need two or three additional parameters to completely determine our model predictions. One is the ratio of eddy turnover time to stopping time on the largest scales $\taustopeddy(\Lmax)$ (or equivalently, ratio $\Teddy(\Lmax)/\Omega^{-1}$), the other is the ratio of mean drift to turbulent velocity $\beta \equiv |\Veddy(\Lmax)|/|\vdrift|$ (or equivalently the disk parameters $\alpha^{1/2}/\Pi$). These are properties of the gas turbulence and mean flow, so in some cases are pre-specified but in other cases are determined in a more complicated manner by other forces. To compare to simulations, we also (in some cases) need to account for their finite resolution, i.e.\ minimum $\scalevar/\Lmax$ or effective Reynolds number. Recall, we assumed a full inertial-range scaling for the turbulence; but many simulations resolve only a very limited (or no) inertial range. As a result, grain clustering (especially on small scales) may be under-estimated, and it would be inappropriate to compare our model assuming a fully-resolved cascade to the limited dynamic range of the simulations. Finally, the one additional parameter which is sometimes important is the fraction of eddies which are anti-cyclonic (anti-aligned with the shear flow in the disk) as opposed to cyclonic, when we compare to simulations with an externally imposed shear flow (e.g.\ a Keplerian gravitational field). For reasons we discuss below, our ``default'' model assumes these are equally likely.

\vspace{-0.5cm}
\subsection{Density PDFs}

\subsubsection{Externally Driven MRI Turbulence}
\label{sec:pred.rhodist.mri}

First consider the simulations in \citet{dittrich:2013.grain.clustering.mri.disk.sims}. These solve the equations of motion for the coupled gas-grain system, in full magnetohydrodynamics (MHD), for a grain population with a single stopping time $\tstop$. The simulations are performed in a three-dimensional, vertically stratified shearing box in a Keplerian potential, and there is a well defined local $\Omega$, $\eta$, $\Pi=0.05$. The simulations develop the magnetorotational instability (MRI), which produces a nearly constant $\alpha\approx 0.004$ (in our units defined here) in the dust layer, and back reaction on gas from grains is ignored so we can take $\rhoratio\rightarrow0$. The authors record a grain density PDF for $\taustop=1$ with large $\rhograin$ fluctuations arising as the MRI develops (their Fig.~11), to which we compare in Fig.~\ref{fig:grain.rho.mri}. Since the disk is vertically stratified, $\langle \rhograin \rangle$ depends on vertical scale height, so we would (ideally) compare our predictions separately in each vertical layer (though they are most appropriate for the disk midplane). Lacking this information, we should compare instead to the local surface density of grains relative to the mean grain surface density $\Sigma_{\rm p}/\langle \Sigma_{\rm p} \rangle$, which is independent of stratification and (as shown therein) closely reflects the distribution of mid-plane grain densities. 

First we compare our exact prediction (computed via the Monte-Carlo method in \S~\ref{sec:pred.rhodist}). Our natural expectation $\deltadim=\Ndim=2$ gives a remarkably accurate prediction of the simulation results! In fact, freeing $\Ndim$ and fitting to the data does not significantly improve the agreement (best-fit $\Ndim\approx1.9\pm0.1$). We also compare with our closed-form analytic approximation to the integrated density distribution, from Table~\ref{tbl:largescale}. This gives a very similar result, indicating that for this case (modest-resolution simulations, so densities are not averaged on extremely small scales, and large $\taustop=1$), the large-scale approximation is good.

In \S~\ref{sec:model.hierarchy}, we adopt the simplest assumption for the effects of an eddy (multiplication by $\langle \deltarhonoabs \rangle$). As noted there, one might extend this model by instead adopting a distribution of multipliers, with characteristic magnitude $\deltarhonoabs$. Here we consider one such example. For each ``event'' in $m$ in the log-Poisson hierarchy, instead of taking $\ln{\rhograin}\rightarrow \ln{\rhograin} + \langle \deltarhonoabs \rangle$, assume the ``multiplier'' is drawn from a Gaussian distribution, so $\ln{\rhograin}\rightarrow\ln{(\rhograin\,[1 + \mathcal{R}])}$ where $\mathcal{R}$ is a Gaussian random variable with dispersion $\langle \mathcal{R}^{2} \rangle^{1/2}=\langle \deltarhonoabs \rangle$. This is a somewhat arbitrary choice, but illustrative and motivated by Gaussian-like distributions in eddy velocities and lifetimes (and it has the advantage of continuously extending the predictions to all finite $\rhograin$, while producing the same change in variance as our fiducial model over small steps in $\deltarhonoabs$);\footnote{Note that we do have to enforce a truncation where $\mathcal{R}>-1$ to prevent an unphysical negative density. However because the $\deltarho$ along individual ``steps'' is small, this has only a small effect on the predictions.} we could instead adopt a $\beta$-model as in \citet{hogan:2007.grain.clustering.cascade.model}, but it would require additional parameters. In either case, this could, for example, represent the known effects of intermittency leading to variations in eddy intensity from region to region within the flow. Here, however, we see that this makes little difference to the predicted PDF. The reason is that the variance predicted in $\rhograin$ is dominated by the variance in the local turbulent field (the ``number of structures'' on different scales), and by the scales on which those structures appear -- not by the variance inherent to an individual structure on a specific scale. 

However, a number of model assumptions are important here, and we illustrate this in Fig.~\ref{fig:grain.rho.mri.examples}. Recall, this simulation includes an external Keplerian field, therefore eddies lead to different outcomes depending on whether they are cyclonic or anti-cyclonic. If we assume all eddies are cyclonic, we obtain too broad a distribution at low densities, and too rapid a cutoff at high densities: this is because cyclonic eddies do not actively concentrate grains, but dispel them, and the concentration is a secondary effect resulting from their trapping in strain regions. But sufficiently large anti-cyclonic eddies lead to large positive concentration effects, hence a stronger tail towards high concentrations. But, assuming all eddies are anti-cyclonic is similarly problematic, predicting too much concentration. Moreover, because we define all eddies relative to co-moving coordinates with the Keplerian disk gas orbits, the total angular momentum in eddies should vanish; if all eddies were cyclonic or anti-cyclonic, it would change the global angular momentum of the system, invalidating our original assumptions. To physically maintain this configuration, the gas would have to develop an addition drift and different orbits as angular momentum is transferred into eddies. 

Remarkably, if we simply assume random eddy orientations (half cyclonic and half anti-cyclonic, with the relative number drawn randomly from a binomial distribution for each number of eddies in the log-Poisson distribution we model), we obtain a good fit to the simulations. This is also the simplest physical configuration which gives negligible net angular momentum in eddies, consistent with our original solution for the gas orbits. For now, we will adopt it as our ``default,'' but wish to stress the caveat that we do not have a full model for what drives the distribution of eddy orientations. 

Interestingly, the role of anti-cyclonic vortices at the high-density tail of the PDF suggests that one shortcoming of our model -- the fact that we do not explicitly treat the strain field, but only vorticity, in modeling this regime -- may be partially ameliorated by the fact that, in cases where large positive-density fluctuations occur, they are dominated by such eddies, rather than by trapping in the strain field around cyclonic eddies.

We can also explore the effects of varying $\Ndim$ and $\deltadim$ in our simple model. At otherwise fixed properties, lower $\Ndim$ leads to lower variance, as expected, since the effects of a single eddy scale $\propto \Ndim$. As noted above, our simplest choice $\Ndim=2$ appears to work very well, while $\Ndim=1$ and $\Ndim=3$ do not. The effects of $\deltadim$ are more subtle. Naively, this also simply multiplies in the variance, and if we keep the product $\Ndim^{2}\,\deltadim$ fixed (the combination which enters our estimate of the variance), we do not see large effects from changing $\deltadim$. In fact a range of $\deltadim\sim1-2$ is permitted, and agrees plausibly with a number of the simulation and experiment metrics to which we compare, if we allow for (relatively) small changes in $\Ndim$ or the cyclonic/anti-cyclonic ratio. 


\begin{figure}
    \centering
    \hspace{-0.2cm}
    \plotonesize{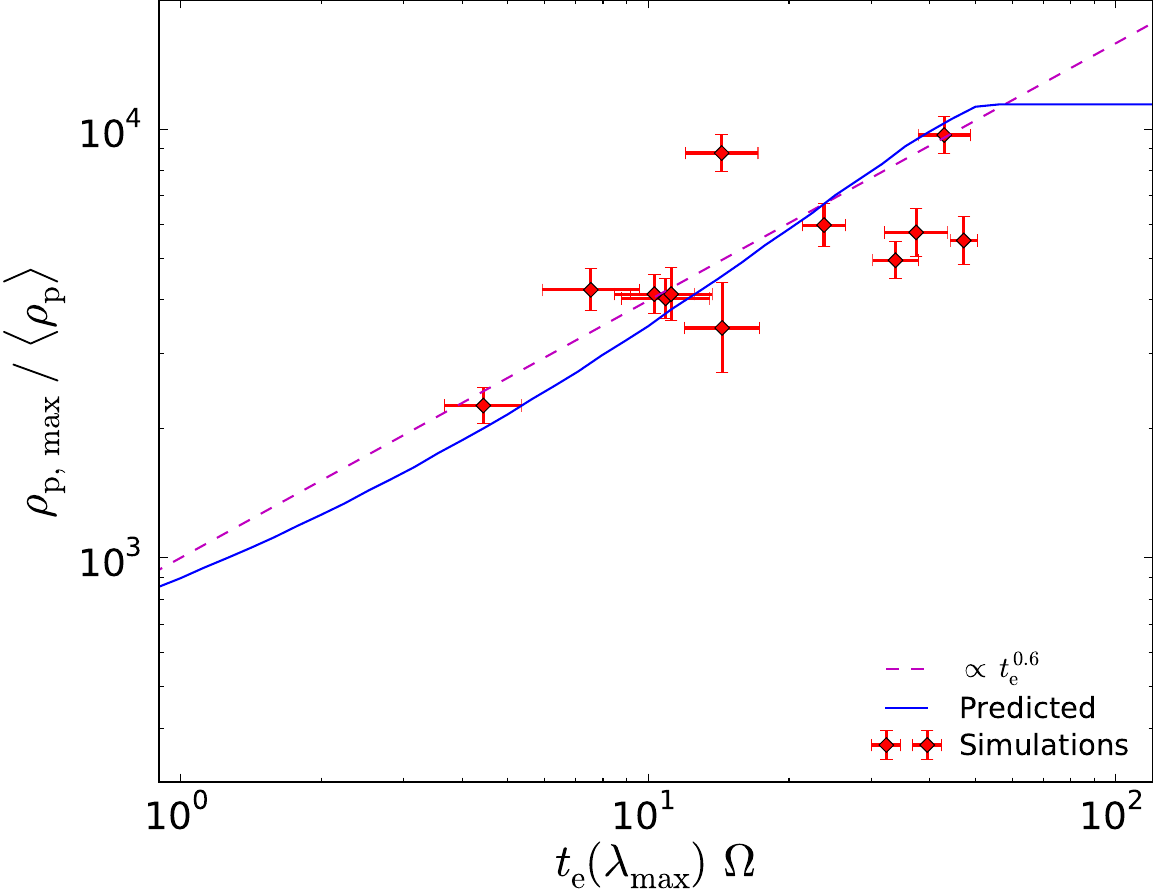}{0.94}
    \caption{Dependence of the maximum grain density in the $\taustop=1$ MRI simulations from Figs.~\ref{fig:grain.rho.mri} \&\ \ref{fig:rho.max} on the correlation time of the largest eddies in the simulation box ($\Teddy(\Lmax)$). Because of differences in the definition of correlation time, and the unknown vertical sedimentation, we treat the normalization of each axis as arbitrary: what matters here is the predicted trend. At fixed numerical resolution, $\rho_{\rm p,\,max}$ increases with $\Teddy(\Lmax)^{0.4-0.7}$ when  $\Teddy(\Lmax)\lesssim1$, until saturating (with the $\Teddy(\Lmax)$-independent scalings in Table~\ref{tbl:largescale}) when $\Teddy(\Lmax)\gg1$.
    \label{fig:rhomax.teddy}}
\end{figure}

\vspace{-0.5cm}
\subsubsection{Self-Driven (Streaming \&\ Kelvin-Helmholtz) Turbulence}
\label{sec:pred.rhodist.streaming}

Next, Fig.~\ref{fig:grain.rho.jy} repeats this comparison, with a different set of simulations from \citet{johansen:2007.streaming.instab.sims} and \citet{bai:2010.grain.streaming.sims.test,bai:2010.streaming.instability}. These are two and three-dimensional simulations,
of non-MHD hydrodynamic shearing boxes but ignoring vertical gravity/stratification (so we can directly take the statistics in $\rhograin/\langle \rhograin \rangle$ as representative of our predictions). The simulations fix $\eta=0.005$, $\Pi=0.05$, $\Omega$, and $\taustop$ for monolithic grain populations, and have no external driving of turbulence. However they do include the grain-gas back-reaction for $\rhoratio = 0.2,\,1.0,\,3.0$, and so develop some turbulence naturally via a combination of streaming and Kelvin-Helmholtz-like shear instabilities, with $\alpha\sim10^{-8}-10^{-2}$ depending on the simulation properties (but recorded for each simulation therein). The two studies adopt entirely distinct numerical methods, so where possible we show the differences owing to numerics. 

In nearly every case, we see good agreement with our simple mathematical model. This is especially true at smaller $\taustop$ and $\rhograin$; at the highest absolute grain densities, our model is less applicable but still performs reasonably well. In particular, the non-linear behavior seen therein, where for example there is a large difference between $\rhograin=0.2-1$ for $\taustop=0.1$ (but little change between $\rhograin=1-3$), and the much larger fluctuations seen for $\rhograin=0.2$ compared to $\rhograin=1-3$ for $\taustop=1$, are all predicted. The large increase with $\rhograin=0.2-1$ for $\taustop=0.1$ follows from $\rhograin$ increasing the ``effective'' stopping time as discussed in \S~\ref{sec:model.highrho}, and then the effect saturates with increasing $\rhograin\gtrsim1$. However much of the difference also owes to the different values of $\alpha$ in each simulation (the case $\taustop=1$, $\rhograin=0.2$ produces a very large $\alpha$, driving much of the very large variance). 

If we repeat the experiments from \S~\ref{sec:pred.rhodist.mri} we come to the same conclusions. For example, adopting the pure log-Poisson model (mean $\langle \deltarhonoabs \rangle$) makes little difference compared to the log-Poisson-Gaussian model discussed in \S~\ref{sec:pred.rhodist.mri} above. However, we see more clearly here that allowing for additional variance in the effects of an individual eddy does, as one might expect, increase the variance at the high-$\rhograin$ tail of the distribution, whereas a ``strict'' log-Poisson model (our default model) has an absolute cutoff at some $\rho_{\rm p,\,max}$. Interestingly, this gives a slightly better fit at high-$\taustop$, and poorer at low-$\taustop$, perhaps indicating the relative importance of in-eddy variance in these two cases.


We should also note that the finite simulation resolution limit is important here when $\taustop=0.1$ and $\rhoratio\ll1$: we will quantify this below.

Interestingly, the streaming instability relies on the cooperation of the rotation (shear terms), radial drift (non-zero $\eta$), and the back-reaction of grains on gas, in order to develop a growing mode. Although we do attempt to include a simplistic treatment of these effects, the qualitative behavior of our model does not directly rely on a drift term -- i.e.\ one could imagine applying it with zero $\eta$ and obtaining a similar result. This is because the model we apply here is really a model for the grain density fluctuations in the non-linear phase of evolution with fully-developed turbulence. We are ``just'' modeling what the turbulence, once present in some saturated amplitude, does to the grains, rather than the cooperation of the grains and gas in driving the turbulence to grow in the first place. As a result, if we took the {\em initial} conditions of the streaming instability experiments above (where there is no turbulence) our {instantaneous} prediction would be that there are no grain density fluctuations. And indeed, this would be instantaneously correct -- but the point is that the model here cannot predict whether such fluctuations {\em should} develop from a smooth initial condition, because we are not attempting to predict or model the time-evolution of the turbulence. Rather, we rely on some other model to tell us the growth history and state of the turbulence, then simply apply this to obtain an estimate of the ensuing grain clustering statistics. As a result, what the agreement here suggests -- although it clearly merits further study in future numerical simulations -- is that the physics unique to the streaming instability may not qualitatively change the key statistical properties of grain density fluctuations, once it powers turbulence at a given level.

\vspace{-0.5cm}
\subsubsection{Turbulent Concentration}
\label{sec:pred.rhodist.turbconcentration}

In Fig.~\ref{fig:grain.rho.tc}, we now compare the density PDFs measured in ``turbulent concentration'' simulations in \citet{hogan:1999.turb.concentration.sims}; these simulations follow a driven turbulent box (no shear or self-gravity), so we should apply the version of the model from \S~\ref{sec:encounters:pure.turb}. We expect the same $\Ndim$ and $\deltadim$, and back-reaction is not included so $\rhoratio\rightarrow0$. We then need to know over what range to integrate the cascade: this is straightforward since each simulation has a well-defined Reynolds number $Re = (\Lmax/\scalevar_{\nu})^{4/3}$. Lacking a model for the dissipation range, we simply truncate the power exponentially when $\scalevar<\scalevar_{\nu}$. The simulation follows particles with Stokes numbers $St\equiv \tstop/\Teddy(\scalevar_{\nu})=1$. 

Perhaps surprisingly, the model agrees fairly well with the simulations. At larger $Re$, the density PDF becomes more broad, because of contributions to fluctuations over a wider range of scales (the response function in Fig.~\ref{fig:response} is broad). However, this does not grow indefinitely -- as $Re\rightarrow\infty$ we predict convergence to a finite PDF width (with $\rho_{\rm p,\,max}\sim300-1000$). This is both because the response function declines, and, as the ``top'' of the cascade becomes larger in velocity scale, the residual (logarithmically growing) offset between grain and eddy velocities becomes larger (Eq.~\ref{eqn:g.timescale.function}), suppressing the added power.

Note, though, that our model is not designed for small $St$, and we see the effect here. The highest-density tail of the PDF is not fully reproduced (the model predictions, especially at $Re=765$, cut off more steeply). We show below that this is because grains with small $St\sim1$ can continue to cluster and experience strong density fluctuations on very small scales $\scalevar \lesssim \scalevar_{\nu}$, which are not accounted for in our calculation. We also do not explicitly account for the role of the strain field in enhancing grain clustering outside of regions of high vorticity; this may be fine in the previous cases where high-density fluctuations are dominated by anti-cyclonic vortices, but needs further investigation in the cases here. In any case, we caution care for particles where the key fluctuations lie outside the inertial range.

\vspace{-0.5cm}
\subsubsection{Experiments and Non-Gaussianity}
\label{sec:pred.rhodist.nongaussian}

In Fig.~\ref{fig:nongaussian}, we extend our comparisons to experimental data. There is a considerable experimental literature for $St\lesssim1$ particles in terrestrial turbulence (see \S~\ref{sec:intro}); unfortunately many of the measurements are either in regimes where our model does not apply or of quantities we cannot predict. However \citet{monchaux:2010.grain.concentration.experiments.voronoi} measure the density PDF in laboratory experiments of water droplets in wind tunnel turbulence,\footnote{Actually they measure the local Voronoi area around each particle: as noted therein, this is strictly equivalent to a local density PDF. We convert between the two as they do, taking the density to be the inverse area.} with $St\sim0.2-6$ and $Re\sim300-1000$ (Taylor $Re_{\lambda}=70-120$). They measure the PDF shape (normalized by its standard deviation) for a large number of experiments with different properties. The range of results, including time variation and variation across experiments, is shown in Fig.~\ref{fig:nongaussian}. We compare this with the predicted log-Poisson distribution: from \S~\ref{sec:pred.rhodist}, the variance in the log-Poisson is $S=\Delta N_{\rm int}\,\deltarho_{\rm int}^{2}$, so normalizing to fixed $\sigma=\sqrt{S}$, the PDF shape varies with the ratio $\deltarho_{\rm int}/\Delta N_{\rm int}$. Taking the predicted (modest) range in this parameter for the same range in simulation properties, we show the predicted PDF shapes. Within this range, the experimental PDF is consistent with the prediction. 

The lower panel makes this more quantitative. For each PDF in \citet{monchaux:2010.grain.concentration.experiments.voronoi} we record the variance in linear $\rhograin$ ($S_{\rho}$) and logarithmic $\ln{\rhograin}$ ($S_{\ln{\rhograin}}$). This scaling for different distributions is discussed in detail in \citet[][see Fig.~4 in particular]{hopkins:2012.intermittent.turb.density.pdfs}. If the distribution of $\rhograin$ were exactly log-normal, then there is a one-to-one relation between the two: $S_{\ln{\rhograin}} = \ln{(1 + S_{\rho})}$. This appears to form an ``upper envelope'' to the experiments. If $\rhograin$ is distributed as a Gaussian in linear-$\rhograin$, there is also a one-to-one relation (straightforward to compute numerically); this predicts relatively small $\ln{\rhograin}$ variation, in conflict with the simulations. For the log-Poisson distribution, the relation depends on the second parameter $\deltarho_{\rm int}/\Delta N_{\rm int}$. As this $\rightarrow0$, the distribution becomes log-normal; for finite values, $S_{\rho}$ is smaller than would be predicted for a log-normal with the same $S_{\ln{\rho}}$; we compare the range predicted for plausible values of $\deltarho_{\rm int}$ in these experiments (similar parameters to the simulations in Fig.~\ref{fig:correlation.functions}).

We could extend this comparison by including the numerical simulations at higher $\tstop$ (and including gravity and shear). But the agreement with our predictions is already discussed, and it is evident by-eye that the distributions in Fig.~\ref{fig:grain.rho.jy} are not exactly lognormal (they are asymmetric in log-space about the median), nor can they be strictly Gaussian in linear $\rho$ (which for such large positive fluctuations would require negative densities). While it is less obvious by eye, the non-trivial fractal spectrum in the turbulent concentration experiments, discussed at length in \citet{cuzzi:2001.grain.concentration.chondrules}, also requires non-Gaussian PDFs.

\vspace{-0.5cm}
\subsection{Density Power Spectra}

Direct measurements of the power spectrum of $\ln{\rho}$ are not available for the simulations we examine here. However, \citet{johansen:2007.streaming.instab.sims} do measure the average one-dimensional power spectra of the linear grain volume density $\rhograin$.\footnote{We exclude their simulation ``AA'' which uses the two-fluid approximation, that the authors note cannot capture the full gas-grain cascade and so predicts an artificially steep power spectrum.} We can compute this as described in \S~\ref{sec:pred.pwrspec}, and show the results in Fig.~\ref{fig:pwrspec}.\footnote{To match what was done in that paper precisely, we calculate the volumetric density PDF and corresponding variance in linear $\rho$ ($S_{\rho}$) explicitly for each scale, use this to obtain the isotropic power spectrum, then use this to realize the density distribution repeatedly on a grid matching the simulations and compute the discrete Fourier transform, and finally plot the mean absolute magnitude of the coefficients as a function of $k$.} Down to the simulation resolution limit (where the simulated power is artificially suppressed) we see good agreement. Because the power spectrum here is in linear $\rhograin$, saturation effects dilute the clarity of the predicted transition near $\Teddy=\tstop$, but it is still apparent. Moreover we confirm the qualitative prediction that for large $\taustop\gtrsim0.1$, most of the power is on relatively large scales where $\Teddy\gtrsim \tdrift$ and $\tstop$. With smaller $\tstop$, there is a larger dynamic range on large scales where $\Teddy\gg \tstop$, over which the power spectrum is flatter. For very small grains, the power would become more concentrated near $\Teddy\sim\tstop$, as in Fig.~\ref{fig:response}.

Freeing $\deltadim$ and $\Ndim$, the simulations with $\taustop=1$ and $\rhoratio\gtrsim1$ try to fit a slightly steeper slope compared to that predicted in the default model, but this is not very significant at present (and no significant change in $\Ndim$ is favored). If confirmed, though, this might imply that the gas turbulence in this regime behaves more like compressible, super-sonic turbulence \citep[see][]{boldyrev:2002.structfn.model,schmidt:2008.turb.structure.fns}; this might be expected when $\taustop$ and $\rhograin$ are large, since the dominant grains can efficiently compress the gas. 

We also compare the power spectra from turbulent concentration simulations. This is the same information as contained in the correlation function, converted via Eq.~\ref{eqn:corrfn.r}, so we discuss it below.

\vspace{-0.5cm}
\subsection{Grain Correlation Functions}

In Fig.~\ref{fig:correlation.functions}, we compare our predictions to published grain correlation functions $\xi$ in turbulent concentration experiments. The same information, represented as the linear density power spectrum, is in Fig.~\ref{fig:pwrspec}. Here we compare the simulations from \citet{pan:2011.grain.clustering.midstokes.sims} and \citet{yoshimoto:2007.grain.clustering.selfsimilar.inertial.range}, with the model appropriate for ``pure turbulence'' (see \S~\ref{sec:pred.rhodist.turbconcentration} above).\footnote{Perhaps because of different definitions, the normalizations for the correlation functions $\xi$, at identical Stokes and Reynolds number, disagree between the authors at the factor $\sim$few level. However the shape of $\xi$ in all cases agrees extremely well in both studies. So we treat the normalization as arbitrary at this level and focus on the shape comparison.} The authors each simulate a range of Stokes numbers; here we only compare with $St>1$ simulations since our model is largely inapplicable to $St\lesssim1$. 

For large $St=43\gg1$, $\xi(\scalevar)\sim$\,constant on small scales (there is no power here since $\Teddy\ll\tstop$). But there is significant power on larger scales, where $\Teddy\sim \tstop$ (for $St=43$, this is when $\scalevar/\scalevar_{\nu}\gtrsim100$). And $\xi(\lambda)$ in all cases truncates at the very largest scales because of the finite box size/driving scale $\Lmax$. For smaller $St=10$, the rising portion of $\xi(\scalevar)$ continues to smaller scales, since $\Teddy\sim\tstop$ at $\scalevar/\scalevar_{\nu}\sim40$. For still smaller $St=5$ this extends to $\scalevar/\scalevar_{\nu}\sim10$. These are all confirmed in the simulations. However, for the smallest $St\sim1$, $\xi(\scalevar)$ (and hence the power in density fluctuations) continues to rise even at $\scalevar\lesssim \scalevar_{\nu}$, where $\Teddy\ll \tstop$. This is well-known, and in fact for $St\sim1$ a power-law rise in $\xi(\scalevar)$ appears to continue to $\scalevar\rightarrow0$, which does not occur when $St\gg1$ \citep[see][]{squires:1991.grain.concentration.experiments,bec:2007.grain.clustering.markovian.flow,yoshimoto:2007.grain.clustering.selfsimilar.inertial.range,pan:2011.grain.clustering.midstokes.sims}. This effect is fundamentally related to the dissipation range and viscous effects not included in our model, so we do not expect to capture it (and, for example, trapping of grains in strain regions outside of vortices). And this is why we do not reproduce the full small-scale power in the density PDFs for $St=1$ in Fig.~\ref{fig:grain.rho.tc}.

For large-scale eddies and grains with $\taustop\gtrsim0.1$, we can compare with the simulations in \citet{carballido:2008.grain.streaming.instab.sims}. The results are consistent, but due to limited resolution the simulations only measure significant clustering in the couple smallest bins/cells quoted (see their Fig.~9); so the constraint is not particularly useful (significantly more information is available in Fig.~\ref{fig:pwrspec}). 

\vspace{-0.5cm}
\subsection{Maximum Grain Densities}

\subsubsection{Dependence on Stopping Time}

Fig.~\ref{fig:rho.max} compares these predictions for the maximum grain concentration to the maximum measured in the MRI-unstable simulations from \S~\ref{sec:pred.rhodist.mri} \citep{dittrich:2013.grain.clustering.mri.disk.sims}. Recall, here $\alpha$ and $\Pi$ are approximately constant in all cases, and grain-gas back-reaction is ignored ($\rhoratio\rightarrow0$), so the only varied parameter is $\taustop$. In the range $\taustop\sim0.01-1$, our predictions are in remarkably good agreement with the simulations, with the maximum grain concentration increasing from tens of percent to factors $\gtrsim 300$, for the simple assumption $\Ndim=2$. Only a small range $\Ndim\sim1.8-2.2$ is allowed if we free this parameter. For $1<\taustop\lesssim5$, the predictions are also reasonably accurate. At very large $\taustop \gtrsim 10$, however, we appear to under-estimate the magnitude of fluctuations; though it also appears that there is some change in the vertical structure in these simulations relative to what is expected (discussed in \citealt{dittrich:2013.grain.clustering.mri.disk.sims}), so the large $\Sigma_{\rm p,\, max}$ may not entirely reflect midplane density fluctuations. 

As noted above, it is important that we account for finite resolution here. We compare the predictions using our best-estimate of $\Lmax$ (the driving scale) relative to the finite resolution limit (factor of $\sim100$ in scale), to the prediction assuming infinite resolution (and $Re\rightarrow\infty$), with density measured on infinitely small scales. For large grains, this makes little difference (most power is on large scales). For small grains $\taustop\ll1$, however, the difference is dramatic. Very small grains with $\taustop\sim0.01$ may still experience factor $\gtrsim100$ fluctuations on small scales. This should not be surprising, however -- this is already evident in the turbulent concentration simulations, which exhibit such large fluctuations (even over limited Reynolds number, but covering the range where $\tstop\sim\Teddy$) despite $\taustop\rightarrow0$, effectively. These $\rhograin$ fluctuations, for small grains, occur on scales where $\tstop\sim \Teddy$, and resolving their full dynamic range (getting convergence here) requires resolution of the broad peak in the response function (Fig.~\ref{fig:response}), crudely we estimate $0.05\,\tstop\lesssim\Teddy\lesssim20\,\tstop$ should be spanned. This translates, even in idealized simulations, to large Reynolds numbers $Re\gtrsim10^{4}-10^{5}$ (not surprising, since the simulations in Fig.~\ref{fig:grain.rho.tc} are not converged yet at $Re\sim1000$). And for the simulations here, which have a fixed box size at of order the dust layer scale height, this would require resolution of a factor $\sim10^{6}$ below the largest eddies scales $\Lmax$ (far beyond present capabilities).

Interestingly, we predict a ``partial'' convergence: because $\deltarho$ is non-monotonic in $\scalevar$, the fluctuations on large scales can converge at reasonable resolution (factor $\sim2$ changes relative to the simulations here make little difference to the predicted curve). Only when the resolution is increased by the much larger factor described above does the additional power manifest. So, if the ``interesting'' fluctuations are those on large scales, such simulations, or the approximations in Table~\ref{tbl:largescale}, are reasonable.

\vspace{-0.5cm}
\subsubsection{Dependence on Scale}

\citet{johansen:2012.grain.clustering.with.particle.collisions} present the maximum density as a function of scale in streaming-instability simulations similar to those in \citet{johansen:2007.streaming.instab.sims}.\footnote{We specifically compare their simulations with no collisions and no grain self-gravity.} Given $\alpha$, $\Pi$, $\rhoratio$ and $\taustop=0.3$ specified in the simulation, it is straightforward to predict $\rho_{\rm p,\,max}(\lambda)$ and compare to their result (using $\rhograin(\Lmax) = \langle \rhograin \rangle$ at the dust scale height). As discussed in \S~\ref{sec:pred.rhomax} and in Table~\ref{tbl:largescale}, on large scales $\rho_{\rm p,\,max}\propto \lambda^{-\gamma}$ with $\gamma\approx \deltadim\,[1-\exp{(-\deltazero)}]$; for $\taustop=0.3$ and $\rhoratio=0.25$, this gives $\gamma\approx1.5$, a power-law like scaling in excellent agreement with the simulations. 

\vspace{-0.5cm}
\subsubsection{Dependence on Eddy Turnover Time}

In the simulations of \citet{dittrich:2013.grain.clustering.mri.disk.sims} from Fig.~\ref{fig:rho.max}, the authors also note that in a separate series of simulations with fixed $\taustop=1$, they see a significant dependence of the maximum $\rhograin$ on the lifetime/coherence time of the largest eddies. They quantify this by comparing $\rho_{\rm p,\,max}$ to (twice) the correlation time of the longest-lived Fourier modes, which should be similar to our $\Teddy(\Lmax)$. As discussed in Appendix~\ref{sec:appendix:largescale}, on sufficiently large scales ($\Teddy\gg \Omega^{-1}$), $\deltarho$ and fluctuation properties asymptote to values independent of $\Teddy$ (the scalings in Table~\ref{tbl:largescale}). However for smaller $\Teddy(\Lmax)\lesssim1$, since the power for large grains $\taustop\sim1$ is concentrated on scales with $\Teddy\sim\tstop\sim \Omega^{-1}$, the integrated power will decline if the top scales only include smaller eddies. Given the asymptotic scaling of $\varpi\propto\taustopeddy^{-1/2}$ for $\taustopeddy\gg\taustop$ ($\Teddy\ll \Omega^{-1}$) and $\taustopeddy\gg1$ ($\Teddy\ll\tstop$), we expect that the power at the largest scales will scale $\propto \taustopeddy(\Lmax)^{-1/2} \propto \Teddy(\Lmax)^{1/2}$. Performing the full calculation for comparable Reynolds number to the simulations, we indeed predict a scaling $\rho_{\rm p,\,max}\propto (\Teddy(\Lmax)\,\Omega)^{0.4-0.7}$ (for a range $\taustop\sim0.5-2$), until saturation. Fig.~\ref{fig:rhomax.teddy} explicitly compares this to the scaling in those simulations; the agreement is good.

\vspace{-0.5cm}
\subsubsection{Effects of Simulation Dimension}

Note that, in \citet{johansen:2007.streaming.instab.sims} and \citet{bai:2010.grain.streaming.sims.test,bai:2010.streaming.instability}, some significant differences are found between two-dimensional and three-dimensional simulations. To first order, the differences are accounted for in our model, not because of a fundamental change in the behavior of grains in response to eddies, but rather because of the different amplitudes of turbulence and/or vertical stratification in the simulations. In three dimensions, it appears more difficult for the streaming instability to generate large-$\alpha$ turbulence. That result itself is not part of our model here, however, for the given $\alpha$ in the different simulations, our predictions appear to agree with the different simulation PDFs. 

\vspace{-0.5cm}
\subsubsection{Effects of Pressure Gradients and Metallicity}

Pressure gradients ($\Pi$) and metallicity ($Z$) enter our model only indirectly, by altering the value of the parameters $\rhoratio$ and $\beta$, and -- under some circumstances such as streaming-instability turbulence -- by altering the bulk turbulent properties (velocities/eddy turnover times). We have already derived the dependence of $\beta$ on $\alpha$, $\rhoratio$, and $\Pi$. And we expect $\rhoratio = (\Sigma_{\rm p}/\Sigma_{\rm gas})\,(h_{\rm gas}/h_{\rm p}) \approx Z\,(\taustop/\alpha)^{1/2}$ \citep[see][]{carballido:2006.grain.sedimentation,youdin:2007.turbulent.grain.stirring}. When other parameters are fixed (for example, if the turbulence is externally driven), it is straightforward to accommodate these parameter variations. However, the dependence of bulk turbulent properties $\alpha$ and $\Teddy(\Lmax)$ on $Z$ and $\Pi$ (or other disk properties) requires some additional model for the driving and generation of turbulence. \citet{bai:2010.grain.streaming.vs.diskparams} consider a survey of these parameters, in the regime where the turbulence is driven by the streaming instability, and find that the dust layer scale height, turbulent dispersion $\alpha$, and largest eddy scales depend in highly non-linear and non-monotonic fashion on $Z$ and $\Pi$. If we adopt some of the simple dimensional scalings they propose therein for these quantities, we qualitatively reproduce the same trends they see: $\rho_{\rm p\,max}$ increases with both increasing $Z$ (increasing $\rhoratio$) and decreasing $\Pi$ (weaker drift, larger $\beta$), but it is difficult to construct a quantitative comparison.

\vspace{-0.5cm}
\section{Discussion}

\subsection{Summary}

We propose a simple, analytic, phenomenological model for the clustering of aerodynamic grains in turbulent media (with or without external shear and gravity). We show that this leads to unique, definite predictions for quantities such as the grain density distribution, density fluctuation power spectrum, maximum grain densities, and correlation functions, as a function of grain stopping/friction time, grain-to-gas volume density ratio, and properties of the turbulence. Our predictions are specifically appropriate for inertial-range turbulence, with large Reynolds numbers and Stokes numbers ($\tstop$ large compared to the eddy turnover time at the viscous scale), the regime of most astrophysical relevance. Within this range, we compare these predictions to numerical simulations and laboratory experiments, with a wide range in stopping times and turbulence properties, and show that they agree well. 

The model assumes that grain density fluctuations are dominated by coherent turbulent eddies, presumably in the form of simple vortices. Such eddies act to accelerate grains and preferentially disperse them away from the eddy center (concentrating grains in the interstices between eddies). Qualitatively, such behavior has been observed in a wide range of simulations and experiments (see \S~\ref{sec:intro}). Quantitatively, our model first adopts a simple calculation of the effects of a vortex with a given eddy turnover time (and lifetime of the same duration) acting on an initially homogeneous, isotropic Lagrangian grain population. We then attach this calculation to a simple assumption for eddy structures on different scales: namely that eddies on different scales are self-similar, statistically independent, and reproduce a Kolmogorov-type scaling.

\vspace{-0.5cm}
\subsection{Key Conclusions and Predictions}

\begin{itemize}

\item {\bf Large grain density fluctuations are expected even in incompressible turbulence:} We predict that even a small aerodynamic de-coupling between gas and grains allows for large (order-of-magnitude) fluctuations in $\rhograin$, even while {\em gas} density fluctuations are negligible.

\item {\bf Grain density fluctuations do not explicitly depend on the {\em driving} mechanisms of turbulence:} Given the simplistic level of detail in our model, it applies equally to simulations with turbulence arising via MRI, Kelvin-Helmholtz, and streaming instabilities, or artificially (numerically) driven. Still, even with limited accounting for the detailed structure of turbulence, we are able to predict some dependence of fluctuations on the stopping time $\tstop$, the ratio of volume-averaged grain-to-gas densities $\rhoratio$, and some basic properties of the turbulence (the Reynolds number and velocity/length/time scale at the driving scale). These can, of course, change depending on the driving.

\item {\bf The grain density distribution $\rhograin$ is log-Poisson:} 
\begin{align}
P_{V}&(\ln{\rhograin})\,{\rm d}\ln{\rhograin} \approx \frac{\Delta N_{\rm int}^{m}\,\exp{(-\Delta N_{\rm int})}}{\Gamma(m+1)}\,\frac{{\rm d}\ln{\rhograin}}{\deltarho_{\rm int}} \\ 
\nonumber
m &= \deltarho_{\rm int}^{-1}\,{\Bigl\{}\Delta N_{\rm int}\,{\Bigl[}1 - \exp{(-\deltarho_{\rm int})} {\Bigr]} - \ln{{\Bigl(} \frac{\rhograin}{\langle \rhograin \rangle}{\Bigr)}} {\Bigr\}}
\end{align}
We predict $\Delta N_{\rm int}$ and $\deltarho_{\rm int}$ as a function of turbulent properties. This arises (in the model here) because the number of eddies is quantized (Poisson), and each produces a multiplicative (logarithmic) effect on the grain density field. 

Generically, we suggest that this can be used as a fitting function, where the best-fit value of $\Delta N_{\rm int} \sim \deltadim\,\ln{(\Lmax/\scalevar_{\rm min})}$ crudely measures the dynamic range of the cascade over which density fluctuations occur, and the value $\deltarho_{\rm int}$ reflects the rms fluctuation amplitude ``per event'' in the turbulence. 

\item {\bf On large scales ($\Teddy\gtrsim\Omega^{-1}$) shear/gravity dramatically enhances density fluctuations.} 

In this model, the fluctuation ``response'' in large ($\Teddy\gtrsim\Omega^{-1}$) eddies is approximately scale-free, with amplitude $\deltarho \sim 2\,\Ndim\,(\taustop + \taustop^{-1})^{-1}$. The variance in $\ln{\rhograin}$, and maximum values of $\rhograin$, increase with $\taustop$ up to a maximum near $\taustop\sim1$ (where maximum $\rhograin$ values can reach thousands of times the mean); much larger grains are too weakly coupled to experience fluctuations and behave in an approximately ``free-steaming'' manner. 

The maximum $\rhograin(r)$ on a smoothing scale $r$ scales $\propto r^{-\gamma}$ with $\gamma\sim\deltadim\,[1-\exp{(-\deltarho)}]$; for small grains with $\deltarho\ll1$ $\gamma\sim2\,\deltarho$ is small, so the scale-dependence is shallow. For large grains with $\deltarho\gtrsim1$, $\gamma$ saturates at $\sim2$ (isothermal-like). 

Most of the {\em power} in $\rhograin$ fluctuations is on large scales for large grains, while for small grains the power spectrum is approximately flat over a range of scales down to a scale $\Lcrit$ where the rms eddy-crossing time becomes shorter than the grain stopping time, below which power is suppressed.

\item {\bf On small scales ($\Teddy\ll \Omega^{-1}$), grain clustering depends only on the ratio $\tstop/\Teddy$.} 

Within the context of our model, the fluctuation amplitude in small eddies is maximized around $\tstop\sim\Teddy$, declining $\propto \tstop/\Teddy$ for $\tstop\ll\Teddy$ (where eddies are ``flung out'' of vortices at speeds limited by the eddy terminal velocity $\propto \tstop$) and $\propto (\tstop/\Teddy)^{-1/2}$ for $\tstop\gg\Teddy$ (where eddies cannot fully trap grains, so their effects add incoherently in a Brownian random walk). 

Integrated over a sufficiently broad cascade (Reynolds number $\rightarrow\infty$), this means that some eddies will always have $\tstop\sim\Teddy$, so the integrated density variance and maximum $\rhograin$ always converge to values only weakly dependent on the absolute value $\tstop$. The maximum $\rhograin$ can reach several hundred times the mean grain density, even in the limit $\rhograin\ll\rhogas$ and $\taustop\ll1$. The variance is concentrated on small scales, however, and the ``resonance region'' of eddy turnover time is broad -- so resolving this in simulations or experiments requires resolved eddies at least over the range $0.05\,\tstop \lesssim \Teddy \lesssim 20\,\tstop$ (Reynolds numbers at least $\gtrsim10^{4}-10^{5}$).

The grain-grain correlation function $\xi(r)$ in this limit scales weakly on the largest scales ($\propto \ln{(1/r)}$), until approaching $\Teddy\sim \tstop$ where it rises as $\xi(r)\propto (\tstop/\Teddy)^{2} \propto r^{-2\,(1-\velslope)}$ (a slope near unity), then converges (flattens to $\xi(r)\rightarrow$\,constant) below $\Teddy\lesssim \tstop$.

\item {\bf Stronger turbulence enhances clustering:} At {\em otherwise identical properties}, larger values of the Mach number, Reynolds number, or driving-scale eddy turnover time $\Teddy(\Lmax)\,\Omega$ give rise to a larger dynamic range of the cascade driving $\rhograin$ fluctuations. Stronger turbulence may decrease $\langle \rhograin \rangle$, so it is not necessarily the case that these lead to larger absolute maximum values of $\rhograin$, but only stronger grain clumping. Conversely, larger {\em drift} (laminar relative grain-gas velocities) weakens the clustering, by suppressing the time grains interact with single eddies.

\item {\bf Higher grain-to-gas density ratios enhance clustering of small grains:} We attempt to consider some very simple approximations to account for the regime where the back-reaction of grains on gas is important. Remarkably, the predictions of the model here appear to reasonably describe simulations in this limit (despite many of our assumptions formally breaking down). Bearing these caveats in mind, this predicts that, up to a saturation level where $\tstop\,(1+\rhoratio) \gtrsim1$, increasing the volume density of grains increases their effective stopping time by dragging gas in a local wake, leading to larger terminal velocities and eddy effects. 

\item {\bf {\em Coherent} eddy structure is critical:} Our predictions rely fundamentally on locally coherent (albeit short-lived) structures in turbulence. We show that a purely Markovian (Gaussian random field) approximation does not produce fluctuations nearly as large (nor with the correct scaling). So in some sense inertial-range $\rhograin$ fluctuations depend intrinsically on structure in the gas turbulence. This may not necessarily be captured in models which treat density perturbations purely as a ``turbulent diffusion'' term or Brownian motion.

\end{itemize}

\vspace{-0.5cm}
\subsection{Limitations of This Model \&\ Areas for Future Work}

This paper is intended as a first step to a model of grain clustering in inertial-range turbulence, and many aspects could be improved. We have {\em intentionally} excluded many important details of turbulent structure, higher-order grain-gas coupling terms, and other effects, in order to construct a simplest-possible phenomenological model which is able to reproduce certain basic statistics of grains in turbulence. As such, many areas a ripe for further investigation.

For example, we fundamentally assume that grains have no effect on the character of gas turbulence statistics (though it may drive that turbulence), which is probably not true when $\rhograin\gtrsim1$. And indeed, we see our predictions do not agree well with the simulations when $\rhoratio\gg1$. Similarly, we appear to predict too rapid a turnover in grain clustering when $\taustop\gg1$. In these limits, it might be more accurate to begin from the statistics of a purely collisionless grain system, and treat the gas perturbatively (essentially the opposite of our approach). Further investigation of this regime is warranted.

We also simplify tremendously by only considering the mean effects of eddies with a given scale; but even at fixed eddy scale there should be a distribution of eddy structure, meaning that $\deltarhonoabs(\scalevar)$ is not simply a number but itself a distribution. More detailed models could generalize our model here to allow this. Such generalizations have been developed for the pure gas statistics \citep[see][]{castaing:1996.thermodynamic.turb.cascade,chainais:2006.inf.divisible.cascade.review}; however, experimental data has been largely unable to distinguish that case from the simplified model. We show one example of such a model, which suggests that the character of the grain density PDF at high-$\rhograin$ may be able to distinguish such higher-order models. Other such models, for example the $\beta$-models proposed in \citet{hogan:1999.turb.concentration.sims}, may be more accurate still.

We have also, of course, assumed a very simplistic model for the eddy statistics, which assumes they are self-similar and statistically independent. In detail, neither of these assumptions is true in real turbulence. A more detailed model for the hierarchy of eddies could be coupled to the results we have derived for the behavior of grains encountering individual eddies; this is straightforward. However it is more complicated to properly account for the non-independent nature of eddies, since if one wishes to treat this case in detail, it is no longer a good approximation to apply the results of our calculation for single eddies to ``each,'' because of non-linear terms in the grain response function. It remains to be seen if the non-linear behavior can be captured outside of direct numerical simulations.

We have also ignored the role of the strain field in determining the clustering of grains; we have essentially considered vorticity as a force ``expelling'' grains (except in anti-cyclonic vortices in a rotating disk, which lead to concentration effects we explicitly model), and modeled their concentration with a simple mass-conservation argument (based on what is expelled being ``spread about''). This is obviously an over-simplification, and will be most significant in the high-grain density regime in small-scale turbulence. Future work should attempt to explicitly model the form and dynamics of grain density fluctuations in a non-linear strain field.

Additionally, our calculation ignores grain-grain collisions, which may significantly alter the statistics on the smallest scales in high-density regions \citep[see][]{johansen:2012.grain.clustering.with.particle.collisions}. And our scalings are derived for inertial-range turbulence; the case appropriate for small Stokes numbers where concentration occurs at/below the viscous scale is much more well-studied in the terrestrial turbulence literature (see \S~\ref{sec:intro}), and may be more relevant for the smallest grains. Thus on the smallest scales where collisions and/or viscous effects dominate, our predictions are expected to break down.

\vspace{-0.5cm}
\subsection{Implications}

Despite these significant limitations, the model here has a range of implications for many important astrophysical questions. An analytic model for grain clustering is particularly important in order to extrapolate to regimes which cannot easily be simulated (large Reynolds numbers and/or small scales). With an analytic description of the grain density power spectrum, it becomes straightforward to apply methods such as those in \citet{hopkins:2013.turb.planet.direct.collapse} to estimate the mass and/or size spectra of grain aggregations meeting various ``interesting'' criteria (such as those aggregations which are self-gravitating). 

Large grain clustering is of central importance to planetesimal formation. In grain overdensities reaching $\sim100-1000$ times the mean, local grain densities in proto-planetary disks can easily exceed gas densities, triggering additional processes such as the streaming instability. It is even possible that such large fluctuations can directly exceed the Roche density and promote gravitational collapse \citep[see][]{cuzzi:2008.chondrule.planetesimal.model.secular.sandpiles}. In future work, we will use the model here to investigate the conditions under which such collapse may be possible.

Grain-grain collisions in proto-planetary disks and the ISM depend sensitively on the small-scale clustering of grains, i.e.\ $\langle n^{2}(\lambda\rightarrow0)\rangle$, which we show can differ dramatically from a homogeneous medium, even for very small grains. Even simple clumping factors $\langle n^{2} \rangle / \langle n \rangle^{2}$ can be large ($\gg 1$). Thus grain clustering can make substantial differences to quantities such as grain collision rates and approach velocities. 

Radiative transfer through the dusty ISM (and consequences such as emission, absorption, and cooling via dust) also depend on dust clumping. Depending on the geometry and details of the problem, this can even extend to extremely small-scale clustering properties within the dust, where inhomogeneities cannot be resolved in current simulations and may depend critically on dust clustering even independent of gas density fluctuations. 

This model should be equally applicable to terrestrial turbulence, in the case of large Reynolds and Stokes numbers. We predict that even relatively large or heavy aerosols may undergo large number density fluctuations in inertial-range turbulence. We specifically provide a theoretical framework for the observations of preferential concentration of large-$St$ grains with amplitudes larger than those corresponding to pure random-phase models \citep{bec:2007.grain.clustering.markovian.flow}, with scale-dependent Stokes number $\taustopeddy=\taustop/\Teddy(\scalevar)$. Measurements of the clustering scales of these particles and their amplitudes can strongly constrain the role of coherent structures in preferential concentration and their geometry/fractal structure.

The intention here is to provide a simple framework in which to interpret simulations and experiments of grain clumping. We provide general fitting functions, which can be used in simulations to quantify important properties of turbulent fluctuations, such as the dynamic range of the cascade contributing to fluctuations and the magnitude of coherent ``events.'' They also provide a guideline for understanding on which scales simulations can resolve clumping, and to understand the regimes to which these results can be generalized.

\vspace{-0.7cm}
\acknowledgments 
We thank Jessie Christiansen and Eugene Chiang for many helpful discussions during the development of this work. We also thank Andrew Youdin, Anders Johansen, and Jeff Cuzzi for several suggestions. We also thank our referee for a number of constructive and useful suggestions, especially pointing out the means to replace some earlier simplified approximations with exact solutions.
Support for PFH was provided by NASA through Einstein Postdoctoral Fellowship Award Number PF1-120083 issued by the Chandra X-ray Observatory Center, which is operated by the Smithsonian Astrophysical Observatory for and on behalf of the NASA under contract NAS8-03060.\\

\vspace{-0.1cm}
\bibliography{/Users/phopkins/Dropbox/Public/ms}

\begin{thebibliography}{70}
\expandafter\ifx\csname natexlab\endcsname\relax\def\natexlab#1{#1}\fi

\bibitem[{{Bai} \&
  {Stone}(2010{\natexlab{a}})}]{bai:2010.streaming.instability}
{Bai}, X.-N., \& {Stone}, J.~M. 2010{\natexlab{a}}, \apj, 722, 1437

\bibitem[{{Bai} \&
  {Stone}(2010{\natexlab{b}})}]{bai:2010.grain.streaming.sims.test}
---. 2010{\natexlab{b}}, \apjs, 190, 297

\bibitem[{{Bai} \&
  {Stone}(2010{\natexlab{c}})}]{bai:2010.grain.streaming.vs.diskparams}
---. 2010{\natexlab{c}}, \apjl, 722, L220

\bibitem[{{Bec} {et~al.}(2009){Bec}, {Biferale}, {Cencini}, {Lanotte}, \&
  {Toschi}}]{bec:2009.caustics.intermittency.key.to.largegrain.clustering}
{Bec}, J., {Biferale}, L., {Cencini}, M., {Lanotte}, A.~S., \& {Toschi}, F.
  2009, eprint arxiv:0905.1192

\bibitem[{Bec {et~al.}(2007)Bec, Cencini, \&
  Hillerbrand}]{bec:2007.grain.clustering.markovian.flow}
Bec, J., Cencini, M., \& Hillerbrand, R. 2007, Phys. Rev. E, 75, 025301

\bibitem[{{Bec} {et~al.}(2008){Bec}, {Cencini}, {Hillerbrand}, \&
  {Turitsyn}}]{bec:2008.markovian.grain.clustering.model}
{Bec}, J., {Cencini}, M., {Hillerbrand}, R., \& {Turitsyn}, K. 2008, Physica D
  Nonlinear Phenomena, 237, 2037

\bibitem[{{Boldyrev}(2002)}]{boldyrev:2002.structfn.model}
{Boldyrev}, S. 2002, \apj, 569, 841

\bibitem[{Bracco {et~al.}(1999)Bracco, Chavanis, Provenzale, \&
  Spiegel}]{bracco:1999.keplerian.largescale.grain.density.sims}
Bracco, A., Chavanis, P.~H., Provenzale, A., \& Spiegel, E.~A. 1999, Physics of
  Fluids, 11, 2280

\bibitem[{Budaev(2008)}]{budaev:2008.tokamak.plasma.turb.pdfs.intermittency}
Budaev, V. 2008, Plasma Physics Reports, 34, 799, 10.1134/S1063780X08100012

\bibitem[{{Burlaga}(1992)}]{burlaga:1992.multifractal.solar.wind.density.velocity}
{Burlaga}, L.~F. 1992, Journal of Geophysical Research, 97, 4283

\bibitem[{{Carballido} {et~al.}(2006){Carballido}, {Fromang}, \&
  {Papaloizou}}]{carballido:2006.grain.sedimentation}
{Carballido}, A., {Fromang}, S., \& {Papaloizou}, J. 2006, \mnras, 373, 1633

\bibitem[{{Carballido} {et~al.}(2008{\natexlab{a}}){Carballido}, {Stone}, \&
  {Turner}}]{carballido:2008.large.grain.clustering.disk.sims}
{Carballido}, A., {Stone}, J.~M., \& {Turner}, N.~J. 2008{\natexlab{a}},
  \mnras, 386, 145

\bibitem[{{Carballido} {et~al.}(2008{\natexlab{b}}){Carballido}, {Stone}, \&
  {Turner}}]{carballido:2008.grain.streaming.instab.sims}
---. 2008{\natexlab{b}}, \mnras, 386, 145

\bibitem[{{Castaing}(1996)}]{castaing:1996.thermodynamic.turb.cascade}
{Castaing}, B. 1996, Journal de Physique II, 6, 105

\bibitem[{{Chainais}(2006)}]{chainais:2006.inf.divisible.cascade.review}
{Chainais}, P. 2006, European Physical Journal B, 51, 229

\bibitem[{{Chiang} \&
  {Youdin}(2010)}]{chiang:2010.planetesimal.formation.review}
{Chiang}, E., \& {Youdin}, A.~N. 2010, Annual Review of Earth and Planetary
  Sciences, 38, 493

\bibitem[{{Cuzzi} {et~al.}(2001){Cuzzi}, {Hogan}, {Paque}, \&
  {Dobrovolskis}}]{cuzzi:2001.grain.concentration.chondrules}
{Cuzzi}, J.~N., {Hogan}, R.~C., {Paque}, J.~M., \& {Dobrovolskis}, A.~R. 2001,
  \apj, 546, 496

\bibitem[{{Cuzzi} {et~al.}(2008){Cuzzi}, {Hogan}, \&
  {Shariff}}]{cuzzi:2008.chondrule.planetesimal.model.secular.sandpiles}
{Cuzzi}, J.~N., {Hogan}, R.~C., \& {Shariff}, K. 2008, \apj, 687, 1432

\bibitem[{{Dittrich} {et~al.}(2013){Dittrich}, {Klahr}, \&
  {Johansen}}]{dittrich:2013.grain.clustering.mri.disk.sims}
{Dittrich}, K., {Klahr}, H., \& {Johansen}, A. 2013, \apj, 763, 117

\bibitem[{{Dubrulle}(1994)}]{dubrulle:logpoisson}
{Dubrulle}, B. 1994, Physical Review Letters, 73, 959

\bibitem[{Elperin {et~al.}(1996)Elperin, Kleeorin, \&
  Rogachevskii}]{elperin:1996:grain.clustering.instability}
Elperin, T., Kleeorin, N., \& Rogachevskii, I. 1996, Phys. Rev. Lett., 77, 5373

\bibitem[{{Elperin} {et~al.}(1998){Elperin}, {Kleeorin}, \&
  {Rogachevskii}}]{elperin:1998.grain.clustering.instability.rotation}
{Elperin}, T., {Kleeorin}, N., \& {Rogachevskii}, I. 1998, Physical Review
  Letters, 81, 2898

\bibitem[{{Falkovich} \&
  {Pumir}(2004)}]{falkovich:2004.intermittent.distrib.heavy.particles}
{Falkovich}, G., \& {Pumir}, A. 2004, Physics of Fluids, 16, L47

\bibitem[{{Federrath}(2013)}]{federrath:2013.intermittency.vs.numerics}
{Federrath}, C. 2013, MNRAS, in press, arXiv:1306.3989

\bibitem[{{Federrath} \&
  {Banerjee}(2015)}]{federrath.2015:density.pdf.and.sfr.in.polytropic.turbulence}
{Federrath}, C., \& {Banerjee}, S. 2015, \mnras, 448, 3297

\bibitem[{{Federrath} {et~al.}(2008){Federrath}, {Klessen}, \&
  {Schmidt}}]{federrath:2008.density.pdf.vs.forcingtype}
{Federrath}, C., {Klessen}, R.~S., \& {Schmidt}, W. 2008, \apjl, 688, L79

\bibitem[{Fessler {et~al.}(1994)Fessler, Kulick, \&
  Eaton}]{fessler:1994.grain.concentration.experiments}
Fessler, J.~R., Kulick, J.~D., \& Eaton, J.~K. 1994, Physics of Fluids, 6, 3742

\bibitem[{{Goodman} \&
  {Pindor}(2000)}]{goodman.pindor:2000.secular.drag.instabilities.grains}
{Goodman}, J., \& {Pindor}, B. 2000, Icarus, 148, 537

\bibitem[{{Gualtieri} {et~al.}(2009){Gualtieri}, {Picano}, \&
  {Casciola}}]{gualtieri:2009.anisotropic.grain.clustering.experiments}
{Gualtieri}, P., {Picano}, F., \& {Casciola}, C.~M. 2009, Journal of Fluid
  Mechanics, 629, 25

\bibitem[{{Gustavsson} {et~al.}(2012){Gustavsson}, {Meneguz}, {Reeks}, \&
  {Mehlig}}]{gustavsson:2012.grain.clustering.randomflow.lowstokes}
{Gustavsson}, K., {Meneguz}, E., {Reeks}, M., \& {Mehlig}, B. 2012, New Journal
  of Physics, 14, 115017

\bibitem[{{Hogan} \& {Cuzzi}(2007)}]{hogan:2007.grain.clustering.cascade.model}
{Hogan}, R.~C., \& {Cuzzi}, J.~N. 2007, Phys. Rev. E, 75, 056305

\bibitem[{{Hogan} {et~al.}(1999){Hogan}, {Cuzzi}, \&
  {Dobrovolskis}}]{hogan:1999.turb.concentration.sims}
{Hogan}, R.~C., {Cuzzi}, J.~N., \& {Dobrovolskis}, A.~R. 1999, Phys. Rev. E,
  60, 1674

\bibitem[{{Hopkins}(2013{\natexlab{a}})}]{hopkins:frag.theory}
{Hopkins}, P.~F. 2013{\natexlab{a}}, \mnras, 430, 1653

\bibitem[{{Hopkins}(2013{\natexlab{b}})}]{hopkins:2012.intermittent.turb.density.pdfs}
---. 2013{\natexlab{b}}, \mnras, 430, 1880

\bibitem[{{Hopkins} \&
  {Christiansen}(2013)}]{hopkins:2013.turb.planet.direct.collapse}
{Hopkins}, P.~F., \& {Christiansen}, J.~L. 2013, \apj, 776, 48

\bibitem[{{Jalali}(2013)}]{jalali:2013.streaming.instability.largescales}
{Jalali}, M.~A. 2013, \apj, in press, arxiv:1301.2064

\bibitem[{{Johansen} \& {Youdin}(2007)}]{johansen:2007.streaming.instab.sims}
{Johansen}, A., \& {Youdin}, A. 2007, \apj, 662, 627

\bibitem[{{Johansen} {et~al.}(2012){Johansen}, {Youdin}, \&
  {Lithwick}}]{johansen:2012.grain.clustering.with.particle.collisions}
{Johansen}, A., {Youdin}, A.~N., \& {Lithwick}, Y. 2012, \aap, 537, A125

\bibitem[{{Kolmogorov}(1941)}]{kolmogorov:turbulence}
{Kolmogorov}, A. 1941, Akademiia Nauk SSSR Doklady, 30, 301

\bibitem[{{Konstandin} {et~al.}(2012){Konstandin}, {Girichidis}, {Federrath},
  \& {Klessen}}]{konstantin:mach.compressive.relation}
{Konstandin}, L., {Girichidis}, P., {Federrath}, C., \& {Klessen}, R.~S. 2012,
  \apj, 761, 149

\bibitem[{{Lee} {et~al.}(2010){Lee}, {Chiang}, {Asay-Davis}, \&
  {Barranco}}]{lee:2010.grain.settling.vs.grav.instability}
{Lee}, A.~T., {Chiang}, E., {Asay-Davis}, X., \& {Barranco}, J. 2010, \apj,
  725, 1938

\bibitem[{{Lyra} {et~al.}(2009){Lyra}, {Johansen}, {Zsom}, {Klahr}, \&
  {Piskunov}}]{lyra:2009.semianalytic.planet.form.model.grain.settling}
{Lyra}, W., {Johansen}, A., {Zsom}, A., {Klahr}, H., \& {Piskunov}, N. 2009,
  \aap, 497, 869

\bibitem[{Marcu {et~al.}(1995)Marcu, Meiburg, \&
  Newton}]{marcu:1995.grain.burgers.vortex}
Marcu, B., Meiburg, E., \& Newton, P.~K. 1995, Physics of Fluids, 7, 400

\bibitem[{{Markiewicz} {et~al.}(1991){Markiewicz}, {Mizuno}, \&
  {Voelk}}]{markiewicz:1991.grain.relative.velocity.calc}
{Markiewicz}, W.~J., {Mizuno}, H., \& {Voelk}, H.~J. 1991, \aap, 242, 286

\bibitem[{{Molina} {et~al.}(2012){Molina}, {Glover}, {Federrath}, \&
  {Klessen}}]{molina:2012.mhd.mach.dispersion.relation}
{Molina}, F.~Z., {Glover}, S.~C.~O., {Federrath}, C., \& {Klessen}, R.~S. 2012,
  \mnras, 423, 2680

\bibitem[{Monchaux {et~al.}(2010)Monchaux, Bourgoin, \&
  Cartellier}]{monchaux:2010.grain.concentration.experiments.voronoi}
Monchaux, R., Bourgoin, M., \& Cartellier, A. 2010, Physics of Fluids, 22,
  103304

\bibitem[{{Monchaux} {et~al.}(2012){Monchaux}, {Bourgoin}, \&
  {Cartellier}}]{monchaux:2012.grain.concentration.experiment.review}
{Monchaux}, R., {Bourgoin}, M., \& {Cartellier}, A. 2012, International Journal
  of Multiphase Flow, 40, 1

\bibitem[{Nakagawa {et~al.}(1986)Nakagawa, Sekiya, \&
  Hayashi}]{nakagawa:1986.grain.drift.solution}
Nakagawa, Y., Sekiya, M., \& Hayashi, C. 1986, Icarus, 67, 375

\bibitem[{{Olla}(2010)}]{olla:2010.grain.preferential.concentration.randomfield.notes}
{Olla}, P. 2010, Phys. Rev. E, 81, 016305

\bibitem[{{Ormel} \&
  {Cuzzi}(2007)}]{ormel:2007.closed.form.grain.rel.velocities}
{Ormel}, C.~W., \& {Cuzzi}, J.~N. 2007, \aap, 466, 413

\bibitem[{{Pan} \& {Padoan}(2010)}]{pan:2010.grain.velocity.sims}
{Pan}, L., \& {Padoan}, P. 2010, Journal of Fluid Mechanics, 661, 73

\bibitem[{{Pan} \& {Padoan}(2013)}]{pan:2013.grain.relative.velocity.calc}
---. 2013, \apj, in press, arXiv:1305.0307

\bibitem[{{Pan} {et~al.}(2011){Pan}, {Padoan}, {Scalo}, {Kritsuk}, \&
  {Norman}}]{pan:2011.grain.clustering.midstokes.sims}
{Pan}, L., {Padoan}, P., {Scalo}, J., {Kritsuk}, A.~G., \& {Norman}, M.~L.
  2011, \apj, 740, 6

\bibitem[{{Peebles}(1993)}]{peebles:1993.cosmology.textbook}
{Peebles}, P.~J.~E. 1993, {Principles of Physical Cosmology}, Vol. ISBN:
  978-0-691-01933-8 (Princeton University Press)

\bibitem[{{Price} {et~al.}(2011){Price}, {Federrath}, \&
  {Brunt}}]{price:2011.density.mach.vs.forcing}
{Price}, D.~J., {Federrath}, C., \& {Brunt}, C.~M. 2011, \apjl, 727, L21

\bibitem[{Rouson \& Eaton(2001)}]{rouson:2001.grain.concentration.experiment}
Rouson, D. W.~I., \& Eaton, J.~K. 2001, Journal of Fluid Mechanics, 428, 149

\bibitem[{{Schmidt} {et~al.}(2008){Schmidt}, {Federrath}, \&
  {Klessen}}]{schmidt:2008.turb.structure.fns}
{Schmidt}, W., {Federrath}, C., \& {Klessen}, R. 2008, Physical Review Letters,
  101, 194505

\bibitem[{{She} \& {Leveque}(1994)}]{sheleveque:structure.functions}
{She}, Z.-S., \& {Leveque}, E. 1994, Physical Review Letters, 72, 336

\bibitem[{{She} \& {Waymire}(1995)}]{shewaymire:logpoisson}
{She}, Z.-S., \& {Waymire}, E.~C. 1995, Physical Review Letters, 74, 262

\bibitem[{{She} \& {Zhang}(2009)}]{shezhang:2009.sheleveque.structfn.review}
{She}, Z.-S., \& {Zhang}, Z.-X. 2009, Acta Mechanica Sinica, 25, 279

\bibitem[{Sigurgeirsson \&
  Stuart(2002)}]{sigurgeirsson:2002.grain.markovian.concentration.toymodel}
Sigurgeirsson, H., \& Stuart, A.~M. 2002, Physics of Fluids, 14, 4352

\bibitem[{{Sorriso-Valvo} {et~al.}(1999){Sorriso-Valvo}, {Carbone}, {Veltri},
  {Consolini}, \&
  {Bruno}}]{sorriso-valvo:1999.solar.wind.intermittency.vs.time}
{Sorriso-Valvo}, L., {Carbone}, V., {Veltri}, P., {Consolini}, G., \& {Bruno},
  R. 1999, Geophysical Research Letters, 26, 1801

\bibitem[{Squires \&
  Eaton(1991)}]{squires:1991.grain.concentration.experiments}
Squires, K.~D., \& Eaton, J.~K. 1991, Physics of Fluids A: Fluid Dynamics, 3,
  1169

\bibitem[{Stewart {et~al.}(2006)Stewart, Strijbosch, Moors, \&
  Batenburg}]{stewart:2006.gamma.function.convolution.approximation}
Stewart, T., Strijbosch, L., Moors, J., \& Batenburg, P.~v. 2006, Tilburg
  University, Center for Economic Research, Discussion Paper,
  http://ideas.repec.org/p/dgr/kubcen/200627.html, 27

\bibitem[{{Voelk} {et~al.}(1980){Voelk}, {Jones}, {Morfill}, \&
  {Roeser}}]{voelk:1980.grain.relative.velocity.calc}
{Voelk}, H.~J., {Jones}, F.~C., {Morfill}, G.~E., \& {Roeser}, S. 1980, \aap,
  85, 316

\bibitem[{{Wilkinson} {et~al.}(2010){Wilkinson}, {Mehlig}, \&
  {Gustavsson}}]{wilkinson:2010.randomfield.correlation.grains.weak}
{Wilkinson}, M., {Mehlig}, B., \& {Gustavsson}, K. 2010, EPL (Europhysics
  Letters), 89, 50002

\bibitem[{Yoshimoto \&
  Goto(2007)}]{yoshimoto:2007.grain.clustering.selfsimilar.inertial.range}
Yoshimoto, H., \& Goto, S. 2007, Journal of Fluid Mechanics, 577, 275

\bibitem[{{Youdin} \&
  {Goodman}(2005)}]{youdin.goodman:2005.streaming.instability.derivation}
{Youdin}, A.~N., \& {Goodman}, J. 2005, \apj, 620, 459

\bibitem[{{Youdin} \& {Lithwick}(2007)}]{youdin:2007.turbulent.grain.stirring}
{Youdin}, A.~N., \& {Lithwick}, Y. 2007, Icarus, 192, 588

\bibitem[{{Zaichik} \&
  {Alipchenkov}(2009)}]{zaichik:2009.grain.clustering.theory.randomfield.review}
{Zaichik}, L.~I., \& {Alipchenkov}, V.~M. 2009, New Journal of Physics, 11,
  103018

\end{thebibliography}

\begin{appendix}

\section{An Exact Solution for the Response Function}
\label{sec:appendix:exact}

\subsection{General Case}
\label{sec:appendix:exact:general}

In the text we discuss the response of grains to a vortex with pure vorticity $1/\Teddy$ ($\delta u_\theta = r/\Teddy$, $\delta u_r = 0$). Here we derive this more exactly. 

Consider a vortex which appears for a time $\delta t = \Teddy$, so $\delta {\bf u}(t) = \delta {\bf u}\,\Theta(0<t<\Teddy)$. 
First note that, for this particular choice of $\delta {\bf u}$, we have $\delta u_{x} = \delta{\bf u}\cdot{\hat x}= -y/\Teddy$, $\delta u_{y} = +x/\Teddy$.\footnote{Here the sign of $\Teddy$ can be positive or negative, reflecting cyclonic or anti-cyclonic vortices, respectively.} In cartesian coordinates, the equations of motion (Eq.~\ref{eqn:eom.1}) in the vortex plane then become
\begin{align}
\delta \dot{v}_{x} &=  2\,\Omega_{R}\,\delta v_{y} - \tstop^{-1}\,(\delta v_{x} + \Teddy^{-1}\,y)\\
\delta \dot{v}_{y} &=  -\frac{1}{2}\,\Omega_{R}\,\delta v_{x} - \tstop^{-1}\,(\delta v_{y} - \Teddy^{-1}\,x)
\end{align}
Note that $\delta v_{x}=\dot{x}$ and $\delta v_{y}=\dot{y}$. For convenience, define the time in units of $\tstop$, so $x^{\prime} \equiv \tstop\,\dot{x}$ and so on. Now the equations become
\begin{align}
x^{\prime\prime} &= 2\,\taustop\,y^{\prime} - x^{\prime} - \taustopeddy\,y \\ 
y^{\prime\prime} &= -\frac{1}{2}\,\taustop\,x^{\prime} - y^{\prime} + \taustopeddy\,x
\end{align}
If we define the vector ${\bf x}\equiv (x,\,y,\,x^{\prime},\,y^{\prime})$, this is a linear system: 
\begin{align}
{\bf x}^{\prime} &= {\bf M}\,\cdot\,{\bf x} \\
{\bf M} &\equiv 
\left( \begin{matrix}
0 & 0 & 1 & 0 \\
0 & 0 & 0 & 1 \\
0 & -\taustopeddy & -1 & 2\,\taustop \\
\taustopeddy & 0 & -\frac{1}{2}\taustop & -1 
\end{matrix} \right)
\end{align}

The eigenvalues $\lambda_{i}$ of ${\bf M}$ solve the characteristic polynomial:
\begin{align}
\lambda_{i}^4 + 2\,\lambda_{i}^3 + (1+\taustop^{2})\,\lambda_{i}^2 - \frac{5}{2}\,\taustop\,\taustopeddy\,\lambda_{i} + \taustopeddy^{2} = 0
\end{align}

Since this is a simple system of linear ordinary differential equations, the solution ${\bf x}(t/\tstop)$ is given by 
\begin{align}
{\bf x}(t/\tstop) = {\bf V}\cdot [{\boldsymbol{\Lambda}} \cdot ({\bf V}^{-1}\,\cdot {\bf x}_{0})]
\end{align}
where ${\bf x}_{0}\equiv {\bf x}(t=0)$, ${\boldsymbol{\Lambda}}$ is the diagonal matrix of eigenfunctions 
\begin{align}
{\boldsymbol{\Lambda}}_{ij} = \delta_{ij}\,\exp{(-\lambda_{i}\,t/\tstop)}
\end{align}
and {\bf V} is the column matrix of eigenvectors ${\bf v}_{i}$ corresponding to each eigenvalue $\lambda_{i}$
\begin{align}
{\bf V} &\equiv 
\left( \begin{matrix}
{\bf v}_{1}  \\
...  \\
...  \\
... \end{matrix}  \right|
\left| \begin{matrix}
{\bf v}_{2}  \\
...  \\
...  \\
... \end{matrix} \right|
\left| \begin{matrix}
{\bf v}_{3}  \\
...  \\
...  \\
... \end{matrix} \right|
\left| \begin{matrix}
{\bf v}_{4}  \\
...  \\
...  \\
... \end{matrix} \right)
\end{align}

At $t=t_{1}=|\Teddy|$, the vortex is removed, so we have 
\begin{align}
{\bf x}_{1} &\equiv x(t=|\Teddy|) = {\bf V}\cdot [{\boldsymbol{\Lambda}_{1}} \cdot ({\bf V}^{-1}\,\cdot {\bf x}_{0})] \\ 
{\boldsymbol{\Lambda}}_{1,\,ij} &= \delta_{ij}\,\exp{(-\lambda_{i}\,|\Teddy|/\tstop)}
\end{align}
and the equations of motion become 
\begin{align}
\delta \dot{v}_{x} &=  2\,\Omega_{R}\,\delta v_{y} - \tstop^{-1}\,\delta v_{x} \\
\delta \dot{v}_{y} &=  -\frac{1}{2}\,\Omega_{R}\,\delta v_{x} - \tstop^{-1}\,\delta v_{y} 
\end{align}
It is straightforward to see that this has the solution $\delta v_{x} = a\,\exp{(-[1+\imath\,\taustop]\,(t-t_{1})/\tstop)} + b\,\exp{(-[1-\imath\,\taustop]\,(t-t_{1})/\tstop)}$, and a corresponding form for $\delta v_{y}$, where the constants are determined by matching to $\delta v_{x}(t=t_{1}) = \tstop^{-1}\,x^{\prime}(t=t_{1})$ and $\delta v_{y}(t=t_{1})=\tstop^{-1}\,y^{\prime}\,(t=t_{2})$. The velocities are exponentially damped (since $\delta {\bf u}(t>\Teddy)=0$, and the final positions $x_{f}$, $y_{f}$ are given by 
\begin{align}
x_{f} &= x(t_{1})+\int_{t_{1}}^{\infty}\delta v_{x}(t)\,{\rm d}t = x(t_{1}) + \frac{x^{\prime}(t_{1}) + 2\,\taustop\,y^{\prime}(t_{1})}{1+\taustop^{2}}  \\ 
y_{f} &= y(t_{1})+\int_{t_{1}}^{\infty}\delta v_{y}(t)\,{\rm d}t = y(t_{1}) + \frac{y^{\prime}(t_{1}) - \frac{1}{2}\,\taustop\,x^{\prime}(t_{1})}{1+\taustop^{2}} 
\end{align}
we can write this as a linear transform 
\begin{align}
{\bf x}_{f} &= {\bf P}\cdot{\bf x}_{1} = {\bf P}\cdot \{ {\bf V}\cdot [{\boldsymbol{\Lambda}_{1}} \cdot ({\bf V}^{-1}\,\cdot {\bf x}_{0})] \} \\
{\bf P} &\equiv 
\left( \begin{matrix}
1 & 0 & (1+\taustop^{2})^{-1} & 2\,\taustop\,(1+\taustop^{2})^{-1} \\
0 & 1 & -\frac{1}{2}\,\taustop\,(1+\taustop^{2})^{-1} & (1+\taustop^{2})^{-1} \\
0 & 0 & 0 & 0 \\
0 & 0 & 0 & 0 \\
\end{matrix} \right)
\end{align}
So the final coordinates are simply a linear transformation of the initial coordinates: 
\begin{align}
{\bf x}_{f} &= {\bf T}\cdot {\bf x}_{0} \\ 
{\bf T} &\equiv  {\bf P}\cdot {\bf V}\cdot {\boldsymbol{\Lambda}_{1}} \cdot {\bf V}^{-1}
\end{align}

Now, consider particles with negligible initial velocities $\delta {\bf v}(t=0)\approx0$ (since we wish to focus on the mean perturbation to their distribution); we will include the effects of non-zero initial velocity below. We can then write 
\begin{align}
\left( \begin{matrix}
x_{f} \\ 
y_{f} 
\end{matrix}  \right)
&= {\bf J}\cdot 
\left( \begin{matrix}
x_{0} \\ 
y_{0} 
\end{matrix}  \right) \\ 
{\bf J} &\equiv 
\left( \begin{matrix}
T_{1,1} & T_{1,2} \\ 
T_{2,1} & T_{2,2} 
\end{matrix}  \right)
\end{align}
i.e.\ the Jacobian matrix {\bf J} is just the upper-left block of the matrix {\bf T}. 
The mapping from $(x_{0},\,y_{0})\rightarrow (x_{f},\,y_{f})$ is just a linear coordinate transformation, so the density distribution $\rho$ of particles must obey 
\begin{align}
\rho_{f}\,{\rm d}x_{f}\,{\rm d}y_{f} &= \rho_{0}\,{\rm d}x_{0}\,{\rm d}y_{0} \\ 
\rho_{f} &= |{\rm Det}[J]|^{-1}\,\rho_{0}
\end{align}
so $f_{f}$ is simply related to $f_{0}$ by the determinant of {\bf J}. Since this is independent of position, we have 
\begin{align}
\langle \delta \ln \rho\rangle &= \delta\ln\rho = -\ln{|{\rm Det}[J]\, |} = -2\,\varpi \\
\varpi &\equiv \frac{1}{2}\,\ln{|{\rm Det}[J]\,|}
\label{eqn:exact}
\end{align}

In general, this must be evaluated numerically, since $\lambda_{i}$ has no closed-form general solution. However, we can exactly evaluate several limiting expressions. 

\vspace{-0.5cm}
\subsection{Large Eddies}
\label{sec:appendix:exact:large}

First, consider the limit of large eddies, $|\Teddy\Omega|\gg1$. While we can derive the limiting expression directly from the above, it is easier to start with our equations of motion, and linearize in terms of $\mathcal{O}(|\Teddy\,\Omega|)^{-1}\ll1$. It is easy to verify this admits a solution of the form $x\propto \exp{(\lambda_{|\Teddy\Omega|\gg1}[\taustop]\,t/|\Teddy|)}$, with eigenvalues 
\begin{align}
\lambda_{|\Teddy\Omega|\gg1} = \frac{5\,\taustop/4 \pm \sqrt{(3\,\taustop/4)^{2}-1}}{1+\taustop^{2}}\,{\rm SIGN}(\Teddy)
\end{align}
where ${\rm SIGN}(\Teddy) \equiv +1$ if $\Teddy>0$, which in these units means the eddy is cylonic (the sign of the vorticity is the same as that of the Keplerian flow), and ${\rm SIGN}(\Teddy) \equiv -1$ if $\Teddy<0$ refers to an anti-cyclonic eddy. For $\Teddy\Omega>0$ this is a growing mode, so the eddy seeds exponential dispersal of the particles (expansion of their Lagrangian radius), while for $\Teddy\Omega<0$ the eddy concentrates particles. 
Note that the second-derivative terms $\mathcal{O}(|\Teddy\,\Omega|)^{-2}$ drop here, so this becomes a first-order linear system of ODEs, and the governing matrices are now $2\times2$, with the form (written in the same notation as our general derivation above) 
\begin{align}
\left( \begin{matrix}
x^{\prime} \\ 
y^{\prime} 
\end{matrix}  \right)
&= {\bf M}\, 
\left( \begin{matrix}
x \\ 
y 
\end{matrix}  \right) \\ 
{\bf M} &\rightarrow \frac{\tstop}{\Teddy}\,\frac{1}{1+\taustop^{2}}\,
\left( \begin{matrix}
2\,\taustop & -1 \\ 
-1 & 2\,\taustop 
\end{matrix}  \right)
\end{align}

After some tedious but straightforward linear algebra, we obtain the Jacobian determinant: 
\begin{align}
\varpi(|\Teddy\,\Omega\gg1|) \rightarrow \frac{5}{4}\,{\rm SIGN}(\Teddy)\,\frac{\taustop}{1+\taustop^{2}}
\end{align}
We see that this is just the linear growth rate ($\lambda/\Teddy$) of the radius (the $\pm\sqrt{(3\,\taustop/4)^{2}-1}$ term in the eigenvalue represents motion of the particles on elliptical trajectories), times the eddy lifetime ($\Teddy$) as we should anticipate for a weak perturbation. Interestingly, this means when $|\Teddy\Omega|\gg1$, the perturbation becomes independent of $\Teddy$ and depends only on $\taustop$. The $5/4$ coefficient here comes from the fact that shear {enhances} the dispersal more strongly in the $x$ direction (the $+2\,\Omega$ term) than in the $y$ direction (the $-{1/2}\,\Omega$ term). Averaged over initial positions, the mean ``stretching'' coefficient for the ellipse along which particles in some initial ring are sheared goes as the mean of the absolute value of the two coefficients: $(1/2)\,(|2| + |-1/2|) = 5/4$.

\vspace{-0.5cm}
\subsection{Small Eddies}
\label{sec:appendix:exact:small}

Now consider small eddies, $|\Teddy\Omega\ll1|$. To leading order, we can now drop the shear terms in {\bf M}, and obtain 
\begin{align}
\lambda_{i} = -\frac{1}{2}\,{\Bigl(}1 \pm \sqrt{1 \pm \imath\,4\,\taustopeddy} {\Bigr)}
\end{align}
The eigenvalues come in two conjugate pairs, a growing pair and a decaying pair. The conjugation simply represents rotation, since the equations without shear are symmetric under rotation. Taking $r^{2}\propto x^{2} + y^{2}$ with $x\sim \sum\,a_{i}\,\exp{(\lambda_{i}\,t/\tstop)}$, it is straightforward to see that this translates to four radial modes: three are damped, one with decay rate (real part of the radial eigenvalue) $=1/\tstop$ and a conjugate pair with decay rate $=1/(2\,\tstop)$. And one is a growing mode, with growth rate $=[-2+\sqrt{2\,(1+\sqrt{1+16\,\taustop^{2}})}]/(4\,\tstop)$, the mode noted in the main text.\footnote{To show this, apply Euler's formula to the $\lambda_{i}$ values above, and use the identities $\cos{(\tan^{-1}{[x]}/2)} = 2^{-1/2}\,[1 + (1+x^{2})^{-1/2}]^{1/2}$ and  $\sin{(\tan^{-1}{[x]}/2)} = 2^{-1/2}\,x\,[1+x^{2}]^{-1/2}\,[1 + (1+x^{2})^{-1/2}]^{-1/2}$.}

With these eigenvalues, we obtain
\begin{align}
\varpi &\rightarrow -\frac{2+a_{+}+a_{-}}{4\,|\taustopeddy|} - \ln{2} - \frac{1}{4}\ln{(1+16\,\taustopeddy^{2})}  \\
\nonumber & + \frac{1}{2}\,\ln{{\Bigl[}-(1+2\,\imath\,\taustopeddy-a_{+}) + (1+2\,\imath\,\taustopeddy+a_{+})\,\exp{(a_{+}\,|\taustopeddy|^{-1})}{\Bigr]}} \\ 
\nonumber & + \frac{1}{2}\,\ln{{\Bigl[}-(1-2\,\imath\,\taustopeddy-a_{-}) + (1-2\,\imath\,\taustopeddy+a_{-})\,\exp{(a_{-}\,|\taustopeddy|^{-1})}{\Bigr]}} \\ 
a_{\pm} &\equiv \sqrt{1\pm \imath\,4\,\taustopeddy}
\end{align}
Here, $\varpi$ is identical under the transformation $\taustopeddy\rightarrow-\taustopeddy$ ($\Teddy\rightarrow-\Teddy$), so cyclonic and anti-cyclonic vortices behave identically. This should be obvious from the symmetry of the problem absent shear. Since $\varpi>0$ for all $\taustopeddy$, this means grains are always dispersed by small eddies, independent of the eddy orientation. 

The expression for $\varpi$ above is rather opaque; however it can be approximated to $\sim1\%$ accuracy at all $\taustopeddy$ by 
\begin{align}
\varpi &\approx |\taustopeddy|\,{\Bigl[}1 + \frac{|\taustopeddy|}{\ln{\sqrt{2}}}\,(1+|\taustopeddy|^{-1})^{-1/5} {\Bigr]}^{-1}
\end{align}
And the limits are easily evaluated: 
\begin{align}
\varpi &\rightarrow |\taustopeddy|\ \ \ \ \ \ \ \ \    (|\taustopeddy|\ll1) \\ 
\varpi &\rightarrow \ln{\sqrt{2}}\ \ \ \ \  (|\taustopeddy|\gg1)
\end{align}

\vspace{-0.5cm}
\section{Approximate Expressions for Large Grains}
\label{sec:appendix:largescale}

For large grains ($\taustop\gtrsim0.1$), and/or large scales ($\Teddy\gtrsim0.1\,\Omega^{-1}$) with appreciable levels of turbulence ($\alpha\gtrsim10^{-8}$), we can simplify our derivations considerably. Over the dynamic range where nearly all the power in density fluctuations is concentrated in Fig.~\ref{fig:response}, we see that the response function is approximately constant, with its asymptotic value $\varpi \approx (5/4)/(\taustop+\taustop^{-1})$. 
Over the same range, the $\beta$ term in $g(\Leddy)$ (Table~\ref{tbl:defns}) is almost always dominant in the function $g$ ($g(\Leddy)\approx \beta^{-2}$), so $h(\Leddy) \approx (\taustopeddy/\taustopeddy(\Lmax))\,(\Leddy/\Lmax)\,g(\Leddy)^{-1/2} \approx \beta\, |\Veddy|/|\Veddy(\Lmax)| = |\Veddy|/|\vdrift|$. Using $|\Veddy|/|\Veddy(\Lmax)| = (\Leddy/\Lmax)^{\velslope}$, we obtain
\begin{align}
\langle\deltarhonoabs\rangle \approx -\frac{5\,\Ndim/4}{\taustop + \taustop^{-1}}\,{\Bigl[}1 + \beta^{-1}\,{\Bigl(} \frac{\Leddy}{\Lmax} {\Bigr)}^{-\velslope} {\Bigr]}^{-1}\,{\rm SIGN}(\Teddy)
\end{align} 

With this approximation, $\Delta_{\ln{\rho}}=\deltadim\,\deltarho^{2}$ follows trivially. The integral quantities $\mu$ and $S_{\ln{\rho}}$ used to estimate the density distribution and $S_{\rho}$ can be evaluated exactly in closed form; these are presented in Table~\ref{tbl:largescale}. The integral $\rho_{\rm p,\,max}$ is straightforward to evaluate numerically, but the closed-form expression is rather unwieldy; using the fact that neither $\deltazero$ (Table~\ref{tbl:largescale}) or $\beta$ are extremely large ($\gtrsim100$), the exact integral can be approximated to very good accuracy by the function (from the logarithmic series expansion) given in Table~\ref{tbl:largescale}.


\end{appendix}

\end{document}